\documentclass[acmsmall, screen, nonacm]{acmart}
\makeatletter
\let\@authorsaddresses\@empty
\makeatother

\setcopyright{rightsretained}
\acmDOI{10.1145/3632868}
\acmYear{2024}
\copyrightyear{2024}
\acmSubmissionID{popl24main-p129-p}
\acmJournal{PACMPL}
\acmVolume{8}
\acmNumber{POPL}
\acmArticle{26}
\acmMonth{1}
\received{2023-07-11}
\received[accepted]{2023-11-07}

\usepackage{preamble}

\usepackage{todonotes}

\newcommand{\appref}[1]{\cref{#1}}

\begin{document}

%% Title information
\title[Asynchronous Probabilistic Couplings in Higher-Order Separation Logic]{Asynchronous Probabilistic Couplings in \\ Higher-Order Separation Logic}
% \title[Short Title]{Full Title}         %% [Short Title] is optional;
%                                         %% when present, will be used in
%                                         %% header instead of Full Title.
% \titlenote{with title note}             %% \titlenote is optional;
%                                         %% can be repeated if necessary;
%                                         %% contents suppressed with 'anonymous'
% \subtitle{Subtitle}                     %% \subtitle is optional
% \subtitlenote{with subtitle note}       %% \subtitlenote is optional;
%                                         %% can be repeated if necessary;
%                                         %% contents suppressed with 'anonymous'

%% Author information
%% Contents and number of authors suppressed with 'anonymous'.
%% Each author should be introduced by \author, followed by
%% \authornote (optional), \orcid (optional), \affiliation, and
%% \email.
%% An author may have multiple affiliations and/or emails; repeat the
%% appropriate command.
%% Many elements are not rendered, but should be provided for metadata
%% extraction tools.

%% Author with single affiliation.
\author{Simon Oddershede Gregersen}
%\authornote{with author1 note}          %% \authornote is optional; %% can be repeated if necessary
\orcid{0000-0001-6045-5232}
\affiliation{
  \institution{Aarhus University}            %% \institution is required
  \country{Denmark}                    %% \country is recommended
}
\email{gregersen@cs.au.dk}          %% \email is recommended

\author{Alejandro Aguirre} 
\email{alejandro@cs.au.dk}
\orcid{0000-0001-6746-2734}
\affiliation{
  \institution{Aarhus University}            %% \institution is required
  \country{Denmark}                    %% \country is recommended
}
\email{alejandro@cs.au.dk}

\author{Philipp~G. Haselwarter}
\orcid{0000-0003-0198-7751}
\affiliation{
  \institution{Aarhus University}            %% \institution is required
  \country{Denmark}                    %% \country is recommended
}
\email{pgh@cs.au.dk}

\author{Joseph Tassarotti}
\orcid{0000-0001-5692-3347}
\affiliation{
  \institution{New York University}            %% \institution is required
  \country{USA}                    %% \country is recommended
}
\email{jt4767@cs.nyu.edu}

\author{Lars Birkedal}
\email{birkedal@cs.au.dk}
\orcid{0000-0003-1320-0098}
\affiliation{
  \institution{Aarhus University}
  \country{Denmark}
}
\email{birke@cs.au.dk}

\begin{abstract}

  Probabilistic couplings are the foundation for many probabilistic relational program logics and arise when relating random sampling statements across two programs.
  In relational program logics, this manifests as dedicated coupling rules that, \eg{}, say we may reason as if two sampling statements return the same value.
  However, this approach fundamentally requires aligning or ``synchronizing'' the sampling statements of the two programs which is not always possible.

  In this paper, we develop \thereflog{}, a higher-order probabilistic relational separation logic that addresses this issue by supporting \emph{asynchronous} probabilistic couplings.
  We use \thereflog{} to develop a logical step-indexed logical relation to reason about contextual refinement and equivalence of higher-order programs written in a rich language with a probabilistic choice operator, higher-order local state, and impredicative polymorphism.
  Finally, we demonstrate our approach on a number of case studies.

  All the results that appear in the paper have been formalized in the Coq proof assistant using the Coquelicot library and the Iris separation logic framework.
\end{abstract}



%% Keywords
%% comma separated list
  %% \keywords are mandatory in final camera-ready submission

%% \maketitle
%% Note: \maketitle command must come after title commands, author
%% commands, abstract environment, Computing Classification System
%% environment and commands, and keywords command.
\maketitle

%key-ideas.tex

%LaTeX code re-use could be better here; a lot of these are copy-pasted in the general figures.

\newcommand{\relcouplerands}{
  \inferH{rel-couple-rands}
  {
  f~\text{bijection} \\
  \All n \leq \tapebound . \refines{\Delta}[\mask]{\fillctx\lctx[n]}{\fillctx\lctx'[f(n)]}{\type}
  }
  {\refines{\Delta}[\mask]{\fillctx\lctx[\Rand(\tapebound)]}{\fillctx\lctx'[\Rand(\tapebound)]}{\type}}
}

\newcommand{\relalloctapel}{
  \inferH{rel-alloc-tape-l}
  { \All \lbl . \progtape{\lbl}{\tapebound}{\nil} \wand \refines{\Delta}[\mask]{\fillctx\lctx[\lbl]}{\expr}{\type}}
  {\refines{\Delta}[\mask]{\fillctx\lctx[\AllocTape(\tapebound)]}{\expr}{\type}}
}

\newcommand{\relcoupletapel}{
  \inferH{rel-couple-tape-l}
  {
    f~\text{bijection} \\
    \expr \not\in \Val \\
    \progtape{\lbl}{\tapebound}{\tape} \\
    \All n \leq \tapebound . \progtape{\lbl}{\tapebound}{\tape \cons n} \wand \refines{\Delta}[\mask]{\expr}{\fillctx\lctx'[f(n)]}{\type}
  }
  {\refines{\Delta}[\mask]{\expr}{\fillctx\lctx'[\Rand(\tapebound)]}{\type}}
}

\newcommand{\relrandtapel}{
  \inferH{rel-rand-tape-l}
  {
    \progtape{\lbl}{\tapebound}{n \cons \tape} \\
    \progtape{\lbl}{\tapebound}{\tape} \wand \refines{\Delta}[\mask]{\fillctx\lctx[n]}{\expr_2}{\type}
  }
  {\refines{\Delta}[\mask]{\fillctx\lctx[\Rand(\tapebound, \lbl)]}{\expr_2}{\type}}
}

\newcommand{\relranderaser}{
  \inferH{rel-rand-erase-r}
  { \spectape{\lbl}{\tapebound}{\nil} \\
    \All n \leq \tapebound . \refines{\Delta}[\mask]{\fillctx\lctx[n]}{\fillctx\lctx'[n]}{\type}
  }
  {\refines{\Delta}[\mask]{\fillctx\lctx[\Rand(\tapebound)]}{\fillctx\lctx'[\Rand(\tapebound, \lbl)]}{\type}}
}

\newcommand{\relranderasel}{
  \inferH{rel-rand-erase-l}
  { \progtape{\lbl}{\tapebound}{\nil} \\
    \All n \leq \tapebound . \refines{\Delta}[\mask]{\fillctx\lctx[n]}{\fillctx\lctx'[n]}{\type}
  }
  {\refines{\Delta}[\mask]{\fillctx\lctx[\Rand(\tapebound, \lbl)]}{\fillctx\lctx'[\Rand(\tapebound)]}{\type}}
}

%model.tex
\newcommand{\figstrucrules}{
\begin{figure*}
  
  \centering
  \begin{mathpar}
    \inferH{wp-wand}
    {\All \val . \Phi(\val) \wand \Psi(\val) \\
      \wpre{\expr}{\Phi}}
    {\wpre{\expr}{\Psi}}
    \and
    \inferH{wp-bind}
    {\wpre{\expr}{\val . \wpre{\fillctx\lctx[\val]}{\Phi}}}
    {\wpre{\fillctx\lctx[\expr]}{\Phi}}
    \and
    \inferH{wp-load}
    {\progheap{\loc}{\val} \\
      \progheap{\loc}{\val} \wand \Phi(\val)}
    {\wpre{\deref\loc}{\Phi}}
    \and
    \inferH{wp-couple-rands}
    { f~\text{bijection} \\
      \specCtx \\
      \spec(\Rand(\tapebound)) \\
      \All n \leq \tapebound . \spec(f(n)) \wand \Phi(n)}
    {\wpre{\Rand(\tapebound)}{\Phi}}
    \and
    \inferH{wp-couple-tape-l}
    { f~\text{bijection} \\
      \expr \not\in \Val \\
      \specCtx \\
      \spec(\Rand(\tapebound)) \\
      \progtape{\lbl}{\tapebound}{\tape} \\
      \All n \leq \tapebound . \left(\spec(f(n)) \sep \progtape{\lbl}{\tapebound}{\tape \cons n}\right) \wand \wpre{\expr}{\Phi}}
    {\wpre{\expr}{\Phi}}
  \end{mathpar}
  \caption{Selected structural rules of the weakest preconditon.\label{fig:struc-rules}}
\end{figure*}
}

%rellog.tex
\newcommand{\figreflogrules}{
\begin{figure*}[t]
  
  \centering
  \begin{mathpar}
    \inferH{rel-pure-l}
    {
      \expr_1 \purestep \expr_1' \\
      \refines{\Delta}[\mask]{\fillctx\lctx[\expr_1']}{\expr_2}{\type}
    }
    {\refines{\Delta}[\mask]{\fillctx\lctx[\expr_1]}{\expr_2}{\type}}
    \and
    \inferH{rel-pure-r}
    {
      \expr_2 \purestep \expr_2' \\
      \refines{\Delta}[\mask]{\expr_1}{\fillctx\lctx[\expr_2']}{\type}
    }
    {\refines{\Delta}[\mask]{\expr_1}{\fillctx\lctx[\expr_2]}{\type}}
    \and
    \inferH{rel-alloc-l}
    {
      \All \loc . \progheap{\loc}{\val} \wand \refines{\Delta}[\mask]{\fillctx\lctx[\loc]}{\expr_2}{\type}
    }
    {\refines{\Delta}[\mask]{\fillctx\lctx[\hspace{-1pt}\Alloc(\val)]}{\expr_2}{\type}}
    \and
    \inferH{rel-alloc-r}
    {
      \All \loc . \specheap{\loc}{\val} \wand \refines{\Delta}[\mask]{\expr_1}{\fillctx\lctx[\loc]}{\type}
    }
    {\refines{\Delta}[\mask]{\expr_1}{\fillctx\lctx[\hspace{-1pt}\Alloc(\val)]}{\type}}
    \and
    \inferH{rel-load-l}
    {
      \progheap{\loc}{\val} \\
      \progheap{\loc}{\val} \wand \refines{\Delta}[\mask]{\fillctx\lctx[\val]}{\expr_2}{\type}
    }
    {\refines{\Delta}[\mask]{\fillctx\lctx[! \loc]}{\expr_2}{\type}} % spacing becomes a bit funny with \deref here
    \and
    \inferH{rel-load-r}
    {
      \specheap{\loc}{\val} \\
      \specheap{\loc}{\val} \wand \refines{\Delta}[\mask]{\expr_1}{\fillctx\lctx[\val]}{\type}
    }
    {\refines{\Delta}[\mask]{\expr_1}{\fillctx\lctx[! \loc]}{\type}} % spacing becomes a bit funny with \deref here
    \and
    \inferH{rel-store-l}
    {
      \progheap{\loc}{\val} \\
      \progheap{\loc}{\valB} \wand \refines{\Delta}[\mask]{\fillctx\lctx[\TT]}{\expr_2}{\type}
    }
    {\refines{\Delta}[\mask]{\fillctx\lctx[\loc \gets \valB]}{\expr_2}{\type}}
    \and
    \inferH{rel-store-r}
    {
      \specheap{\loc}{\val} \\
      \specheap{\loc}{\valB} \wand \refines{\Delta}[\mask]{\expr_1}{\fillctx\lctx[\TT]}{\type}
    }
    {\refines{\Delta}[\mask]{\expr_1}{\fillctx\lctx[\loc \gets \valB]}{\type}}
    \and
    \inferH{rel-pack}
    {
      \All \val_1, \val_2 . \persistent{R(\val_1, \val_2)} \\
      \refines{\Delta, \alpha \mapsto R}[\top]{\expr_1}{\expr_2}{\type}
    }
    {\refines{\Delta}[\top]{\Pack \expr_1}{\Pack \expr_2}{\Exists \alpha . \type}}
    \and
    \inferH{rel-rec}
    {
      \always\left(\All \val_1, \val_2 . \semInterp{\type}{\Delta}{\val_1, \val_2} \wand
      \refines{\Delta}[\top]{(\Rec {f_1} \var_1 = \expr_1)\, \val_1}{(\Rec {f_2} \var_2 = \expr_2)\, \val_2}{\type \to \typeB}\right)
    }
    {\refines{\Delta}[\top]{\Rec {f_1} \var_1 = \expr_1}{\Rec {f_2} \var_2 = \expr_2}{\type \to \typeB}}
    \and
    \inferH{rel-return}
    {
      \semInterp{\type}{\Delta}{\val_1,\val_2}
    }
    {\refines{\Delta}[\top]{\val_1}{\val_2}{\type}}
    \and
    \inferH{rel-bind}
    {
      \refines{\Delta}[\mask]{\expr_1}{\expr_2}{\type} \\
      \All \val_1, \val_2 . \semInterp{\type}{\Delta}{\val_1, \val_2} \wand
      {\refines{\Delta}[\top]{\fillctx\lctx[\val_1]}{\fillctx\lctx'[\val_2]}{\typeB}}
    }
    {\refines{\Delta}[\mask]{\fillctx\lctx[\expr_1]}{\fillctx\lctx'[\expr_2]}{\typeB}}
  \end{mathpar}
  \caption{Selected structural and symbolic execution rules for the \thereflog{} refinement judgment.}
  \label{fig:reflog-rules}
\end{figure*}
}

\newcommand{\figreflogtaperules}{
\begin{figure}
  
  \centering
  \begin{mathpar}
    \inferH{rel-rand-l}
    {
      \All n \leq \tapebound . \refines{\Delta}[\mask]{\fillctx\lctx[n]}{\expr_2}{\type}
    }
    {\refines{\Delta}[\mask]{\fillctx\lctx[\Rand(\tapebound)]}{\expr_2}{\type}}
    \and
    \inferH{rel-rand-r}
    { \expr_1 \not\in \Val \\
      \All n \leq \tapebound. \refines{\Delta}[\mask]{\expr_1}{\fillctx\lctx[b]}{\type}
    }
    {\refines{\Delta}[\mask]{\expr_1}{\fillctx\lctx[\Rand(\tapebound)]}{\type}}
    \and
    \inferH{rel-alloc-tape-l}
    { \All \lbl . \progtape{\lbl}{\tapebound}{\nil} \wand \refines{\Delta}{\fillctx\lctx[\lbl]}{\expr}{\type} }
    {\refines{\Delta}{\fillctx\lctx[\AllocTape(\tapebound)]}{\expr}{\type}}
    \and
    \inferH{rel-alloc-tape-r}
    { \All \lbl . \spectape{\lbl}{\tapebound}{\nil} \wand \refines{\Delta}{\expr}{\fillctx\lctx[\lbl]}{\type}}
    {\refines{\Delta}{\expr}{\fillctx\lctx[\AllocTape(\tapebound)]}{\type}}
    \and
    \inferH{rel-rand-tape-l}
    {
      \progtape{\lbl}{\tapebound}{n \cons \tape} \\
      \progtape{\lbl}{\tapebound}{\tape} \wand \refines{\Delta}[\mask]{\fillctx\lctx[n]}{\expr_2}{\type}
    }
    {\refines{\Delta}[\mask]{\fillctx\lctx[\Rand(\tapebound, \lbl)]}{\expr_2}{\type}}
    \and
    \inferH{rel-rand-tape-r}
    {
      \spectape{\lbl}{\tapebound}{n \cons \tape} \\
      \spectape{\lbl}{\tapebound}{\tape} \wand \refines{\Delta}[\mask]{\expr_1}{\fillctx\lctx[n]}{\type}
    }
    {\refines{\Delta}[\mask]{\expr_1}{\fillctx\lctx[\Rand(\tapebound, \lbl)]}{\type}}
    \and
    \inferH{rel-rand-tape-empty-l}
    {
      \progtape{\lbl}{\tapebound}{\nil} \\
      \All n \leq \tapebound . \progtape{\lbl}{\tapebound}{\nil} \wand \refines{\Delta}[\mask]{\fillctx\lctx[n]}{\expr_2}{\type}
    }
    {\refines{\Delta}[\mask]{\fillctx\lctx[\Rand(\tapebound, \lbl)]}{\expr_2}{\type}}
    \and
    \inferH{rel-rand-tape-empty-r}
    { \expr_1 \not\in \Val \\
      \spectape{\lbl}{\tapebound}{\nil} \\
      \All n \leq \tapebound . \spectape{\lbl}{\tapebound}{\nil} \wand \refines{\Delta}[\mask]{\expr_1}{\fillctx\lctx[n]}{\type}
    }
    {\refines{\Delta}[\mask]{\expr_1}{\fillctx\lctx[\Rand(\tapebound, \lbl)]}{\type}}
  \end{mathpar}
  \caption{Rules for non-relational probabilistic choices and tapes for the \thereflog{} refinement judgment.}
  \label{fig:reflog-tape-rules}
\end{figure}
}

\newcommand{\figreflogcouplingrules}{
\begin{figure}
  
  \centering
  \begin{mathpar}
    \relcouplerands{}
    \and
    \inferH{rel-couple-tape-l}
    {
      f~\text{bijection} \\
      \expr_1 \not\in \Val \\
      \progtape{\lbl}{\tapebound}{\tape} \\
      \All n \leq \tapebound . \progtape{\lbl}{\tapebound}{\tape \cons n} \wand \refines{\Delta}[\mask]{\expr_1}{\fillctx\lctx[f(n)]}{\type}
    }
    {\refines{\Delta}[\mask]{\expr_1}{\fillctx\lctx[\Rand(\tapebound)]}{\type}}
    \and
    \inferH{rel-couple-tape-r}
    {
      f~\text{bijection} \\
      \spectape{\lbl}{\tapebound}{\tape} \\
      \All n \leq \tapebound . \spectape{\lbl}{\tapebound}{\tape \cons f(n)} \wand \refines{\Delta}[\mask]{\fillctx\lctx[n]}{\expr_2}{\type}
    }
    {\refines{\Delta}[\mask]{\fillctx\lctx[\Rand(\tapebound)]}{\expr_2}{\type}}
    \and
    \inferH{rel-couple-tapes}
    {
      f~\text{bijection} \\
      \expr_1 \not\in \Val \\
      \progtape{\lbl}{\tapebound}{\tape} \\
      \spectape{\lbl'}{\tapebound}{\tape'} \\
      \All n \leq \tapebound . \progtape{\lbl}{\tapebound}{\tape \cons n} \sep \spectape{\lbl'}{\tapebound}{\tape' \cons f(n)} \wand
      \refines{\Delta}[\mask]{\expr_1}{\expr_2}{\type}
    }
    {\refines{\Delta}[\mask]{\expr_1}{\expr_2}{\type}}
  \end{mathpar}
  \caption{Coupling rules for the \thereflog{} refinement judgment.}
  \label{fig:reflog-coupling-rules}
\end{figure}
}

\newcommand{\figrefloginvariantrules}{
\begin{figure}[t]
  
  \centering
  \begin{mathpar}
    \inferH{rel-na-inv-open}
    {
      \namesp \in \mask \\
      \nInv{\namesp}{\prop} \\
      \later \prop \sep \nClose{\namesp}{\mask}{\prop} \wand \refines{\Delta}[\mask \setminus \namesp]{\expr_1}{\expr_2}{\type}
    }
    {\refines{\Delta}[\mask]{\expr_1}{\expr_2}{\type}}
    \and
    \inferH{rel-na-inv-close}
    {
      \later \prop \\
      \nClose{\namesp}{\mask}{\prop} \\
      \refines{\Delta}[\mask]{\expr_1}{\expr_2}{\type}
    }
    {\refines{\Delta}[\mask \setminus \namesp]{\expr_1}{\expr_2}{\type}}
    \and
    \inferH{rel-na-inv-alloc}
    {
      \later \prop \\
      \nInv{\namesp}{\prop} \wand \refines{\Delta}[\mask]{\expr_1}{\expr_2}{\type}
    }
    {\refines{\Delta}[\mask]{\expr_1}{\expr_2}{\type}}
  \end{mathpar}
  \caption{Non-atomic invariant access rules for the \thereflog{} refinement judgment.}
  \label{fig:reflog-invariant-rules}
\end{figure}
}

%%% Local Variables:
%%% mode: latex
%%% TeX-master: "clutch-popl24"
%%% End:

\section{Introduction}
\label{sec:intro}

Relational reasoning is a useful technique for proving properties of probabilistic programs.
By relating a complex probabilistic program to a simpler one, we can often reduce a challenging verification task to an easier one.
In addition, certain important properties of probabilistic programs are naturally expressed in a relational form, such as stability of machine learning algorithms \cite{10.1162/153244302760200704}, differential privacy \cite{dwork_algorithmic_2013}, and provable security \cite{DBLP:journals/jcss/GoldwasserM84}.
Consequently, a number of relational program logics and models have been developed for probabilistic programs, \eg{}, pRHL \cite{DBLP:conf/lpar/BartheEGHSS15}, approximate pRHL \cite{DBLP:conf/popl/BartheKOB12, DBLP:conf/ccs/BartheFGGHS16, DBLP:conf/lics/BartheGGHS16}, EpRHL \cite{DBLP:journals/pacmpl/BartheEGHS18}, HO-RPL \cite{DBLP:journals/pacmpl/AguirreBGGKS21}, Polaris \cite{DBLP:journals/pacmpl/TassarottiH19}, logical relations \cite{DBLP:conf/fossacs/BizjakB15, DBLP:journals/pacmpl/WandCGC18, DBLP:conf/lics/JohannSV10}, and differential logical relations \cite{DBLP:journals/pacmpl/LagoG22}.

Many probabilistic relational program logics make use of \emph{probabilistic couplings} \cite{thorisson/2000, lindvall_lectures_2002, Villani2008OptimalTO}, a mathematical tool for reasoning about pairs of probabilistic processes.
Informally, couplings correlate outputs of two processes by specifying how corresponding sampling statements are correlated.
To understand how couplings work in such logics, let us consider a pRHL-like logic.
In pRHL and its variants, we prove Hoare \emph{quadruples} of the form $\hoare{P}{e_1 \sim e_2}{Q}$, where $e_1$ and $e_2$ are two probabilistic programs,
and $P$ and $Q$ are pre and post-\emph{relations} on states of the two programs.
Couplings arise when reasoning about random sampling statements in the two programs, such as in the following rule:
\begin{mathpar}
  \inferH{prhl-couple}
  { }
  {\hoare{P[v/x_{1}, v/x_{2}]}{x_{1} \xleftarrow{\$} d \sim x_{2} \xleftarrow{\$} d}{P}}
\end{mathpar}
Here, the two programs both sample from the same distribution $d$ and store the result in variable $x_1$ and $x_2$, respectively. 
The rule says that we may reason \emph{as if} the two sampling statements return the same value $v$ in both programs, and one says
that the sample statements have been ``coupled''.
This is a powerful method that integrates well with existing reasoning principles from relational program logics.
However, this kind of coupling rules require aligning or ``synchronizing'' the sampling statements of the two programs: both programs have to be executing the sample statements we want to couple for their next step when applying the rule.
To enable this alignment, pRHL has various rules that enable taking steps on one side of the quadruple at a time or commuting statements in a (first-order) program.
Nevertheless, with the rules from existing probabilistic relational logics, it is not always possible to synchronize sampling statements.

For example, consider the following program written in an ML-like language that \emph{eagerly} performs a probabilistic coin flip and returns the result in a thunk:
\newcommand{\eagerprog}{\mathit{eager}}
\newcommand{\lazyprog}{\mathit{lazy}}
\begin{align*}
  \eagerprog \eqdef{}
  \Let b = \Flip () in
  \Lam \_ . b
\end{align*}
An indistinguishable---but \emph{lazy}---version of the program only does the coin flip when the thunk is invoked for the first time but stores the result in a reference that is read from in future invocations:
\begin{align*}
  \lazyprog \eqdef{}
  &\Let r = \Alloc(\None) in \\
  &\Lam \_ .
    \MatchML \deref r with
      \Some(b) => b
    | \None => {\begin{array}[t]{l}
      \Let b = \Flip \TT in \\
      r \gets \Some(b); \\
      b
    \end{array}}
    end {}
\end{align*}
The usual symbolic execution rules of relational logics will allow us to progress the two sides independently according to the program execution, but they will \emph{not} allow us to line up the $\Flip()$ expression in $\eagerprog$ with that in $\lazyprog$.
Consequently, the coupling rule \ruleref{prhl-couple} cannot be applied.
Intuitively, the $\Flip()$ expression in $\eagerprog$ is evaluated immediately but the $\Flip()$ expression in $\lazyprog$ only gets evaluated when the thunk is invoked---to relate the two thunks one is forced to first evaluate the eager sampling, but this then makes it impossible to couple it with the lazy sampling.

While the example may seem contrived, these kinds of transformations of eager and lazy sampling are widely used, \eg{}, in proofs in the Random Oracle Model \cite{DBLP:conf/ccs/BellareR93} and in game playing proofs \cite{BellareR04,BellareR06}.
For this reason, systems like EasyCrypt \cite{DBLP:conf/fosad/BartheDGKSS13} and CertiCrypt \cite{DBLP:conf/itp/BartheGB10, DBLP:conf/popl/BartheGB09} support reasoning about lazy/eager sampling through special-purpose rules for swapping statements that allows alignment of samplings; the approach is shown to work for a first-order language with global state and relies on syntactic criteria and assertions on memory disjointness.
However, in rich enough languages (\eg{} with general references and closures) these kinds of swapping-equivalences are themselves highly non-trivial, even in the non-probabilistic case \cite{awkward:pitts, DBLP:journals/jfp/DreyerNB12}.

In this paper we develop \emph{\thereflog{}}, a higher-order probabilistic relational separation logic that addresses this issue by enabling \emph{asynchronous} probabilistic couplings.
To do so, \thereflog{} introduces a novel kind of ghost state, called \emph{presampling tapes}.
Presampling tapes let us reason about sampling statements as if they executed ahead of time and stored their results for later use.
This converts the usual alignment problem of coupling rules into the task of reasoning about this special form of state.
Fortunately, reasoning about state is well-addressed with modern separation logics.

\thereflog{} provides a ``logical'' step-indexed logical relation \cite{DBLP:journals/corr/abs-1103-0510} to reason about \emph{contextual refinement and equivalence} of probabilistic higher-order programs written in $\thelang{}$, a rich language with a probabilistic choice operator, higher-order local state, recursive types, and impredicative polymorphism.
Intuitively, expressions $\expr_{1}$ and $\expr_{2}$ of type $\tau$ are contextually equivalent if no well-typed context $\ctx$ can distinguish them, \ie{}, if the expression $\fillctx\ctx[\expr_{1}]$ has the same observable behaviors as $\fillctx\ctx[\expr_{2}]$.
Contextual equivalence can be decomposed into contextual refinement: we say $\expr_{1}$ refines $\expr_{2}$ at type $\type$, written $\ctxrefinesS{\expr_{1}}{\expr_{2}}{\type}$, if, for all contexts $\ctx$ expecting something of type $\type$, if $\fillctx\ctx[\expr_{1}]$ has some observable behavior, then so does $\fillctx\ctx[\expr_{2}]$.
As our language is probabilistic, here ``observable behavior'' means the \emph{probability} of observing an outcome, such as termination.
Using the \emph{logical~approach}~\cite{iris-logrel-journal}, in \thereflog{}, types are interpreted as relations expressed in separation logic.
The resulting model allows us to prove, among other examples, that the $\eagerprog$ program above is contextually equivalent to the $\lazyprog$ program.

The work presented in this paper is \emph{foundational} \cite{DBLP:conf/lics/Appel01} in the sense that all results, including the semantics, the logic, the necessary mathematical analysis results, the relational model, and all the examples are formalized\footnote{\href{https://github.com/logsem/clutch}{https://github.com/logsem/clutch}} in the Coq proof assistant \cite{coq} using the Coquelicot library \cite{DBLP:journals/mics/BoldoLM15} and the Iris separation logic framework \cite{iris, iris2, iris3, irisjournal}.

In summary, we make the following contributions:
\begin{itemize}
\item A higher-order probabilistic relational separation logic, \thereflog{}, for reasoning about probabilistic programs written in \thelang{}, an ML-like programming language with higher-order local state, recursive types, and impredicative polymorphism.
\item A proof method for relating asynchronous probabilistic samplings in a program logic; a methodology that allows us to reason about sampling as if it were state and to exploit existing separation logic mechanisms such as \emph{ghost state} and \emph{invariants} to reason about probabilistic programs.
  We demonstrate the usefulness of the approach with a number of case studies.
% \item The first \emph{logical} step-indexed logical relation for a language with probabilistic choice.
\item The first coupling-based relational program logic to reason about contextual refinement and equivalence of programs in a higher-order language with local state, recursive types, and impredicative polymorphism.
\item Novel technical ideas, namely, \emph{left-partial couplings}, a \emph{coupling modality}, and an \emph{erasure} argument, that allow us to prove soundness of the relational logic.
\item Full mechanization in Coq using Coquelicot and the Iris separation logic framework.
\end{itemize}

%%% Local Variables:
%%% mode: latex
%%% TeX-master: "../clutch-popl24"
%%% End:

\section{Key Ideas}
\label{sec:key-ideas}
The key conceptual novelties of the \thereflog{} logic are twofold: a \emph{logical probabilistic refinement judgment} and a novel kind of ghost resource, called \emph{presampling tapes}.

\paragraph{Logical refinement.} The refinement judgment $\refines{\Delta}[\mask]{\expr_{1}}{\expr_{2}}{\type}$ should be read as ``the expression $\expr_{1}$ refines the expression $\expr_{2}$ at type $\type$'' and it satisfies a range of structural and symbolic execution rules as showcased in \cref{fig:reflog-rules} and further explained in \cref{sec:rellog}.
Just like contextual refinement, the judgment is indexed by a type $\type$---the environment $\Delta$ assigns semantic interpretations to type variables in $\type$ and $\mask$ is an \emph{invariant mask} as elaborated on in \cref{sec:rellog}.
Both are safely ignored in this section.
The meaning of the judgment is formally reflected by the following soundness theorem.
\begin{theorem}[Soundness]\label{thm:soundness}
  If $\refines{\emptyset}{\expr_{1}}{\expr_{2}}{\type}$ is derivable in \thereflog{} then $\ctxrefinesS{\expr_{1}}{\expr_{2}}{\type}$.
\end{theorem}

The refinement judgment is \emph{internal} to the ambient \thereflog{} separation logic.
This means that we can combine the judgment in arbitrary ways with other logical connectives: \eg{}, the \emph{separating conjunction} $\prop \sep \propB$ and its adjoint \emph{separating implication} (magic wand) $\prop \wand \propB$.
All inference rules that we present can be internalized as propositions in the logic and we will use an inference rule with premises $\prop_{1}, \ldots, \prop_{n}$ and conclusion $\propB$ as notation for $(\prop_{1} \sep \ldots \sep \prop_{n}) \proves Q$.

The language \thelang{} contains a single probabilistic primitive $\Rand(\tapebound)$ that reduces uniformly at random to some $n \in \{0,1,\ldots,\tapebound\}$:
\begin{align*}
  \Rand(\tapebound), \state &\step^{1/(\tapebound+1)} n, \state && n \in \{0, 1, \ldots, \tapebound\}
\end{align*}
where $\state$ is the current program state and $\step \subseteq \Conf \times [0,1] \times \Conf$ is a small-step transition relation, annotated with the probability that the transition occurs.
By defining $\Flip() \eqdef{} \If \Rand(1) = 0 then \False \Else \True $ we recover the Boolean fair coin flip operator used in the motivating example.
To reason relationally about probabilistic choices that \emph{can} be synchronized, \thereflog{} admits a classical coupling rule that allows us to continue reasoning as if the two sampled values are related by a bijection $f$ on the sampling space $\{0, \dots, \tapebound\}$:
\begin{align*}
  \relcouplerands{}
\end{align*}
where $\lctx$ and $\lctx'$ are arbitrary evaluation contexts.

\paragraph{Asynchronous couplings.} To support \emph{asynchronous} couplings we introduce \emph{presampling tapes}.
Reminiscent of how \emph{prophecy variables} \cite{DBLP:conf/lics/AbadiL88, DBLP:journals/tcs/AbadiL91, DBLP:journals/pacmpl/JungLPRTDJ20} allow us to talk about the future, presampling tapes give us the means to talk about the outcome of probabilistic choices \emph{in the future}.\footnote{As showcased in \cref{sec:counterexamples}, however, prophecy variables as previously developed in Iris are unsound for the coupling- logic.}
Tapes manifest both in the operational semantics and in the logic.

Operationally, a tape consists of an upper bound $\tapebound \in \nat$ and a finite sequence of natural numbers less than or equal to $\tapebound$, representing future outcomes of $\Rand(\tapebound)$ commands.
Each tape is labeled with an identifier $\lbl \in \Lbl$, and a program's state is extended with a finite map from labels to tapes.
Tapes can be dynamically allocated using a $\AllocTape$ primitive:
\begin{align*}
  \AllocTape(\tapebound), \state &\step^{1} \lbl, \state[\lbl \mapsto (\tapebound, \nil)] && \text{if}~\lbl = \fresh(\sigma)
\end{align*}
which extends the mapping with an empty tape and the upper bound $\tapebound$, and it returns its fresh label $\lbl$.
The $\Rand$ primitive can then optionally be annotated with a tape label $\lbl$.
If $\state(\lbl) = (\tapebound, \nil)$, \ie{}, the corresponding tape is empty, $\Rand(\tapebound, \lbl)$ reduces to any $n \leq \tapebound$ with equal probability:
\begin{align*}
  \Rand(\tapebound, \lbl), \state &\step^{1/(\tapebound+1)} n, \state && \text{if } \state(\lbl) = (\tapebound, \nil) \text{ and } n \leq \tapebound
\end{align*}
but if the tape is \emph{not} empty, the $\Rand(\tapebound, \lbl)$ primitive reduces \emph{deterministically} by taking off the first element of the tape and returning it:
\begin{align*}
  \Rand(\tapebound,\lbl), \state &\step^{1} n, \state[\lbl \mapsto (\tapebound, \tape)\,] && \text{if}~\state(\lbl) = (\tapebound, n \cons \tape)
\end{align*}
If the tape bounds do not match, then $\Rand(\tapebound, \lbl)$ reduces as if the tape was empty:
\begin{align*}
  \Rand(\tapebound,\lbl), \state &\step^{1/(\tapebound + 1)} n, \state && \text{if}~\state(\lbl) = (\tapeboundB, \tape)~\text{and}~\tapebound \neq \tapeboundB~\text{and}~n \leq \tapebound
\end{align*}
However, \emph{no} primitives in the language add values to the tapes!
Instead, values are added to tapes as part of presampling steps that will be \emph{ghost operations} appearing only in the relational logic.
That is, presampling will purely be a proof-device that has no operational effect: in the end, tapes can in fact be erased entirely through refinement as will be clear by the end of this section.
\begin{figure}
  \centering

  \usetikzlibrary{chains, decorations.pathmorphing}
  \tikzset{snake it/.style={decorate, decoration=snake, segment length=5mm}}
  \scalebox{1}{
  \begin{tikzpicture}[xscale=1.25, yscale=1.05]

\edef\sizetape{0.5cm}
\tikzstyle{tape}=[draw,minimum size=\sizetape]

    % left
    \draw (0, 3) node (cone) {{\scriptsize{$\vdots$}}};
    \draw (0, 2) node (ctwo) {$K_1[\expr],$};
    \draw (0.6, 2) node (ctwoL) {$\lbl\mapsto$};

    \foreach \i\j in {0/$x_1$, 1/$x_2$, 2/$\dots$, 3/$x_k$} 
    {
      \draw (1.1+ \i*0.4, 2) node {\j};
      \draw (1.1+ \i*0.4, 2) +(-0.2,-.2)rectangle +(0.2,0.2);
    }

    \draw (0, 1) node (cthree) {$K_1[\expr],$};
    \draw (0.6, 1) node (ctwoL) {$\lbl\mapsto$};
    \foreach \i\j in {0/$x_1$, 1/$x_2$, 2/$\dots$, 3/$x_k$, 4/$\color{red}{n}$} 
    {
      \draw (1.1+ \i*0.4, 1) node {\j};
      \draw (1.1+ \i*0.4, 1) +(-0.2,-.2)rectangle +(0.2,0.2);
    }

    \draw (0, -0.5) node (cfour)  {$K_1'[\Rand(\lbl,\tapebound)],$};
    \draw (1.15, -0.5) node  {$\lbl\mapsto$};
    \foreach \i\j in {0/$\color{red}{n}$, 1/$y_1$, 2/$\dots$} 
    {
      \draw (1.65+ \i*0.4, -0.5) node {\j};
      \draw (1.65+ \i*0.4, -0.5) +(-0.2,-.2)rectangle +(0.2,0.2);
    }

    \draw (0, -1.5) node (cfive) {$K_1'[{\color{red}{n}}],$};
    \draw (0.65, -1.5) node  {$\lbl\mapsto$};
    \foreach \i\j in {0/$y_1$, 1/$\dots$} 
    {
      \draw (1.15+ \i*0.4, -1.5) node {\j};
      \draw (1.15+ \i*0.4, -1.5) +(-0.2,-.2)rectangle +(0.2,0.2);
    }

    % upper edges
    \draw (cone) edge[->] (ctwo);
    \draw (ctwo) edge[->, red, dashed] (cthree);
    \draw (cthree) edge[->, dotted] node[right]{\small\textit{(after $k$ samples from $\iota$)}} (cfour);
    \draw (cfour) edge[->] (cfive);

    % lower
    \draw (5, 3) node (mone) {$\vdots$};
    \draw (5, 2) node (mtwo) {$\fillctx K_2[\Rand(\tapebound)], \state'$};
    \draw (5, 1) node (mthree) {$\fillctx\lctx[f({\color{red}{n}})], \state'$};
    \draw (5, 0) node[] (mdots) {$\vdots$};

    % lower edges
    \draw (mone) edge[->] (mtwo);
    \draw (mtwo) edge[->] (mthree);
    \draw (mthree) edge[->] (mdots);

    % crossing edge
    \draw (0.1, 1.5) edge[snake it, red] (4.9, 1.5);
  \end{tikzpicture}%
}
  \caption{Illustration of an asynchronous coupling established through the rule \ruleref{rel-couple-tape-l}.}%
  \label{fig:async-coupling}%
\end{figure}

At the logical level, \thereflog{} comes with a $\progtape{\lbl}{\tapebound}{\tape}$ assertion that denotes \emph{ownership} of the label $\lbl$ and its contents $(\tapebound, \tape)$, analogously to how the traditional points-to-connective $\progheap{\loc}{\val}$ of separation logic denotes ownership of the location $\loc$ and its contents on the heap.
When a tape is allocated, ownership of the fresh empty tape is acquired, \ie{},
\begin{mathpar}
  \relalloctapel{}
\end{mathpar}
Asynchronous couplings between probabilistic choices can be established in the refinement logic by coupling ghost presamplings with program steps.
For example, the rule below allows us to couple an (unlabeled) probabilistic choice on the right with a presampling on the $\lbl$ tape on the left:
\begin{mathpar}
  \relcoupletapel{}
\end{mathpar}
Intuitively, as illustrated in \cref{fig:async-coupling}, the rule allows us to couple a logical \emph{ghost} presampling step on the left (illustrated using a red dashed arrow) with a \emph{physical} sampling on the right.
A symmetric rule holds for the opposite direction and two ghost presamplings can be coupled as well.
When we---at some point in the future---reach a presampled $\Rand(\tapebound, \lbl)$, we simply read off the presampled values from the $\lbl$ tape deterministically in a first-in-first-out order, \ie{},
\begin{mathpar}
  \relrandtapel{}
\end{mathpar}
If we do not perform any presamplings, tapes and labels can be ignored and we can couple labeled sampling commands as if they were unlabeled:
\begin{mathpar}
  \relranderaser{}
\end{mathpar}
Here the assertion $\spectape{\lbl}{\tapebound}{\nil}$ denotes ownership of an empty tape $\lbl$ of the right-hand side program (the program on the ``specification'' side).

\paragraph{Example.} Using presampling tapes, we can show that $\lazyprog$ is a contextual refinement of $\eagerprog$ from \cref{sec:intro}, that is, $\ctxrefinesS{\lazyprog}{\eagerprog}{\tunit \to \tbool}$.
We first define an intermediate labeled version of $\lazyprog$, using $\Flip(\lbl) \eqdef \If \Rand(1, \lbl) = 0 then \False \Else \True$:
\begin{align*}
  \lazyprog' \eqdef{}
  &\Let \lbl = \AllocTape(1) in \\
  &\Let r = \Alloc(\None) in \\
  &\Lam \_ .
    \MatchML \deref r with
      \Some(b) => b
    | \None => {\begin{array}[t]{l}
      \Let b = \Flip(\lbl) in \\
      r \gets \Some(b); \\
      b
    \end{array}}
    end {}
\end{align*}
By transitivity of contextual refinement and \cref{thm:soundness} it suffices to show $\refines{}{\lazyprog}{\lazyprog'}{\tunit \to \tbool}$ and $\refines{}{\lazyprog'}{\eagerprog}{\tunit \to \tbool}$.
The former follows straightforwardly using symbolic execution rules and \ruleref{rel-rand-erase-r}.
To show the latter we allocate a tape $\lbl$ and a reference $\loc$ on the left by symbolic execution and couple the presampling of a $b \in \{0,1\}$ on the $\lbl$ tape with the $\Flip\TT$ on the right using \ruleref{rel-couple-tape-l}. This establishes an \emph{invariant}
\begin{align*}
  (\progtape{\lbl}{1}{b} \sep \progheap{\loc}{\None}) \lor \progheap{\loc}{\Some(b)}
\end{align*}
that expresses how either $b$ is on the $\lbl$ tape and the location $\loc$ is empty \emph{or} $\loc$ contains the value $b$.
Invariants are particular kinds of propositions in \thereflog{} that, in this particular case, are guaranteed to always hold at the beginning and at the end of the function evaluation.
Under this invariant, we show that the two thunks are related by symbolic execution and rules for accessing invariants that we detail in \cref{sec:rellog}.
Symmetric arguments allow us to show the refinement in the other direction and consequently the contextual equivalence.

This example shows how presampling tapes are simple and powerful, yet merely a proof-device: the final equivalence holds for programs without any mention of tapes.
Intuitively, tapes allow us to separate the process of building a coupling from the operational semantics of the program.
One might be tempted to believe, though, that as soon as the idea of presampling arises, the high-level proof rules as supported by \thereflog{} are straightforward to state and prove.
This is \emph{not} the case. As we will show throughout the paper, a great deal of care goes into defining a system that supports presampling while being sound.
In \cref{sec:counterexamples} we discuss two counterexamples that illustrate some of the subtleties involved in defining a sound system.

%%% Local Variables:
%%% mode: latex
%%% TeX-master: "../clutch-popl24"
%%% End:

\section[Preliminaries and the Language F\_(μ,ref,rand)]{Preliminaries and the Language \thelang{}}
\label{sec:lang}

To account for non-terminating behavior, we will define our operational semantics using probability sub-distributions which we recall below.
\begin{definition}[Sub-distribution]
  A (discrete) \emph{sub-distribution} over a countable set $A$ is a function $\mu : A \to [0,1]$ such that $\sum_{a \in A} \mu(a) \leq 1$.
  We write $\Distr{A}$ for the set of all sub-distributions over $A$.
\end{definition}

\begin{definition}[Support]
  The \emph{support} of $\mu \in \Distr{A}$ is the set of elements
  \[\supp(\mu) \eqdef{} \set{ a \in A \mid \mu(a) > 0 }\]
\end{definition}

%Discrete probability subdistributions form a monad.
\begin{lemma}[Probability Monad]
  Let $\mu \in \Distr{A}$, $a \in A$, and $f : A \to \Distr{B}$.
  Then
  \begin{enumerate}
  \item $\mbind(f,\mu)(b) \eqdef{} \sum_{a \in A} \mu(a) \cdot f(a)(b)$
  \item $\mret(a)(a') \eqdef{}
    \begin{cases}
      1 & \text{if } a = a' \\
      0 & \text{otherwise}
    \end{cases}$
  \end{enumerate}
  gives monadic structure to $\DDistr$.
  We write $\mu \mbindi f$ for $\mbind(f, \mu)$.
\end{lemma}

The syntax of the language $\thelang{}$ is defined by the grammar below.
\begin{align*}
  \val, \valB \in \Val \bnfdef{}
  & z \in \integer \ALT
    b \in \bool \ALT
  \TT \ALT
  \loc \in \Loc \ALT
  \lbl \in \Lbl \ALT
  \Rec \lvarF \lvar = \expr \ALT
  (\val,\valB) \ALT
  \Inl(\val) \ALT
  \Inr(\val)
  \\
  \expr \in \Expr \bnfdef{}&
  \val \ALT
  \lvar \ALT
  % \RecE \lvarF \lvar= \expr \ALT
  \expr_1(\expr_2) \ALT
%  \HLOp_1 \expr \ALT
%  \expr_1 \HLOp_2 \expr_2 \ALT \\ &
  \If \expr then \expr_1 \Else \expr_2 \ALT
  % (\expr_1,\expr_2)_\exprForm \ALT
  \Fst(\expr) \ALT
  \Snd(\expr) \ALT
  % \InlE(\expr) \ALT
  % \InrE(\expr) \ALT
  \Alloc(\expr) \ALT
  \deref \expr \ALT
  \expr_1 \gets \expr_2 \ALT
  \\&
  \Match \expr with \Inl(\val) => \expr_1 | \Inr(\valB) => \expr_2 end \ALT
  \fold \expr \ALT
  \unfold \expr \ALT
  \Lambda \expr \ALT
  \expr~\_ \ALT
  \\&
  \Pack \expr \ALT
  \Unpack \expr as \var in \expr \ALT
  \AllocTape(\expr) \ALT
  \Rand(\expr_{1}, \expr_{2}) \ALT
  \expr_1 + \expr_2 \ALT
  \expr_1 - \expr_2 \ALT \cdots
  \\
% \HLOp_1 \bnfdef{}& - \ALT \neg \\
% \HLOp_2 \bnfdef{}& + \ALT - \ALT \mathop{=} \ALT \ldots \\
  \lctx \in \Ectx \bnfdef{} &
 -
 % \ALT \lctx \binop e
 % \ALT v \binop \lctx
 % \ALT \If{\lctx}then{e_{1}}\Else{e_{2}}
 % \ALT (\lctx,e)
 % \ALT (v,\lctx) 
 % \ALT& \Proj{1}{\lctx}
 % \ALT \Proj{2}{\lctx} 
 % \ALT \Inj{1}{\lctx}
 % \ALT \Inj{2}{\lctx}
 % \ALT \MatchS \lctx with \Inj{i} => \expr_{i} end
 \ALT \expr\,\lctx
 \ALT \lctx\,\val
 % \ALT \Alloc(\lctx)
 \ALT \deref \lctx
 \ALT \expr \gets \lctx
 \ALT \lctx \gets \val
 % \ALT \fold \lctx
 % \ALT \unfold \lctx \\
 % \ALT& \Pack \lctx
 % \ALT \Unpack \lctx as \var in \expr
\ALT \AllocTape(\lctx)
 \ALT \Rand(\expr,\lctx)
\ALT \Rand(\lctx,\val)
\ALT \ldots
\\
  \state \in \State \eqdef{}& (\Loc \fpfn \Val) \times (\Lbl \fpfn \Tape) \\
  t \in \Tape \eqdef{}& \{ (\tapebound, \tape) \mid \tapebound \in \mathbb{N} \wedge \tape \in \mathbb{N}_{\leq \tapebound}^{\ast} \} \\
  \cfg \in \Conf \eqdef{}& \Expr \times \State \\
  \type \in \Type \bnfdef{} & \alpha \ALT
                              \tunit \ALT
                              \tbool \ALT
                              \tnat \ALT
                              \tint \ALT
                              \type \times \type \ALT
                              \type + \type \ALT
                              \type \to \type \ALT
                              \All \alpha . \type \ALT
                              \Exists \alpha . \type \ALT
                              \tmu \alpha . \type \ALT
                              \tref{\type} \ALT
                              \ttape
\end{align*}
The term language is mostly standard but note that there are no types in terms; we write $\Lambda \expr$ for type abstraction and $\expr~\_$ for type application.
$\fold \expr$ and $\unfold \expr$ are the special term constructs for iso-recursive types.
$\Alloc(\expr)$ allocates a new reference, $\deref \expr$ dereferences the location $\expr$ evaluates to, and $\expr_{1} \gets \expr_{2}$ assigns the result of evaluating $\expr_{2}$ to the location that $\expr_{1}$ evaluates to.
We introduce syntactic sugar for lambda abstractions $\Lam \var . \expr$ defined as $\Rec {\_} \var = \expr$,
let-bindings $\Let \var = \expr_{1} in \expr_{2}$ defined as $(\Lam \var . \expr_{2})(\expr_{1})$, and sequencing $\expr_{1} ; \expr_{2}$ defined as $\Let \_ = \expr_{1} in \expr_{2}$.
We write $\Rand(\tapebound)$ for $\Rand(\tapebound, \TT)$, \ie{} an unlabeled probabilistic choice.

We implicitly coerce from $\state \in \State$ to heaps and tapes, \eg{}, $\state(\loc) = \pi_{1}(\state)(\loc)$ and $\state(\lbl) = \pi_{2}(\state)(\lbl)$.
Tapes are formally pairs $(\tapebound, \tape)$ of $\tapebound \in \mathbb{N}$ and a finite sequence $\tape$ of natural numbers less than or equal to $\tapebound$.
The language has a call-by-value single-step-reduction relation $\step \subseteq \Conf \times [0, 1] \times \Conf$ defined using evaluation contexts $\lctx \in \Ectx$.
The relation is mostly standard: all the non-probabilistic constructs reduce as usual with weight $1$ and $\Rand(\expr_{1}, \expr_{2})$ reduces as discussed in~\cref{sec:key-ideas}.

To define full program execution, let $\stepdistr(\cfg) \in \Distr{\Conf}$ denote the distribution induced by the single step reduction of configuration $\cfg \in \Conf$.
First, we define a stratified execution probability $\execVal_{n}\colon \Conf \to \Distr{\Val}$ by induction on $n$:
\begin{align*}
  \execVal_{n}(\expr, \state) \eqdef{}
  \begin{cases}
    \mathbf{0} & \text{if}~\expr \not\in\Val~\text{and}~n = 0 \\
    \mret(\expr) & \text{if}~\expr \in \Val \\
    \stepdistr(\expr, \state) \mbindi \execVal_{(n - 1)} & \text{otherwise}
  \end{cases}
\end{align*}
where $\mathbf{0}$ denotes the everywhere-zero distribution.
That is, $\execVal_{n}(\expr, \state)(\val)$ denotes the probability of stepping from the configuration $(\expr, \state)$ to a value $\val$ in less than $n$ steps.
The probability that a full execution, starting from configuration $\cfg$, reaches a value $\val$ is the limit of its stratified approximations, which exists by monotonicity and boundedness:
\begin{align*}
  \execVal(\cfg)(\val) \eqdef{} \lim_{n \to \infty} \execVal_{n}(\cfg)(\val)
\end{align*}
The probability that a full execution from a starting configuration $\cfg$ terminates then becomes $\execTerm(\cfg) \eqdef{} \sum_{\val \in \Val} \execVal(\rho)(\val)$.

Typing judgments have the form $\pfctx \mid \vctx \proves \expr : \type$ where $\vctx$ is a context assigning types to program variables, and $\pfctx$ is a context of type variables that may occur in $\vctx$ and $\type$.
The inference rules for the typing judgments are standard (see, \eg{}, \citet{DBLP:journals/lmcs/FruminKB21} or the Coq formalization) and omitted, except for the straightforward rules for typing tapes and samplings shown below:
\begin{mathpar}
  \inferH{t-tape}
  {\pfctx \mid \vctx \proves \expr : \tnat}
  {\pfctx \mid \vctx \proves \AllocTape(\expr) : \ttape}
  \and
  \inferH{t-rand}
  { \pfctx \mid \vctx \proves \expr_1 : \tnat \\
    \pfctx \mid \vctx \proves \expr_2 : \type \\
    \type = \tunit \lor \type = \ttape}
  {\pfctx \mid \vctx \proves \Rand(\expr_1, \expr_2) : \tnat}
\end{mathpar}

The notion of contextual refinement that we use is also mostly standard and uses the termination probability $\execTerm$ as observation predicate.
Since we are in a typed setting, we consider only typed contexts.
A program context is well-typed, written $\ctx : (\pfctx \mid \vctx \proves \type) \Ra (\pfctx' \mid \vctx' \proves \type')$, if for any term $\expr$ such that $\pfctx \mid \vctx \proves \expr : \type$ we have $\pfctx' \mid \vctx' \proves \fillctx\ctx[\expr] : \type'$.
We say expression $\expr_{1}$ \emph{contextually refines} expression $\expr_{2}$ if for all well-typed program contexts $\ctx$ resulting in a closed program then the termination probability of $\fillctx\ctx[\expr_{1}]$ is bounded by the termination probability of $\fillctx\ctx[\expr_{2}]$:
\begin{align*}
  \ctxrefines{\pfctx \mid \vctx}{\expr_{1}}{\expr_{2}}{\type} \eqdef{}
  &\All \type', (\ctx : (\pfctx \mid \vctx \proves \type) \Ra (\emptyset \mid \emptyset \proves \type')), \state . \\
  &\quad \execTerm(\fillctx \ctx [\expr_{1}], \state) \leq \execTerm(\fillctx \ctx [\expr_{2}], \state)
\end{align*}
Note that contextual refinement is a precongruence, and that the statement itself is in the meta-logic (\eg{}, Coq) and makes no mention of \thereflog{} or Iris.
Contextual equivalence $\ctxeq{\pfctx \mid \vctx}{\expr_{1}}{\expr_{2}}{\type}$ is defined as the symmetric interior of refinement: $(\ctxrefines{\pfctx \mid \vctx}{\expr_{1}}{\expr_{2}}{\type}) \land (\ctxrefines{\pfctx \mid \vctx}{\expr_{2}}{\expr_{1}}{\type})$.

%%% Local Variables:
%%% mode: latex
%%% TeX-master: "../clutch-popl24"
%%% End:

\section{The \thereflog{} Refinement Logic}
\label{sec:rellog}

In the style of ReLoC \cite{DBLP:journals/lmcs/FruminKB21}, we define a \emph{logical refinement judgment} $\refines{\Delta}[\mask]{\expr_{1}}{\expr_{2}}{\type}$ as an internal notion in the \thereflog{} separation logic by structural recursion over the type $\type$.
The fundamental theorem of logical relations will then show that logical refinement implies contextual refinement.
This means proving contextual refinement can be reduced to proving logical refinement, which is generally much easier.
When defining and proving logical refinement, we can leverage the features of modern separation logic, \eg{}, (impredicative) invariants and (higher-order) ghost state as inherited from Iris, to model and reason about complex programs and language features.

\thereflog{} is based on higher-order intuitionistic separation logic and the most important propositions are shown below.
\begin{align*}
  \prop,\propB \in \iProp \bnfdef{}
  & \TRUE \ALT \FALSE \ALT \prop \land \propB \ALT \prop \lor \propB \ALT \prop \Ra \propB \ALT
   \All \var . \prop \ALT \Exists \var . \prop \ALT \prop \sep \propB \ALT \prop \wand \propB \ALT \\
  & \always \prop \ALT
    \later \prop \ALT
    \MU \var . \prop \ALT
    % \pvs[\mask_{1}][\mask_{2}] \prop
    \ulcorner \pprop \urcorner \ALT
    \knowInv\namesp\prop \ALT
    \nInv\namesp\prop \ALT
    \progheap{\loc}{\val} \ALT
    \specheap{\loc}{\val} \ALT \\
  & \progtape{\lbl}{\tapebound}{\tape} \ALT
    \spectape{\lbl}{\tapebound}{\tape} \ALT
    \semInterp{\type}{\Delta}{\val_{1}, \val_{2}} \ALT
    \refines{\Delta}[\mask]{\expr_{1}}{\expr_{2}}{\type} \ALT
    \ldots
\end{align*}
As \thereflog{} is built upon the base logic of Iris \cite{irisjournal}, it includes all its connectives such as the \emph{persistence} modality~$\always$, the \emph{later} modality~$\later$, fixpoints $\MU \var . \prop$, \emph{invariants}~$\knowInv\namesp\prop$, and \emph{non-atomic invariants} \cite{irisdoc}, written $\nInv\namesp\prop$, which we will introduce as needed.
The proposition $\ulcorner \pprop \urcorner$ embeds a meta-logic (\eg, Coq) proposition $\pprop$ (\eg{}, equality or a coupling) into \thereflog{} but we will omit the brackets whenever the type of $\pprop$ is clear from the context.

Like ordinary separation logic, \thereflog{} has heap points-to assertions. Since the logic is relational, these come in two forms: $\progheap{\loc}{\val}$ for the left-hand side program's state and $\specheap{\loc}{\val}$ for the right-hand side's state (the ``specification'' side).
For the same reason, tape assertions come in two forms as well, $\progtape{\lbl}{\tapebound}{\tape}$ and $\spectape{\lbl}{\tapebound}{\tape}$ respectively.

\subsection{Refinement judgments}\label{sec:judgments}

The refinement judgment $\refines{\Delta}[\mask]{\expr_{1}}{\expr_{2}}{\type}$ should be read as ``in environment $\Delta$, the expression $\expr_{1}$ refines the expression $\expr_{2}$ at type $\type$ under the invariants in $\mask$''.
We refer to $\expr_{1}$ as the \emph{implementation} and to $\expr_{2}$ as the \emph{specification}.
The environment $\Delta$ assigns \emph{interpretations} to type variables occurring in $\type$.
These interpretations are \thereflog{} relations of type~$\Val \times \Val \to \iProp$.
One such relation is the binary interpretation $\semInterp{\type}{\Delta}{-, -}$ of a syntactic type $\type \in \Type$ which is used to define the refinement judgment, as discussed in \cref{sec:model-refines}.
For example, for base types such as $\tbool$ and $\tint$, the value interpretation asserts equality between the values.

\cref{fig:reflog-rules} showcases a selection of the type-directed structural and computational rules for proving logical refinement for deterministic reductions.
Our computational rules resemble the typical forward-symbolic-execution-style rules from, \eg{}, the weakest precondition calculus in Iris \cite{irisjournal}, but come in forms for both the left-hand side and the right-hand side.
For example, \ruleref{rel-pure-l} and \ruleref{rel-pure-r} symbolically execute ``pure'' reductions, \ie{} reductions that do not depend on the state, such as $\beta$-reductions.
\ruleref{rel-store-l} and \ruleref{rel-store-r} on the other hand depend on the heap and require ownership of a location to store values at it.
We remark that all the rules for the deterministic fragment of the \thereflog{} refinement judgment are identical to the rules for the sequential fragment of the non-probabilistic relational logic ReLoC \cite{DBLP:journals/lmcs/FruminKB21}---even though the underlying semantics and model are very different.
This is one of the key reasons behind the support for modular reasoning.

\figreflogrules{}

The rules in \cref{fig:reflog-tape-rules} showcase the computational rules for non-coupled probabilistic reductions and for interactions with presampling tapes.
The rules \ruleref{rel-rand-tape-l} and \ruleref{rel-rand-tape-r} allow us to read off values from a tape as explained in \cref{sec:key-ideas}; if the tapes are empty, \ruleref{rel-rand-tape-empty-l} and \ruleref{rel-rand-tape-empty-r} continue with a fresh sampling just like for unlabeled rands in \ruleref{rel-rand-l} and \ruleref{rel-rand-r}.
Notice how the rules resemble the rules for interacting with the heap.

\figreflogtaperules{}

The main novelty of \thereflog{} is the support for both synchronous and asynchronous couplings for which rules are shown in \cref{fig:reflog-coupling-rules}.
\ruleref{rel-couple-rands} is a classical coupling rule that relates two samplings that can be aligned, just like \ruleref{prhl-couple} as we saw in \cref{sec:intro}.
The rules \ruleref{rel-couple-tape-l} and \ruleref{rel-couple-tape-r}, on the other hand, are asynchronous coupling rules; they both couple a sampling reduction with an arbitrary expression on the opposite side by presampling a coupled value to a tape, as discussed in \cref{sec:key-ideas}.
Finally, \ruleref{rel-couple-tapes} couples two ghost presamplings to two tapes, and hence offers full asynchrony.

\figreflogcouplingrules{}

\subsection{Persistence and invariants}\label{sec:invariants}

% \joe{Should we just absorb this whole paragraph into \cref{sec:invariants}? For one who does not know what Iris's invariant mechanism is, I do not think this will make sense; on the other hand if they do know then the discussion about masks is not needed.}

As mentioned above, the environment $\Delta$ in \thereflog{}'s refinement judgement provides an interpretation of types as relations in the logic.
However, \thereflog{} is a substructural separation logic, while the type system of \thelang{} is \emph{not} substructural.
% For example, closures can be invoked arbitrarily many times and hence ephemeral resources that may change over time cannot always be guaranteed to be available.
% For this reason, the \emph{persistence modality}~$\always$ plays a crucial role in, for example, the rule \ruleref{rel-rec} to guarantee that only \emph{persistent} resources are used to verify the closure's body.
To account for the non-substructural nature of \thelang{}'s types, we make use of the \emph{persistence modality}~$\always$.
We say $\prop$ is persistent, written $\persistent{\prop}$ if $\prop \proves \always \prop$; otherwise, we say that $\prop$ is \emph{ephemeral}.
Persistent resources can freely be duplicated ($\always \prop \provesIff \always \prop \sep \always \prop$) and eliminated ($\always \prop \proves \prop$).
For example, invariants and non-atomic invariants are persistent: once established, they will remain true forever.
On the contrary, ephemeral propositions like the points-to connective $\progheap{\loc}{\val}$ for the heap may be invalidated in the future when the location is updated.
For exactly this reason, the rule \ruleref{rel-pack} also requires the interpretation of the type variable to be persistent, to guarantee that it does not depend on ephemeral resources.

To reason about, \eg{}, functions that make use of ephemeral resources, a common pattern is to ``put them in an invariant'' to make them persistent, as sketched in \cref{sec:key-ideas} for the lazy/eager example.
Since our language is sequential, when a function is invoked, no other code can execute before the function returns.
This means that we can soundly keep invariants ``open'' and temporarily invalidate them for the entire duration of a function invocation---as long as the invariants are reestablished before returning.
Non-atomic invariants allow us to capture exactly this intuition.

Invariants are annotated with invariant names $\namesp \in \InvName$ and the refinement judgment is annotated by \emph{invariant masks} $\mask \subseteq \InvName$ that indicates which non-atomic invariants that are currently \emph{closed}.
This is needed for bookkeeping of the invariant mechanism in order to avoid reentrancy issues, where invariants are opened in a nested (and unsound) fashion.

% Ordinary Iris invariants can only be opened around atomic expressions (that evaluate to a value in a single step); this is crucial for soundness in a concurrent setting.
% Since \thelang{} is a sequential language, we will make use of the more flexible non-atomic invariants, which can be kept open during multiple execution steps---our use of non-atomic invariants will only require invariants to be closed at the end of proofs.

\cref{fig:reflog-invariant-rules} shows structural rules for the refinement judgment's interaction with non-atomic invariants.
An invariant $\nInv{\namesp}{\prop}$ can be allocated (\ruleref{rel-na-inv-alloc}) by giving up ownership of $\prop$.
When opening an invariant (\ruleref{rel-na-inv-open}) one obtains the resources $\prop$ together with a resource $\nClose{\namesp}{\mask}{\prop}$ that allows one to close the invariant again (\ruleref{rel-na-inv-close}) by reestablishing $\prop$.
% The mask annotation $\mask$ on the refinement judgment keeps track of which invariants have been opened to avoid opening invariants in an (unsound) nested fashion.
We guarantee that all invariants are closed by the end of evaluation by requiring $\top$, the set of all invariant names, as mask annotation on the judgment in all value cases (see, \eg{}, \ruleref{rel-rec}, \ruleref{rel-pack}, and \ruleref{rel-return} in \cref{fig:reflog-rules}).

\thereflog{} invariants are inherited from Iris and hence they are \emph{impredicative} \cite{DBLP:conf/esop/SvendsenB14} which means that the proposition $\prop$ in $\nInv{\namesp}{\prop}$ is \emph{arbitrary} and can, \eg{}, contain other invariant assertions.
To ensure soundness of the logic and avoid self-referential paradoxes, invariant access guards $\prop$ by the later modality $\later$.
When invariants are not used impredicatively, the later modality can mostly be ignored as we have done and will do throughout the paper.
The later modality is essential for the soundness of the logical relation and taking guarded fixpoints~$\MU \var . \prop$ that require the recursive occurrence $\var$ to appear under the later modality, but our use is entirely standard.
We refer to \citet{irisjournal} for more details on the later modality and how it is generally used in Iris.

\figrefloginvariantrules{}

%%% Local Variables:
%%% mode: latex
%%% TeX-master: "../clutch-popl24"
%%% End:

\section{Model of \thereflog{}}\label{sec:model}

In this section we show how the connectives of \thereflog{} are modeled through a shallow embedding in the \emph{base logic} of the Iris separation logic \cite{irisjournal}.
First, we describe how we define a relational \emph{coupling logic} (\cref{sec:coupling-logic}) that is used to establish couplings between programs.
Next, we show how the coupling logic in combination with a binary interpretation of types is used to define the refinement logic (\cref{sec:model-refines}).
Finally, we summarize how the final soundness theorem is proven (\cref{sec:model-soundness}).

The general structure and skeleton of our model mimics the construction of several \emph{non}-probabilistic logical relations found in prior work~\cite{DBLP:conf/popl/TuronTABD13, DBLP:conf/icfp/TuronDB13, DBLP:conf/popl/KrebbersTB17, DBLP:journals/lmcs/FruminKB21}.
A key contribution and benefit of \thereflog{} is that that same structure can be adapted to handle \emph{probabilistic} refinements through the right choice of intermediate definitions and abstractions, as we will highlight throughout this section.
% but that the soundness theorem, how we prove it, and multiple technical facets are novel as we will highlight throughout this section.
While some aspects of the model are Iris-specific, the key ideas are general and should apply to other frameworks as well.
% There are many details and layers to keep track of and for this reason, we will present the model using a top-down approach to not lose track of the bigger picture when working through the model.

\subsection{Coupling logic}\label{sec:coupling-logic}

We recall that probabilistic couplings are used to prove relations between distributions by constructing a joint distribution that relates two distributions in a particularly desirable way:
\begin{definition}[Coupling]
  Let $\mu_{1} \in \Distr{A}$, $\mu_{2} \in \Distr{B}$. % for countable sets $A$ and $B$
  A sub-distribution $\mu \in \Distr{A \times B}$ is a \emph{coupling} of $\mu_{1}$ and $\mu_{2}$ if
  \begin{enumerate}
  \item $\All a . \sum_{b \in B} \mu(a, b) = \mu_{1} (a)$
  \item $\All b . \sum_{a \in A} \mu(a, b) = \mu_{2} (b)$
  \end{enumerate}
  Given a relation $R \subseteq A \times B$ we say $\mu$ is an $R$-coupling if furthermore $\supp(\mu) \subseteq R$.
  We write $\Rcoupl{\mu_1}{\mu_2}{R}$ if there exists an $R$-coupling of $\mu_{1}$ and $\mu_{2}$.
\end{definition}
Couplings can be constructed and composed along the monadic structure of sub-distributions.
\begin{lemma}[Composition of couplings]\label{lem:coupling-comp}
  Let $R \subseteq A \times B$, $S \subseteq A' \times B'$, $\mu_{1} \in \Distr{A}$, $\mu_2 \in \Distr{B}$, $f_{1} : A \to \Distr{A'}$, and $f_{2} : B \to \Distr{B'}$.
  \begin{enumerate}
  \item If $(a, b) \in R$ then $\Rcoupl{\mret(a)}{\mret(b)}{R} $.
  \item If $\Rcoupl{\mu_{1}}{\mu_{2}}{R}$ and for all $(a, b) \in R$ it is the case that $\Rcoupl{f_{1}(a)}{f_{2}(b)}{S}$ then $\Rcoupl{\mu_{1} \mbindi f_{1}}{\mu_{2} \mbindi f_{2}}{S}$
  \end{enumerate}
\end{lemma}
Once a coupling has been established, we can often extract a concrete relation from it between the probability distributions.
In particular, for $(=)$-couplings, we have the following result.
\begin{lemma}
  If $\Rcoupl{\mu_{1}}{\mu_{2}}{(=)}$ then $\mu_{1} = \mu_{2}$.
\end{lemma}

The \thereflog{} coupling logic can be seen as a higher-order separation logic analogue of \citet{DBLP:conf/lpar/BartheEGHSS15}'s pRHL logic.
However, unlike pRHL, which uses the four-part Hoare \emph{quadruples} that we saw in \cref{sec:intro} to do relational reasoning, the coupling logic instead follows CaReSL~\citep{DBLP:conf/icfp/TuronDB13} and encodes one of the programs as a separation logic \emph{ghost resource}.
In particular, the coupling logic consists of two components: (1) a \emph{unary} weakest precondition theory $\wpre{\expr}{\Phi}$; and (2) a \emph{specification resource} $\spec(\expr')$ with \emph{specification context} $\specCtx$.
We think of the program $\expr$ in the weakest precondition predicate as representing the program that occurs on the left side of a quadruple, while the specification program $\expr'$ represents the right side program.
The specification context assertion $\specCtx$ will be used to connect the weakest precondition to the specification resource.
Ultimately, by showing
\begin{align*}
  \specCtx \sep \spec(\expr') \proves \wpre{\expr}{\val . \Exists \val' . \spec(\val') \sep \varphi(\val, \val')}
\end{align*}
in the logic, we will have established a $\varphi$-coupling of the executions of the programs $\expr$ and $\expr'$.

\paragraph{The weakest precondition.}
The weakest precondition connective $\wpre{\expr}{v.\Phi}$ is a new probabilistic weakest precondition that we formally define below.
In isolation it simply means that the execution of $\expr$ is safe (\ie{}, the probability of crashing is zero), and for every possible return value $\val$ of $\expr$, the postcondition $\Phi(\val)$ holds.
Note however, that it encodes \emph{partial} correctness, as it does not imply that the probability of termination is necessarily one, meaning the program may diverge.

In most Iris-style program logics, the weakest precondition $\wpre{\expr}{\Phi}$ is a predicate stating that either the program $\expr$ is a value satisfying $\Phi$ or it is reducible such that for any other term $\expr'$ that it reduces to, then $\wpre{e'}{\Phi}$ must hold as well.
This guarantees safety of the full execution of the program $\expr$.
The weakest precondition that we define in this section has---in isolation---the same intuition but it is fundamentally different.
It is still a \emph{unary} predicate, but in order to do relational reasoning, the weakest precondition pairs up the probability distribution of individual program steps of the left-hand side with the probability distribution of individual steps of \emph{some} other program in such a way that there exists a \emph{probabilistic coupling} among them.
Through the $\specCtx$ we will guarantee that this ``other'' program is tied to the program tracked by the $\spec(\expr')$ resource.
The weakest precondition itself satisfies all the usual structural rules such as \ruleref{wp-wand} and \ruleref{wp-bind} found in \cref{fig:struc-rules} as well as language-level primitive rules such as \ruleref{wp-load}, but in combination with the $\specCtx$ and $\spec(\expr')$ resources, the coupling logic satisfies rules like \ruleref{wp-couple-rands} and \ruleref{wp-couple-tape-l}. Notice the resemblance between \ruleref{wp-couple-rands} and \ruleref{prhl-couple} from \cref{sec:intro}.

\figstrucrules{}

% The goal of the weakest precondition connective is to construct couplings between individual program steps that can then be composed using \cref{lem:coupling-comp} in the soundness theorem to form a coupling for the entire execution of the two programs.

The weakest precondition connective is given by a guarded fixpoint of the equation below---the fixpoint exists because the recursive occurrence appears under the later modality.\footnote{We omit from the definition occurrences of the Iris \emph{fancy update modality} needed for resource updates and necessary book-keeping related to Iris invariants---these matters are essential but our use is entirely standard. For the Iris expert we refer to \appref{sec:app-model} for the full definition.}
\begin{align*}
  \wpre{\expr_{1}}{\Phi} \eqdef{}
  & (\expr_{1} \in \Val \mathrel{\land} \Phi(\expr_{1})) \lor{} \\
  & (\expr_{1} \not\in \Val \mathrel{\land} \All \state_{1}, \cfg'_{1} . \stateinterp(\state_{1}) \sep \specinterp(\cfg'_{1}) \wand \\
  & \quad \execCoupl((\expr_{1}, \state_{1}), \cfg'_{1})(\Lam (\expr_{2}, \state_{2}), \cfg'_{2} .
    \later \stateinterp(\state_{2}) \sep \specinterp(\cfg'_{2}) \sep \wpre{\expr_{2}}[\mask]{\Phi}))
\end{align*}
The base case says that if the expression $\expr_{1}$ is a value then the postcondition $\Phi(\expr_{1})$ must hold.
On the other hand, if $\expr_{1}$ is \emph{not} a value, we get to assume two propositions $\stateinterp(\state_{1})$ and $\specinterp(\cfg'_{1})$ for any $\state_{1} \in \State, {\cfg_{1}}' \in \Conf$, and then we must prove $\execCoupl((\expr_{1}, \state_{1}), \cfg'_{1})(\ldots)$.
The $\stateinterp : \State \to \iProp$ predicate is a \emph{state interpretation} that interprets the state (the heap and the tapes) of the language as resources in \thereflog{} and gives meaning to the $\progheap{\loc}{\val}$ and $\progtape{\lbl}{\tapebound}{\tape}$ connectives.
The $\specinterp : \Conf \to \iProp$ predicate is a \emph{specification interpretation} that allows us to interpret and track the ``other'' program that we are constructing a coupling with---we return to its instantiation momentarily.

The key technical novelty and the essence of the weakest precondition is the \emph{coupling modality}: Intuitively, the proposition $\execCoupl(\cfg_{1}, \cfg'_{1})(\Lam \cfg_{2}, \cfg'_{2} .
\prop)$ says that there exists a series of (composable) couplings starting from configurations $\cfg_{1}$ and $\cfg'_{1}$ that ends up in some configurations $\cfg_{2}$ and $\cfg'_{2}$ such that the proposition $\prop$ holds.
With this intuition in mind, the last clause of the weakest precondition says that the execution of $(\expr_{1}, \state_{1})$ can be coupled with the execution of $\cfg'_{1}$ such that the state and specification interpretations still hold for the end configurations, and the weakest precondition holds recursively for the continuation $\expr_{2}$.

\paragraph{Coupling modality}
% \simon{Rework, present as an inductive definition?}

The coupling modality is an inductively defined proposition in \thereflog, formally defined as a least fixpoint of an equation with six different disjuncts found in \appref{sec:app-model}.
The modality supports both synchronous and asynchronous couplings on both sides while ensuring that the left program takes at least one step.
As it is inductively defined, we can chain together multiple couplings but it always ends in base cases that couple a single step of the left-hand side program---this aligns with the usual intuition that each unfolding of the recursively defined weakest precondition corresponds to one physical program step.

For instance, we can couple two physical program steps through the following constructor:
\begin{mathpar}
  \infer
  {\red(\cfg_1) \\
    \Rcoupl{\stepdistr(\cfg_1)}{\stepdistr(\cfg_1')}{R} \\
    \All \cfg_{2}, {\cfg_{2}}' . R(\cfg_{2}, {\cfg_{2}}') \wand Z(\cfg_{2}, {\cfg_{2}}')
  }
  {\execCoupl(\cfg_1, \cfg_1')(Z)}
\end{mathpar}
Intuitively, this says that to show $\execCoupl(\cfg_{1}, {\cfg_{1}}')(Z)$ we (1) have to show that the configuration $\cfg_{1}$ is \emph{red}ucible  which means that the program \emph{can} take a step (this is to guarantee safety of the left-hand side program), (2) pick a relation $R$ and show that there exists an $R$-coupling of the two program steps, and (3) for all configurations $\cfg_{2}, {\cfg_{2}}'$ in the support of the coupling, the logical predicate $Z(\cfg_{2}, {\cfg_{2}}')$ holds.
This rule is used to justify the classical coupling rule \ruleref{wp-couple-rands} that (synchronously) couples two program samplings.

The coupling modality also allows to construct a coupling between a program step and a trivial (Dirac) distribution; this is used to validate proof rules that symbolically execute just one of the two sides.
Indeed, the rule below allows us to progress the right-hand side independently from the left-hand side, but notice the occurrence of the coupling modality in the premise---this allows us to chain multiple couplings together in a single coupling modality.
\begin{mathpar}
  \infer
  {\Rcoupl{\mret(\cfg_1)}{\stepdistr(\cfg'_{1})}{R} \\
    \All {\cfg_{2}}' . R(\cfg_{1}, {\cfg_{2}}') \wand \execCoupl(\cfg_{1}, {\cfg_{2}}')(Z)
  }
  {\execCoupl(\cfg_{1}, {\cfg_{1}}')(Z)}
\end{mathpar}

To support \emph{asynchronous} couplings, we introduce a \emph{state step} reduction relation $\step_{\lbl} \subseteq \State \times [0, 1] \times \State$ that uniformly at random samples a natural number $n$ to the end of the tape $\lbl$:
\begin{align*}
  \state\step_{\lbl}^{1/(\tapebound + 1)} \state[\lbl \to (\tapebound, \tape \cdot n)]
  && \text{if } \state(\lbl) = (\tapebound, \tape) \text{ and } n \leq \tapebound
\end{align*}
Let $\statestepdistr_{\lbl}(\state)$ denote the induced distribution of a single state step reduction of $\state$.
The coupling modality allows us to introduce couplings between $\statestepdistr_{\lbl}(\state)$ and a sampling step:
\begin{mathpar}
  \infer
  {\Rcoupl{\statestepdistr_{\lbl}(\state_{1})}{\stepdistr({\cfg_{1}}')}{R} \\
    \All \state_{2}, {\cfg_{2}}' . R(\state_{2}, {\cfg_{2}}') \wand \execCoupl((\expr_{1}, \state_{2}), {\cfg_{2}}')(Z)
  }
  {\execCoupl((\expr_{1}, \state_{1}), {\cfg_{1}}')(Z)}
\end{mathpar}
Note that here the left-hand side program does not take a physical step, thus the coupling modality appears in the premise as well.
This particular rule is key to the soundness of the asynchronous coupling rule \ruleref{wp-couple-tape-l} that couples a sampling to a tape on the left with a program sampling on the right.
We use similar constructors of $\execCoupl$ to prove, \eg{} \ruleref{rel-couple-tape-r}.
The crux is, however, that the extra state steps that we inject in the coupling modality to prove the asynchronous coupling rules \emph{do not matter} (!)
in the sense that they can be entirely erased as part of the coupling logic's adequacy theorem (\cref{thm:adequacy}).

\paragraph{A specification resource and context with run ahead}\label{sec:model-specctx}
% The ghost specification connective $\spec(\expr)$, together with the $\specCtx$ proposition, satisfies a number of symbolic execution rules following the operational semantics.
% For brevity, we elide these rules---they correspond directly to all the right-hand side rules (ending in \rulenamestyle{-R}) shown in \cref{fig:reflog-rules} and \cref{fig:reflog-tape-rules}.
% The $\specCtx$ proposition is an Iris invariant that ties together the ghost specification resource $\spec(\expr)$ and the heap and tape assertions, $\specheap{\loc}{\val}$ and $\spectape{\lbl}{\tapebound}{\tape}$, with an execution on the right-hand side as we discuss further in \cref{sec:model-specctx}.
% \simon{Expand the paragraph above}

We will encode a relational specification into a unary specification by proving a unary weakest precondition about $\expr$ (the \emph{implementation}), in which $\expr'$ (the \emph{specification}) is tracked using a ghost resource $\spec(\expr')$ that can be updated to reflect execution steps.
The ghost specification connective $\spec(\expr')$, together with the $\specCtx$ proposition, satisfies a number of symbolic execution rules following the operational semantics.

The $\specCtx$ proposition is an Iris invariant and its purpose is twofold: (1) it gives meaning to the ghost specification resource $\spec(\expr)$ and the heap and tape assertions, $\specheap{\loc}{\val}$ and $\spectape{\lbl}{\tapebound}{\tape}$, and (2) it connects the $\spec(\expr)$ resource to the program $\expr'$ that we are constructing a coupling with in the weakest precondition.
We keep track of $\expr'$ through the specification interpretation $\specinterp$.
When constructing a final closed proof we will want $\expr$ to be equal to $\expr'$, however, during proofs they are not always going to be the same---we will allow $\expr$ to \emph{run ahead} of $\expr'$.
As a consequence, it will be possible to reason \emph{independently} about the right-hand side without consideration of the left-hand side as exemplified by the rules below\footnote{Technically, the consequence of the rules is under a fancy update modality that we omit for the sake of presentation.}, that allow us to progress the specification program but without considering the weakest precondition or the left-hand side program.
\begin{mathpar}
  \inferH{spec-pure}
  { \specCtx \\
    \spec(\fillctx\lctx[\expr]) \\
    \expr \purestep \expr'}
  { \spec(\fillctx\lctx[\expr'])}
  \and
  \inferH{spec-store}
  { \specCtx \\
    \spec(\fillctx\lctx[\loc \gets \valB]) \\
    \specheap{\loc}{\val}}
  {\spec(\fillctx\lctx[\TT]) \sep \specheap{\loc}{\valB}}
\end{mathpar}
Similarly looking rules exists for all the deterministic right-hand side reductions.

To define $\specCtx$ we will use two instances of the \emph{authoritative resource algebra} \cite{iris} from the Iris ghost theory.
It suffices to know that an instance $F$ gives us two resources $F_{\authfull}(a)$ and $F_{\authfrag}(a)$ satisfying $F_{\authfull}(a) \sep F_{\authfrag}(b) \proves a = b$ and that $F_{\authfull}(a) \sep F_{\authfrag}(b)$ can be updated to $F_{\authfull}(a') \sep F_{\authfrag}(a')$.
To connect the two parts we will keep $\specinterpauth(\cfg)$ in the specification interpretation $G$ (that ``lives'' in the weakest precondition), and the corresponding $\specinterpfrag(\cfg)$ in $\specCtx$:
\begin{align*}
  \specinterp(\cfg) &\eqdef \specinterpauth(\cfg) \\
  \textlog{specInv} &\eqdef \Exists \cfg, \expr, \state, n .
                      \begin{aligned}[t]
                        &\specinterpfrag(\cfg) \sep \specauth(\expr) \sep \specstate(\state) \sep \exec_{n}(\cfg)(\expr, \state) = 1
                      \end{aligned} \\
  \specCtx &\eqdef \knowInv{\namesp.\textlog{spec}}{\textlog{specInv}}
\end{align*}
This ensures that the configuration $\cfg$ tracked in the weakest precondition is the same as the configuration $\cfg$ tracked in $\specCtx$.
On top of this, $\specCtx$ contains resources $\specauth(\expr)$ and $\specstate(\state)$ while guaranteeing that the configuration $(\expr, \state)$ can be reached in $n$ deterministic program steps from $\cfg$.
The $\specstate(\state)$ resource gives meaning---using standard Iris ghost theory---to the heap and tape assertions, $\specheap{\loc}{\val}$ and $\spectape{\lbl}{\tapebound}{\tape}$, just like the state interpretation in the weakest precondition.
$\exec_{n} : \Conf \to \Distr{\Conf}$ denotes the distribution of $n$-step partial execution.
By letting $\spec(\expr) \eqdef{} \specfrag(\expr)$ this construction permits the right-hand side program to progress (with deterministic reduction steps) without consideration of the left-hand side as exemplified by \ruleref{spec-pure} and \ruleref{spec-store}.
However, when applying coupling rules that actually need to relate the two sides, the proof first ``catches up'' with $\spec(\expr)$ using the $\execCoupl$ rule that progresses the right-hand side independently, before constructing the coupling of interest.

\subsection{Refinement logic}\label{sec:model-refines}
Contextual refinement is a typed relation and hence logical refinement must be typed as well.
To define the refinement logic, we first define a binary value interpretation $\semInterpS{\type}{\Delta}$ that characterizes the set of pairs of closed values $(\val_1,\val_2)$ of type $\type$ such that $\val_1$ contextually refines $\val_2$.
The definition follows the usual structure of (``logical'') logical relations, see, \eg{}, \citet{DBLP:journals/lmcs/FruminKB21, iris-logrel-journal}, by structural recursion on $\type$ and uses corresponding logical connectives.
Functions are interpreted via (separating) implication, universal types are interpreted through universal quantification, \etc{}, as found in \appref{sec:app-model}.
The only novelty is the interpretation of the new type of tapes shown below:
\begin{align*}
  \semInterp{\ttape}{\Delta}{\val_{1}, \val_{2}} \triangleq{} \Exists \lbl_{1}, \lbl_{2}, \tapebound .
  & (\val_{1} = \lbl_{1}) \sep (\val_{2} = \lbl_{2}) \sep{}
    \knowInv{\namesp.\lbl_{1}.\lbl_{2}}{\progtape{\lbl_{1}}{\tapebound}{\nil} \sep \spectape{\lbl_{2}}{\tapebound}{\nil}}
\end{align*}
The interpretation requires that the values are tape labels, \ie{}, references to tapes, and  that they are always empty as captured by the invariant.
Intuitively, this guarantees through coupling rules and the symbolic execution rules from \cref{fig:reflog-tape-rules} that we always can couple samplings on these tapes as needed in the compatibility lemma for \ruleref{t-rand} as discussed in \cref{sec:model-soundness}.
Point-wise equality of the two tapes would also have been sufficient for the compatibility lemma but by requiring them to be empty we can prove general equivalences such as $\ctxeq{\lbl : \ttape}{\Rand(\tapebound)}{\Rand(\tapebound, \lbl)}{\tnat}$.

The refinement judgment is defined using the coupling logic in combination with the binary value interpretation.
Recall how the intuitive reading of the refinement judgment $\refines{\Delta}[\mask]{\expr_{1}}{\expr_{2}}{\type}$ is that the expression $\expr_{1}$ refines the expression $\expr_{2}$ at type $\type$ under the invariants in the mask $\mask$ with interpretations of type variables in $\type$ taken from $\Delta$.
Besides the coupling logic and the binary value interpretations, we will also make use of the resource $\nTok{\mask}$ that keeps track of the set of non-atomic invariants that are currently \emph{closed}.

Putting everything together, the refinement judgment is formally defined as follows:
\begin{align*}
  \refines{\Delta}[\mask]{\expr_{1}}{\expr_{2}}{\type} \eqdef{}
   \All \lctx .
    & \specCtx \wand \spec(\fillctx\lctx[\expr_{2}]) \wand \nTok{\mask} \wand \\
    & \wpre{\expr_{1}}{\val_{1} . \Exists \val_{2} . \spec(\fillctx\lctx[\val_{2}]) \sep \nTok{\top} \sep \semInterp{\type}{\Delta}{\val_{1}, \val_{2}}}
\end{align*}
%
% This definition has multiple components and we will go over them one by one.
% First, note that we encode a relational specification into a unary specification by proving a unary weakest precondition about $e_1$ (the \emph{implementation}), in which $e_2$ (the \emph{specification}) is tracked using a ghost resource $\spec(\expr_{2})$ that can be updated to reflect execution steps.
% The weakest precondition connective $\wpre{\expr}{v.\Phi}$ is a new probabilistic weakest precondition that we formally define and discuss in \cref{sec:model-weakestpre}.
% In isolation it simply means that the execution of $\expr$ is safe (\ie{}, the probability of crashing is zero), and for every possible return value $\val$ of $\expr$, the postcondition $\Phi(\val)$ holds.
% Note however, that it does not imply that the probability of termination is itself one.
% Finally, it involves the ghost specification connective $\spec(\expr_{2})$ that states that the right-hand side program is currently $\expr_{2}$.
%
The definition assumes that the right-hand side program is executing $\expr_{2}$ and that the invariants in $\mask$ are closed, and it concludes that the two executions can be aligned so that if $\expr_{1}$ reduces to some value $\val_{1}$ then there exists a corresponding execution of $\expr_{2}$ to a value $\val_{2}$ and all invariants have been closed.
Moreover, the values $\val_{1}$ and $\val_{2}$ are related via the binary value interpretation $\semInterp{\type}{\Delta}{\val_{1}, \val_{2}}$.
By quantifying over $\lctx$, we close the definition under evaluation contexts on the right-hand side.
For the left-hand side this is not needed as the weakest precondition already satisfies \ruleref{wp-bind}.

\subsection{Soundness}\label{sec:model-soundness}

The soundness of the refinement judgment hinges on the soundness of the coupling logic.
The goal of the coupling logic is to show a \emph{coupling} of the execution of the two programs, but to establish a coupling of two distributions they must have the same mass.
Intuitively, due to the approximative nature of step-indexed logics like \thereflog{}, we need to show---at every logical step-index---that a coupling exists, even when the left-hand side program has not yet terminated.
This means we might not have enough mass on the left-hand side to cover all of the mass on the right-hand side.
% However, contextual refinement is not defined as an equality between distributions, but rather as a pointwise inequality.
For this reason we introduce a new notion of \emph{left-partial~coupling}.
\begin{definition}[Left-Partial Coupling]
  Let $\mu_{1} \in \Distr{A}, \mu_{2} \in \Distr{B}$. % for countable sets $A$ and $B$.
  A sub-distribution $\mu \in \Distr{A \times B}$ is a \emph{left-partial~coupling} of $\mu_{1}$ and $\mu_{2}$ if
  \begin{enumerate}
  \item $\All a . \sum_{b \in B} \mu(a, b) = \mu_{1} (a)$
  \item $\All b . \sum_{a \in A} \mu(a, b) \leq \mu_{2} (b)$
  \end{enumerate}
  Given a relation $R \subseteq A \times B$ we say $\mu$ is an $R$-left-partial-coupling if furthermore $\supp(\mu) \subseteq R$.
  We write $\refRcoupl{\mu_1}{\mu_2}{R}$ if there exists an $R$-left-partial-coupling of $\mu_{1}$ and $\mu_{2}$.
\end{definition}
This means that, for any $\mu\in\Distr{B}$ and any $R \subseteq A \times B$, the zero distribution $\mathbf{0}$ trivially satisfies $\refRcoupl{\mathbf{0}}{\mu}{R}$.
This reflects the asymmetry of both contextual refinement and our weakest precondition---it allows us to show that a diverging program refines any other program of appropriate type.

Left-partial couplings can also be constructed and composed along the monadic structure of the sub-distribution monad and are implied by regular couplings:
\begin{lemma}
  If $\Rcoupl{\mu_{1}}{\mu_{2}}{R}$ then $\refRcoupl{\mu_{1}}{\mu_{2}}{R}$.
\end{lemma}
Additionally, proving a $(=)$-left-partial-coupling coincides with the point-wise inequality of distributions that will allow us to reason about contextual refinement.
\begin{lemma}
  If $\refRcoupl{\mu_{1}}{\mu_{2}}{(=)}$ then $\All a . \mu_{1}(a) \leq \mu_{2}(a)$.
\end{lemma}
The \emph{adequacy} theorem of the coupling logic is stated using left-partial couplings.
\begin{theorem}[Adequacy]\label{thm:adequacy}
  Let $\varphi : \Val \times \Val \to \mProp$ be a predicate on values in the meta-logic.
  If
  \begin{align*}
    \specCtx \sep \spec(\expr') \proves \wpre{\expr}{\val . \Exists \val' . \spec(\val') \sep \varphi(\val, \val')}
  \end{align*}
 is provable in \thereflog{} then $\All n . \refRcoupl{\execVal_{n}(\expr, \state)}{\execVal(\expr', \state')}{\varphi}$.
\end{theorem}
As a simple corollary, contextual refinement follows from continuity of $\execVal_{n}$.

The proof of the adequacy theorem goes by induction in both $n$ and the $\execCoupl$ fixpoint, followed by a case distinction on the big disjunction in the definition of $\execCoupl$.
Most cases are simple coupling compositions along the monadic structure except the cases where we introduce state step couplings that rely on \emph{erasure} in the following sense:
\begin{lemma}[Erasure]\label{lem:erasure}
  If $\state_{1}(\lbl) \in \dom(\state_{1})$ then
  \begin{align*}
    \Rcoupl{\execVal_{n}(\expr_{1}, \state_{1})}{(\statestepdistr_{\lbl}(\state_{1}) \mbindi \Lam \state_{2} . \execVal_{n}(\expr_{1}, \state_{2}))}{(=)}
  \end{align*}
\end{lemma}
Intuitively, this lemma tells us that we can prepend any program execution with a state step reduction and it will not have an effect on the final result.
The idea behind the proof is that if we append a sampled value $n$ to the end of a tape, and if we eventually consume $n$, then we obtain the same distribution as if we never appended $n$ in the first place.
This is a property that one should \emph{not} take for granted: the operational semantics has been carefully defined such that reading from an empty tape reduces to a value as well, and none of the other program operations can alter or observe the contents of the tape.
This ensures that presampled values are untouched until consumed and that the proof and the execution is independent.

To show the soundness theorem of the refinement logic, we extend the interpretation of types to typing contexts---$\semInterp{\Gamma}{\Delta}{\vec{v},\vec{w}}$ iff for every $x_i\colon\typeB_i$ in $\Gamma$ then $\semInterp{\typeB_i}{\Delta}{v_i,w_i}$ holds---and the refinement judgment to open terms by closing substitutions as usual:
\begin{align*}
  \refines{\Delta \mid \Gamma}{\expr_1}{\expr_2}{\type} \eqdef{}
  \All \vec{\val}, \vec{\valB} . \semInterp{\Gamma}{\Delta}{\vec{\val},\vec{\valB}} \wand
  \refines{\Delta}{e_1[\vec{\val}/\Gamma]}{e_2[\vec{\valB}/\Gamma]}{\type}
\end{align*}
where $e_1[\vec{\val}/\Gamma]$ denotes simultaneous substitution of every $x_i$ from $\Gamma$ in $e_1$ by the value $\val_i$.

We then show, using the structural and symbolic execution rules of the refinement judgment, that the typing rules are \emph{compatible} with the relational interpretation: for every typing rule, if we have a pair of related terms for every premise, then we also have a pair of related terms for the conclusion.
See for instance the compatibility rule for \ruleref{t-rand} below in the case $\type = \ttape$ that follows using \ruleref{rel-bind} and \ruleref{rel-couple-tapes}.
\[
    \inferH{rand-compat}
    {
      \refines{\Delta \mid \Gamma}{\expr_1}{{\expr_1}'}{\tnat} \\
    \refines{\Delta \mid \Gamma}{\expr_2}{{\expr_2}'}{\ttape}
    }
    {\refines{\Delta \mid \Gamma}{\Rand(\expr_1, \expr_{2})}{\Rand({\expr_{1}}', {\expr_2}')}{\tnat}}
\]
As a consequence of the compatibility rules, we obtain the fundamental theorem of logical relations.
\begin{theorem}[Fundamental theorem]
  Let $\Xi \mid \Gamma \proves \expr : \type$ be a well-typed term, and let $\Delta$ assign a relational interpretation to every type variable in $\Xi$.
  Then $\refines{\Delta\mid\Gamma}{e}{e}{\type}{}$.
\end{theorem}
The compatibility rules, moreover, yield that the refinement judgment is a congruence, and together with \cref{thm:adequacy} we can then recover contextual refinement:
\begin{theorem}[Soundness]
  Let $\Xi$ be a type variable context, and assume that, for all $\Delta$ assigning a relational interpretation to all type variables in $\Xi$, we can derive $\refines{\Delta\mid\Gamma}{e_1}{e_2}{\type}{}$.
  Then $\ctxrefines{\Xi\mid\Gamma}{e_1}{e_2}{\type}{}$
\end{theorem}

%%% Local Variables:
%%% mode: latex
%%% TeX-master: "../clutch-popl24"
%%% End:

\section{Case Studies}\label{sec:case-studies}
In the coming sections, we give an overview of some of the example equivalences we have proven with \thereflog.
Further details are found in \appref{sec:app-additional-examples} and our Coq development.
In particular, in \appref{sec:app-additional-examples} we discuss an example by \citet{DBLP:conf/popl/SangiorgiV16}, which previous probabilistic logical relations without asynchronous couplings could not prove \cite[Sec.~1.5]{Bizjak:phd}.

\subsection{Lazy/eager Coin}
In this section we give a more detailed proof of the lazy-eager coin example from \cref{sec:intro}.
We will go through the proof step by step but omit the use of \ruleref{rel-pure-l} and \ruleref{rel-pure-r} which should be interleaved with the application of most of the mentioned proof rules.

Recall the definitions of $\lazyprog$ and $\eagerprog$ from \cref{sec:intro}.
% \begin{align*}
%   \eagerprog \eqdef{}
%   &\Let b = \Flip () in
%   \Lam \_ . b \\
%   \lazyprog \eqdef{}
%   &\Let r = \Alloc(\None) in \\
%   &\Lam \_ .
%     \MatchML \deref r with
%       \Some(b) => b
%     | \None => {\begin{array}[t]{l}
%       \Let b = \Flip \TT in \\
%       r \gets \Some(b); \\
%       b
%     \end{array}}
%     end {}
% \end{align*}
%
The goal is to show $\ctxeq{}{\lazyprog}{\eagerprog}{\tunit \to \tbool}$ by first showing $\ctxrefinesS{\lazyprog}{\eagerprog}{\tunit \to \tbool}$ and then $\ctxrefinesS{\eagerprog}{\lazyprog}{\tunit \to \tbool}$.

To show $\ctxrefinesS{\lazyprog}{\eagerprog}{\tunit \to \tbool}$, we first define an intermediate labeled version $\lazyprog'$ of $\lazyprog$ (found in \cref{sec:key-ideas}).
%
% \begin{align*}
%   \lazyprog' \eqdef{}
%   &\Let \lbl = \AllocTape(1) in \\
%   &\Let r = \Alloc(\None) in \\
%   &\Lam \_ .
%     \MatchML \deref r with
%       \Some(b) => b
%     | \None => {\begin{array}[t]{l}
%       \Let b = \Flip(\lbl) in \\
%       r \gets \Some(b); \\
%       b
%     \end{array}}
%     end {}
% \end{align*}
%
By transitivity of contextual refinement and \cref{thm:soundness} it is sufficient to show $\refines{}{\lazyprog}{\lazyprog'}{\tunit \to \tbool}$ and $\refines{}{\lazyprog'}{\eagerprog}{\tunit \to \tbool}$.

The first refinement $\refines{}{\lazyprog}{\lazyprog'}{\tunit \to \tbool}$ is mostly straightforward.
By applying \ruleref{rel-alloc-l} followed by \ruleref{rel-alloc-tape-r} and \ruleref{rel-alloc-r} we are left with the goal of proving that the two thunks are related, given $\spectape{\lbl}{1}{\nil}$, $\progheap{\loc}{\None}$ and $\specheap{\loc'}{\None}$ for some fresh label $\lbl$ and fresh locations on the heap $\loc$ and $\loc'$.
Using \ruleref{rel-na-inv-alloc} we allocate the invariant
\begin{align*}
  \spectape{\lbl}{1}{\nil} \sep (& (\progheap{\loc}{\None} \sep \specheap{\loc'}{\None}) \\
  &\lor{} (\Exists b. \progheap{\loc}{\Some(b)} \sep \specheap{\loc'}{\Some(b)}))
\end{align*}
with some name $\namesp$ that expresses how the $\lbl$ tape is always empty and that \emph{either} both $\loc$ and $\loc'$ contain $\None$ \emph{or} both contain $\Some(b)$ for some $b$.
We continue by \ruleref{rel-rec} after which we open the invariant and do a case distinction on the disjunction in the invariant.
If $\loc$ and $\loc'$ are empty, this is the first time we invoke the function.
We continue using \ruleref{rel-load-l} and \ruleref{rel-load-r} after which we are left with the goal
\begin{align*}
  \refines{}[\top \setminus \namesp]
    {
  \begin{aligned}
    &\Let b = \Flip\TT in \\
    &r \gets \Some(b); b
  \end{aligned}
  }
  {
  \begin{aligned}
    &\Let b = \Flip(\lbl) in \\
    &r \gets \Some(b); b
  \end{aligned}
  }
  {\tunit \to \tbool}
\end{align*}
We continue using \ruleref{rel-rand-erase-r} to couple the two $\Flip$s, we follow by \ruleref{rel-store-l} and \ruleref{rel-store-r} to store the fresh bit on the heaps, we close the invariant (now showing the right disjunct as the locations have been updated) using \ruleref{rel-na-inv-close}, and we finish the case using \ruleref{rel-return} as the program returns the same Boolean $b$ on both sides.

If $\loc$ and $\loc'$ were \emph{not} empty, this is not the first time the function is invoked and we straightforwardly load the same Boolean on both sides using \ruleref{rel-load-l} and \ruleref{rel-load-r} and finish the proof using \ruleref{rel-na-inv-close} and \ruleref{rel-return}.

For the second refinement $\refines{}{\lazyprog'}{\eagerprog}{\tunit \to \tbool}$ we start by allocating the tape on the left using \ruleref{rel-alloc-tape-l} which gives us ownership of a fresh tape $\progtape{\lbl}{1}{\nil}$.
We now couple the $\lbl$ tape with the unlabeled $\Flip()$ on the right using \ruleref{rel-couple-tape-l}.
This gives us that for some $b$ then $\progtape{\lbl}{1}{b}$ and the $\Flip()$ on the right returned $b$ as well.
We continue by allocating the reference on the left using \ruleref{rel-alloc-l} which gives us some location $\loc$ and $\progheap{\loc}{\None}$.
Now, we allocate the invariant
\begin{align*}
  (\progtape{\lbl}{1}{b} \sep \progheap{\loc}{\None}) \lor \progheap{\loc}{\Some(b)}
\end{align*}
which expresses that \emph{either} the location $\loc$ is empty but $b$ is on the $\lbl$ tape, \emph{or} $b$ has been stored at $\loc$.
We are now left with proving that the two thunks are related under this invariant.
We continue using \ruleref{rel-rec} after which we open the invariant using \ruleref{rel-na-inv-open}, do a case distinction on the disjunction, and continue using \ruleref{rel-load-l}.
If the location $\loc$ is empty, we have to show
\begin{align*}
  \refines{}[\top \setminus \namesp]
    {
  \begin{aligned}
    &\Let b = \Flip(\lbl) in \\
    &r \gets \Some(b); b
  \end{aligned}
  }
  { b }
  {\tunit \to \tbool}
\end{align*}
But as we own $\progtape{\lbl}{1}{b}$ we continue using \ruleref{rel-rand-tape-l}, \ruleref{rel-store-l}, \ruleref{rel-na-inv-close} (now establishing the right disjunct as $\loc$ has been updated), and \ruleref{rel-return} as the return value $b$ is the same on both sides.
If the location $\loc$ was \emph{not} empty, we know $\progheap{\loc}{\Some(b)}$ which means \ruleref{rel-load-l} reads $b$ from $\loc$ and we finish the proof using \ruleref{rel-na-inv-close} and \ruleref{rel-return}.

The proof of $\ctxrefinesS{\eagerprog}{\lazyprog}{\tunit \to \tbool}$ is analogous and we have shown the contextual equivalence of the programs $\eagerprog$ and $\lazyprog$.

\subsection{ElGamal public key encryption}
\label{sec:elgamal}
An encryption scheme is seen as secure if no probabilistic polynomial-time (PPT) adversary $\adv$ can break it with non-negligible probability.
A common pattern in cryptographic security proofs are security reductions.
To perform a reduction, one assumes that such an adversary $\adv$ exists, and constructs another PPT adversary $\mathcal{B}$ that, using $\adv$, solves a computational problem $P$ that is believed to be hard.
By contradiction, this means the construction is secure under the assumption that the problem $P$ is hard.
A crucial proof step is showing that $\mathcal{B}$ together with $P$ corresponds to the original construction which can be thought of as the ``soundness'' of the security reduction.
In this section, we use \thereflog{} to show the soundness of a security reduction of the ElGamal public key encryption scheme \cite{DBLP:journals/tit/Elgamal85} to the decisional Diffie-Hellman (DDH) computational assumption.

\newcommand{\keygen}{\mathit{keygen}}
\newcommand{\encrypt}{\mathit{enc}}
\newcommand{\decrypt}{\mathit{dec}}
\newcommand{\query}{\mathit{query}}
\newcommand{\dhreal}{\mathit{DH_{real}}}
\newcommand{\dhrand}{\mathit{DH_{rand}}}
\newcommand{\pkots}{\mathit{PK}}
\newcommand{\tpkots}{\tau{}_{\mathit{\pkots}}}
\newcommand{\tdh}{\tau{}_{\mathit{DH}}}
\newcommand{\pkotsreal}{\pkots\mathit{_{real}}}
\newcommand{\pkotsrealtape}{\pkots^{\mathit{tape}}_\mathit{real}}
\newcommand{\pkotsrand}{\pkots\mathit{_{rand}}}
\newcommand{\pkotsrandtape}{\pkots^{\mathit{tape}}_\mathit{rand}}

\begin{figure}

  \begin{minipage}[t]{0.5\linewidth}
    \begin{align*}
      \keygen &\eqdef{} \begin{aligned}[t]
        \Lam\,\_\,. &\Let sk = \Rand(n) in \\
                                     &\Let pk = g^{sk} in \\
                                     &(sk, pk)
      \end{aligned} \\
      \decrypt &\eqdef{} \Lam\ sk\ (B,X).\ X \cdot{} B^{-sk}
    \end{align*}
  \end{minipage}\hfill
  \begin{minipage}[t]{0.5\linewidth}
    \begin{align*}
      \encrypt \eqdef{} \Lam\ pk\ msg. &\Let b = \Rand(n) in \\
                                &\Let B = g^b in \\
                                &\Let X = msg \cdot{} pk^b in \\
                                &(B, X)
    \end{align*}
  \end{minipage}
  \caption{The ElGamal public key scheme.}
  \label{fig:elgamal}
  \vspace{-1em}
\end{figure}

% \simon{CPA security says you cannot distinguish encryptions of an adversarially chosen message from a random ciphertext. In public key encryption, this property even holds for adversaries that produces ciphertexts for whatever they want (since they have access to the public key) }

The ElGamal construction is a public key encryption scheme consisting a tuple of algorithms $(\keygen, \encrypt, \decrypt)$ whose implementation in \thelang{} is shown in \cref{fig:elgamal}.
The implementation is parameterized by a group $G$ which serves to represent messages, ciphertexts, and keys.
We write $G = (1, \,\cdot{}\,,{-}^{-1})$ for a finite cyclic group of order $|G|$, generated by $g$, and let $n = |G|-1$.
Intuitively, to show that ElGamal encryption is secure it suffices to show that, given the DDH assumption holds for the group $G$, an adversary $\adv$ cannot distinguish an encrypted message from a random ciphertext~(see, \eg{}, \cite[\S{}15.3]{Rosulek:Joy:2020}).
The DDH assumption for a group $G$ says that the two games $\dhreal$ and $\dhrand$ in \cref{fig:ddh} are PPT-indistinguishable which intuitively means that the value $g^{ab}$ looks random, even to someone who has seen $g^{a}$ and $g^{b}$.

% The ElGamal security statement will say that if the DDH assumption is true, then an encrypted message is indistinguishable from random under chosen plaintext attacks, also called CPA\$-secure (see, \eg{}, \cite[\S{}7.2, \S{}15.3]{Rosulek:Joy:2020}.

% $\adv$ cannot distinguish an encrypted message from a random ciphertext

% \simon{only groups for which DDH holds}
% \simon{We're talking about randomized algorithms}
% \simon{hint that we're capturing CPA\$ security }

% We recall that a public key encryption scheme is said to have \emph{one-time chosen-plaintext security in the real-random paradigm} if an adversary $\adv$ cannot distinguish an encrypted message from a random ciphertext~\cite[\S{}15.3]{Rosulek:Joy:2020}.

%
% The adversary gets to examine the public key and an ``encryption oracle'', \ie, a partial application of the encryption function specialized to a particular key.

% invoke the $\query$ function.
% \simon{call the query function---they can also just encrypt themselves with the pk}
%
% This is captured by giving the adversary access to an ``encryption oracle'', \ie, a partial application of the encryption function specialized to a particular key.
%
% \simon{not really preventing brute-force attacks: an attack on the Elgamal scheme reduces to an attack on the DDH scheme/assumption}
% In order to prevent brute-force attacks, $\adv$ is assumed to have limited computational power.

\begin{figure}

  \begin{subfigure}[t]{0.66\linewidth}
    \begin{minipage}[t]{0.5\linewidth}
      \begin{align*}
        &\smash{\pkotsreal \eqdef{}}\\
        &\smash{\Let (sk, pk) = \keygen{} () in }\\
        &\smash{\Let count = \Alloc 0 in }\\
        &\smash{\DumbLet \query = \Lam\ msg. }\\
        &\smash{\quad\If \deref{count} \neq 0 then }\\
        &\smash{\quad\quad\None }\\
        &\smash{\quad\Else }\\
        &\smash{\quad\quad count \gets 1 ; }\\
        \\
        \\
        &\smash{\quad\quad \DumbLet (B, X) = \mhl{\encrypt\ pk\ msg} \In }\\
        &\smash{\quad\quad \Some\ (B, X) }\\
        &\smash{\In (pk, \query)}
      \end{align*}
    \end{minipage}\hfill
    \begin{minipage}[t]{0.5\linewidth}
      \begin{align*}
        &\smash{\pkotsrand \eqdef{}}\\
        &\smash{\Let (sk, pk) = \keygen{} () in }\\
        &\smash{\Let count = \Alloc 0 in }\\
        &\smash{\DumbLet \query = \Lam\ msg. }\\
        &\smash{\quad \If \deref{count} \neq 0 then }\\
        &\smash{\quad\quad \None }\\
        &\smash{\quad \Else }\\
        &\smash{\quad\quad count \gets 1 ; }\\
        &\smash{\quad\quad \Let b = \Rand(n) in }\\
        &\smash{\quad\quad \Let x = \Rand(n) in }\\
        &\smash{\quad\quad \smash{\Let (B, X) = \mhl{(g^b, g^x)} in } }\\
        &\smash{\quad\quad \Some\ (B, X) }\\
        &\smash{\In (pk, \query)}
      \end{align*}
    \end{minipage}
    \caption{The security games.}
  \end{subfigure}
  \begin{subfigure}[t]{0.33\linewidth}
    \begin{minipage}[t]{1.0\linewidth}
      \begin{align*}
        &\smash{C[-] \eqdef{}}\\
        &\smash{\Let (pk,B,C) = - in }\\
        &\smash{\Let count = \Alloc 0 in }\\
        &\smash{\DumbLet \query = \Lam\ msg. }\\
        &\smash{\quad\If \deref{count} \neq 0 then }\\
        &\smash{\quad\quad\None }\\
        &\smash{\quad\Else }\\
        &\smash{\quad\quad count \gets 1 ; }\\
        \\
        \\
        &\smash{\quad\quad \DumbLet X = msg \cdot{} C \In }\\
        &\smash{\quad\quad \Some\ (B, X) }\\
        &\smash{\In (pk, \query)}
      \end{align*}
    \end{minipage}
    \caption{The DH reduction context.\label{fig:redctx}}
  \end{subfigure}
  \caption{Public key security.}
  \label{fig:pksec}
  \vspace{-1em}
\end{figure}

The intuitive notion of encryption scheme security can be made precise\footnote{Several formulations exist in the literature; we take inspiration from the textbook presentation of \citet{Rosulek:Joy:2020}.} as the indistinguishability of two security games, \ie{}, stylized interactions, $\pkotsreal$ and $\pkotsrand$ shown in \cref{fig:pksec}, by a PPT\footnote{Polynomial-time with respect to the security parameter, \ie{} the logarithm of the size of the group for ElGamal.} adversary.
Here we interpret the notion of an ``adversary'' as a program context.
Both security games are initialised by generating a secret/public-key pair $(sk,pk)$, of which $pk$ is returned to the adversary (the context).
The adversary gets to examine the public key and an ``encryption oracle'' $\query$, \ie, a partial application of the encryption function specialized to a particular key.
The difference between $\pkotsreal$ and $\pkotsrand$ lies in the $\query$ function.
While $\pkotsreal$ encrypts the message $msg$ provided as input, $\pkotsrand$ instead returns a randomly sampled ciphertext.
Both games use a counter $count$ to ensure that the $\query$ oracle can be called only once.
One attempt at distinguishing the security games will thus correspond exactly to one attempt at distinguishing $\dhreal$ from $\dhrand$.

\begin{figure}

  \begin{minipage}[t]{0.5\linewidth}
    \begin{align*}
      \dhreal \eqdef{} &\Let a = \Rand(n) in \\
                       &\Let b = \Rand(n) in \\
                       &(g^a, g^b, g^{a b})
    \end{align*}
  \end{minipage}\hfill
  \begin{minipage}[t]{0.5\linewidth}
    \begin{align*}
      \dhrand \eqdef{} &\Let a = \Rand(n) in \\
                       &\Let b = \Rand(n) in \\
                       &\Let c = \Rand(n) in \\
                       &(g^a, g^b, g^c)
    \end{align*}
  \end{minipage}
  \vspace{-0.5em}
  \caption{The Decisional Diffie-Hellman game.}
  \label{fig:ddh}
  \vspace{-0.5em} %sorry...
\end{figure}

% We now use \thereflog{} to prove that the ElGamal scheme is secure under the Decisional Diffie-Hellman assumption (DDH), which states that the $\dhreal$ and $\dhrand$ games in \cref{fig:ddh} are PPT-indistinguishable.
% \simon{Say in text what the DDH assumption is?}
% \simon{Highlight fig. 9}
% %
% Note that the third element in $\dhreal$ is computed using $g^{ab}$, whereas in $\dhrand$ the third element is computed using a randomly chosen exponent.

The idea now is to use \thereflog{} as a step towards reducing indistinguishability of $\pkotsreal$ and $\pkotsrand$ to the indistinguishability of $\dhreal$ and $\dhrand$.
Specifically, we will exhibit a context $\ctx$ and show
\begin{align}
  &\ctxeq{}{\pkotsreal}{\ctx[\dhreal]}{\tpkots}\label{eq:redreal} \\
  &\ctxeq{}{\pkotsrand}{\ctx[\dhrand]}{\tpkots}\label{eq:redrand}
\end{align}
Then we can complete the reduction on paper (outside of \thereflog{}) as follows.\footnote{To mechanize the argument one would need to formalize a notion of PPT and a proof that the context is in fact PPT which is out of scope for the work at hand.}
To prove that the DDH assumption implies public key security, we assume the contrapositive, \ie, that there exists an adversarial context $\adv$ that can distinguish $\pkotsreal$ from $\pkotsrand$.
Using \labelcref{eq:redreal,eq:redrand} we then get that $\adv$ can distinguish $\ctx[\dhreal]$ from $\ctx[\dhrand]$.
But this means that $\adv[\ctx[-]]$ is a context that can distinguish the DDH games, and hence contradicts our assumption, if $\adv[\ctx[-]]$ is PPT.
The context $\ctx$ for \labelcref{eq:redreal,eq:redrand} is given by \cref{fig:pksec}\subref{fig:redctx} (note that the hole is in the first line).
The proof that $\adv[\ctx[-]]$ is PPT if $\adv[-]$ is PPT is outside of the scope of \thereflog{}.

We will only focus on the first equation \eqref{eq:redreal}, since the proof of \eqref{eq:redrand} is similar.
The proof proceeds via an intermediate program, $\pkotsrealtape$, which differs from $\pkotsreal$ only in that the random sampling in \emph{query} is labelled with the tape $\beta$.
By transitivity, it suffices to show that $\ctxeq{}{\pkotsreal}{\pkotsrealtape}{\tau}$ and $\ctxeq{}{\pkotsrealtape}{\ctx[\dhreal]}{\tau}$, as displayed in \cref{fig:real-reduction}.
The first equivalence is trivial.
The essential difference between $\pkotsrealtape$ and $\ctx[\dhreal]$ is that the $\query$ function in $\pkotsrealtape$ samples $b$ lazily, whereas in $\ctx[\dhreal]$, the sampling of $b$ occurs eagerly in the beginning.
The proof now proceeds in a manner similar to the lazy-eager coin example; details can be found in the formalization.

\thereflog{} is well-suited for proving the soundness of the reduction for two reasons.
Firstly, any public key encryption scheme can only be secure if it employs randomized encryption \cite{DBLP:journals/jcss/GoldwasserM84}.
Dealing with randomization is thus unavoidable.
Secondly, reasoning about the encryption oracle involves moving the random sampling used in the encryption across a function boundary (the $\query$ oracle) as we saw.
This part of the argument crucially relies on asynchronous couplings.
Systems like EasyCrypt and CertiCrypt handle this part of the argument through special-purpose rules for swapping statements that allows moving the random sampling outside the function boundary.
However, it crucially relies on the fact that these works consider first-order languages with global state and use syntactic criteria and assertions on memory disjointness.

Note moreover that our security formulation makes crucial use of the fact that \thelang{} is higher-order, randomized, and supports local state to return the $\query$ closure as a first class value.
This allows us to capture the textbook cryptographic notion of adversaries and of a (closed-box) ``oracle'' precisely using standard notions such as higher-order functions and contextual equivalence, without introducing special linguistic and logical categories of adversaries parameterized by a set of oracles.

\begin{figure}

  \begin{minipage}[t]{0.33\linewidth}
    \makebox[.8\textwidth]{$\pkotsreal$} $\ctxeqrel$
    \begin{align*}
      \\
      &\smash{\Let sk = \Rand(n) in} \\
      &\smash{\Let pk = g^{sk} in} \\
      \\
      &\smash{\Let count = \Alloc 0 in} \\
      &\smash{\DumbLet \query = \Lam\ msg.} \\
      &\smash{\quad\If \deref{count} \neq 0 then} \\
      &\smash{\quad\quad\None} \\
      &\smash{\quad\Elsenospac} \\
      &\smash{\quad\quad count \gets 1 ;} \\
      &\smash{\quad\quad \DumbLet b = \Rand(n) \In} \\
      &\smash{\quad\quad \DumbLet B = g^b \In} \\
      \\
      &\smash{\quad\quad \DumbLet X = msg \cdot{} pk^b \In} \\
      &\smash{\quad\quad \Some\ (B, X)} \\
      &\smash{\In (pk, \query)}
    \end{align*}
  \end{minipage}\hfill
  \begin{minipage}[t]{0.33\linewidth}
    \makebox[.8\textwidth]{$\pkotsrealtape{}$} $\ctxeqrel$
    \begin{align*}
      &\smash{\mhl{\Let \beta = \AllocTape(n) in}} \\
      &\smash{\Let sk = \Rand(n) in} \\
      &\smash{\Let pk = g^{sk} in} \\
      \\
      &\smash{\Let count = \Alloc 0 in} \\
      &\smash{\DumbLet \query = \Lam\ msg.} \\
      &\smash{\quad \If \deref{count} \neq 0 then} \\
      &\smash{\quad\quad \None} \\
      &\smash{\quad \Elsenospac} \\
      &\smash{\quad\quad count \gets 1 ;} \\
      &\smash{\quad\quad \DumbLet b = \smash{\mhl{\Rand(n, \beta)}} \In} \\
      &\smash{\quad\quad \DumbLet B = g^b \In} \\
      &\smash{\quad\quad \DumbLet C = pk^b \In} \\
      &\smash{\quad\quad \DumbLet X = msg \cdot{} C \In} \\
      &\smash{\quad\quad \Some\ (B, X)} \\
      &\smash{\quad \In (pk, \query)}
    \end{align*}
  \end{minipage}\hfill
  \begin{minipage}[t]{0.33\linewidth}
    \makebox[.8\textwidth]{$\ctx[\dhreal]$}
    \begin{align*}
      &\smash{\DumbLet (pk, \mhl{B, C}) =} \\
      &\smash{\quad\Let a = \Rand(n) in} \\
      &\smash{\quad\mhl{\Let b = \Rand(n) in}} \\
      &\smash{\quad(g^a, \mhl{g^b, g^{a b}}) \In} \\
      &\smash{\Let count = \Alloc 0 in} \\
      &\smash{\DumbLet \query = \Lam\ msg. } \\
      &\smash{\quad \If \deref{count} \neq 0 then} \\
      &\smash{\quad\quad \None} \\
      &\smash{\quad\Elsenospac} \\
      &\smash{\quad\quad count \gets 1 ;} \\
      \\
      \\
      \\
      &\smash{\quad\quad \Let X = msg \cdot{} C in} \\
      &\smash{\quad\quad \Some\ (B, X)} \\
      & \In (pk, \query)
    \end{align*}
  \end{minipage}
  \caption{The ``real'' direction of the security reduction.}
  \label{fig:real-reduction}
  \vspace{-1em}
\end{figure}

\subsection{Hash functions}

\newcommand{\eagerhash}{\mathit{eager\_hash}}
\newcommand{\lazyhash}{\mathit{lazy\_hash}}
\newcommand{\initmap}{\mathit{init\_map}}
\newcommand{\getmap}{\mathit{get}}
\newcommand{\setmap}{\mathit{set}}
\newcommand{\sampleall}{\mathit{sample\_all}}
\newcommand{\alloctapes}{\mathit{alloc\_tapes}}
\newcommand{\inithashrng}{\mathit{init\_hash\_rng}}
\newcommand{\maxint}{\texttt{MAX}}
\newcommand{\initboundedrng}{\mathit{init\_bounded\_rng}}

\newcommand{\hashcodeshort}{
  \begin{minipage}[t]{0.4\linewidth}
    \centering
    \begin{align*}
      &\eagerhash \eqdef{} \\
      &\quad \Lam n.
        \begin{aligned}[t]
          &\Let m = \initmap\ () in \\
          &\sampleall\ m\ (n + 1); \\
          &(\Lam k.
            \MatchML \getmap\ m\ k with
            \Some(b) => b
            | \None => \False
            end) {}
        \end{aligned}
    \end{align*}
  \end{minipage}
  \begin{minipage}[t]{0.5\linewidth}
    \begin{align*}
      &\lazyhash \eqdef{} \\
      &\quad \Lam n.
        \begin{aligned}[t]
          &\Let vm = \initmap\ \TT in \\
          &\Let tm = \initmap\ \TT in \\
          &\alloctapes\ tm\ (n + 1); \\
          &(\Lam k.
            \MatchML \getmap\ vm\ k with
            \Some(b) => b
            |\None =>
            {\MatchMLnarrow \getmap\ tm\ k with
            \Some(\lbl) =>
            {\begin{array}[t]{l}
               \Let b = \Flip(\lbl) in \\
               \setmap\ vm\ b; b
             \end{array}}
            | \None => \False
            end {}}
            end)
        \end{aligned}
    \end{align*}
  \end{minipage}
}
When analyzing data structures that use hash functions, one commonly
models the hash function under the
\emph{uniform hash assumption} or the \emph{random oracle model} \cite{DBLP:conf/ccs/BellareR93}.
That is, a hash function $h$ from a set of keys $K$ to values $V$
behaves as if, for each key $k$, the hash $h(k)$ is randomly sampled
from a uniform distribution over $V$, independently of all the other
keys.  Of course, hash functions are not known to satisfy
this assumption perfectly, but it can nevertheless be a useful modeling
assumption for analyzing programs that use hashes.

The function $\eagerhash$ in \cref{fig:eager-hash} encodes
such a model of hash functions in $\thelang{}$. (We explain
the reason for the ``eager'' name later.) Given a
non-negative integer $n$, executing $\eagerhash\ n$ returns a hash
function with $K = \{0, \dots, n\}$ and $V = \bool$.
To do so, it initializes a mutable map $m$ and then calls $\sampleall$,
which samples a Boolean $b$ with $\Flip$ for each key $k$ and stores the results in $m$.
These Booleans serve as the hash values.
On input $k$, the hash function returned by $\eagerhash$ looks up $k$ in the map $m$
and returns the result, with a default value of $\False$ if $k \not\in K$.

\begin{figure}
  \centering
  \hashcodeshort
  \caption{Eager and lazy models of hash functions.}
  \label{fig:eager-hash}
  \vspace{-1em}
\end{figure}

However, this model of uniform hash functions can be inconvenient for
proofs because all of the random hash values are sampled
\emph{eagerly} when the function is initialized. To overcome this, an important
technique in pencil-and-paper proofs is to show that the hash values
can be sampled \emph{lazily}~(see, \eg{}, \citet{MittelbachF21}).  That is, we only sample a
key $k$'s hash value when it is hashed for the first time. This lets
us more conveniently couple that sampling step with some step in
another program.

Motivated by applications to proofs in cryptography, \citet{AlmeidaBBBDGL0S19} formalized in EasyCrypt a proof of equivalence between an eager and lazy random oracle.
Although sufficient for their intended application, this proof was done in the context of a language that uses syntactic restrictions to model the hash function's private state.
To the best of our knowledge, no such equivalence proof between lazy and eager sampling has previously been given for a language with higher-order state and general references.

% Although this models the uniform hash assumption,
% this version turns out to be inconvenient for relational reasoning. The issue
% is that the random hash values for all keys are sampled \emph{eagerly}
% when the hash function is initialized. Therefore, to use the usual coupling
% rules, one must couple the random hash values for every key during initialization.
%
% \simon{Do we have a reference to ``long standing proof technique''?}
% \joe{I don't have original reference, just textbooks. I think it's folklore so I can't find an attribution.}
% To avoid this issue, a long standing proof technique in
% security proofs using the random oracle model of hashes is to show that the random oracle can be
% sampled \emph{lazily}. That is, we only sample a key $k$'s hash value
% when it is hashed for the first time. This lets us more conveniently
% couple that sampling step with some step in another program.

As an application of \thereflog{}, we prove such an equivalence in \thelang{}.
The function $\lazyhash$ shown in \cref{fig:eager-hash} encodes
the lazy sampling version of the random hash generator.  For its
internal state, the lazy hash uses two mutable maps: the tape map $tm$ stores
tapes to be used for random sampling, and the value map $vm$ stores
the previously sampled values for keys that have been hashed. After
initializing these maps, it calls $\alloctapes$, which allocates a
tape for each key $k \in K$ and stores the associated tape in $tm$,
but does not yet sample hashes for any keys.  The hash function
returned by $\lazyhash$ determines the hash for a key $k$ in two
stages. It first looks up $k$ in $vm$ to see if $k$ already has a
previously sampled hash value, and if so, returns the found
value. Otherwise, it looks up $k$ in the tape map $tm$. If no tape is
found, then $k$ must not be in $K$, so the function returns
$\False$. If a tape $\lbl$ is found, then the code samples a Boolean
$b$ from this tape with $\Flip$, stores $b$ for the key $k$ in $vm$,
and then returns $b$.

We prove that the eager and lazy versions are contextually equivalent, that is,
$\ctxeq{}{\eagerhash\ n}{\lazyhash\ n}{\tint \rightarrow \tbool}$.
The core idea behind this contextual equivalence proof is to maintain
an invariant between the internal state of the two hash
functions.  Let $m$ be the internal map used by the eager hash and let
$tm$ and $vm$ be the tape and value maps, respectively, for the lazy
hash. Then, at a high level, the invariant maintains the following
properties:
\begin{enumerate}
  \item $\dom(m) = \dom(tm) = \{0, \dots, n\}$.
  \item For all $k \in \{0, \dots, n\}$, if $m[k] = b$ then either
    \begin{enumerate}
    \item $vm[k] = b$, or
    \item $vm[k] = \bot$ and  $tm[k] = \lbl$ for some tape label $\lbl$ such that $\progtape{\lbl}{1}{b}$.
    \end{enumerate}
\end{enumerate}
Case (a) and (b) of the second part of this invariant capture the two possible states each key $k$ can be in.
Either the hash of $k$ has been looked up before (case a), and so the sampled value stored in $vm$ must match that of $m$, or it has not been looked up (case b) and the tape for the key must contain the same value as $m[k]$ for its next value.

To establish this invariant when the hashes are initialized, we
asynchronously couple the eager hash function's $\Flip$ for key $k$
with a tape step for the tape $\lbl$ associated with $k$ in the lazy
table.  The invariant ensures that the values returned by the two hash
functions will be the same when a key $k$ is queried.  The cases of
the invariant correspond to the branches of the lazy function's
match statements: if the key $k$ is in $K$ and has been queried
before, the maps will return the same values found in $m$ and $vm$. If
it has not been queried before, then $\Flip$ in the lazy
version will sample the value on the tape for the key, which matches
$m[k]$. Moreover, the update that writes this sampled value to $vm$
preserves the invariant, switching from case (b) to case (a) for
the queried key.

We have used this more convenient lazy encoding to verify examples that use hash functions.
For instance, one scheme to implement random number generators is to use a cryptographic hash function~\cite{Barker2015}.
The program $\inithashrng$ in \cref{fig:hashrng} implements a simplified version of such a scheme.

\begin{figure}

  \centering
  \begin{subfigure}[t]{0.4\linewidth}
    \begin{align*}
      & \inithashrng \eqdef{} \\
      & \quad \Lam \_.
      \begin{aligned}[t]
        &\Let f = \lazyhash\ \maxint in \\
        &\Let c = \Alloc{0} in \\
        &(\Lam \_.
          \begin{aligned}[t]
            &\Let n = \deref{c} in \\
            &\Let b = f\ n in \\
            &c \gets n + 1; b)
          \end{aligned}
      \end{aligned}
    \end{align*}
    \caption{Hashing random number generator.\label{fig:hashrng}}
  \end{subfigure}
  \begin{subfigure}[t]{0.5\linewidth}
    \begin{align*}
      & \initboundedrng \eqdef{} \\
      & \quad \Lam \_.
        \begin{aligned}[t]
          &\Let c = \Alloc{0} in \\
          &(\Lam \_.
            \begin{aligned}[t]
              &\Let n = \deref{c} in \\
              &\Let b =
                \begin{aligned}[t]
                  & \If n \leq \maxint then \Flip \TT \\
                  & \Else \False in \\
                \end{aligned} \\
              &c \gets n + 1; b)
            \end{aligned}
        \end{aligned}
    \end{align*}
    \caption{Bounded random number generator.\label{fig:bndrng}}
  \end{subfigure}
  \caption{Random number generators.}
  \vspace{-1em}
\end{figure}

When run, $\inithashrng$ generates a lazy hash function $f$ for the key space $K = \{0, \dots, \maxint\}$
for some fixed constant $\maxint$. It also allocates a counter $c$ as a reference initialized to $0$.
It returns a sampling function, let us call it $h$, that uses $f$ and $c$ to generate random Booleans. Each time
$h$ is called, it loads the current value $n$ from $c$ and hashes $n$ with $f$ to get
a Boolean $b$. It then increments $c$ and returns the Boolean $b$.
Repeated calls to $h$ return independent, uniformly sampled Booleans, so long as we make no more than $\maxint$ calls.

We prove that $\inithashrng$ is contextually equivalent to a ``bounded'' random number generator $\initboundedrng$ in \cref{fig:bndrng} that directly calls $\Flip$.
The proof works by showing that, so long as $n \leq \maxint$, then each time a sample is generated,
the value of $n$ will not have been hashed
before. Thus, we may couple the random hash value with the $\Flip$ call in $\initboundedrng$.
This argument relies on the fact that the counter $c$ is private, encapsulated
state, which is easy to reason about using the relational judgment since \thereflog{} is a separation logic.

\subsection{Lazily sampled big integers}

\newcommand{\samplelazy}{\mathit{sample\_lazy\_int}}
\newcommand{\sampleint}{\mathit{sample32}}
\newcommand{\sampleeager}{\mathit{sample256}}
\newcommand{\getnext}{\mathit{get\_next}}
\newcommand{\cmplist}{\mathit{cmp\_list}}
\newcommand{\cmplazy}{\mathit{cmp\_lazy}}
\newcommand{\cmpZ}{\mathit{cmp}}
% \begin{figure}
%   \begin{align*}
%     \samplelazy \eqdef{} & \Lam \_. (\AllocTape, \Alloc \None) \\
%     \\
%     \getnext \eqdef{}
%     &\Lam\ \lbl\ r. \\
%     &\MatchML \deref{r} with
%       \Some\ v => v
%       | \None =>
%       {\begin{array}[t]{l}
%       \Let z = \sampleint\ \lbl in \\
%       \Let next = \Alloc \None in \\
%       r \gets \Some\ (z, next); \\
%       (z, next)
%       \end{array}}
%       end {} \\
%     \\
%     \cmplist \eqdef{}
%     &\Rec f n\ \lbl_1\ l_1\ \lbl_2\ l_2 = \\
%     &\If n = 0 then 0 \\
%     &\Elsenospac \\
%     &\quad\Let (z_1, l_1') = \getnext\spac \lbl_1\spac l_1 in \\
%     &\quad\Let (z_2, l_2') = \getnext\spac \lbl_2\spac l_2 in \\
%     &\quad\Let res = \cmpZ\ (z_1, z_2) in \\
%     &\quad\If res = 0 then \\
%     &\quad\quad f\ (n-1)\ \lbl_1\ l_1'\ \lbl_2\ l_2' \\
%     &\quad\Elsenospac\ res\\
%     \\
%     \cmplazy \eqdef{}
%     &\Lam\ (x_1, x_2). \\
%     &\Let (\lbl_1, l_1) = x_1 in \\
%     &\Let (\lbl_2, l_2) = x_2 in \\
%     &\If l_1 = l_2 then 0 \\
%     &\Elsenospac \cmplist\ 8\ \lbl_1\ l_1\ \lbl_2\ l_2
%   \end{align*}
%   \caption{Implementation of lazily-sampled integers.}
%   \label{fig:lazy-int}
% \end{figure}
%

% Our last example is motivated by the need to generate random numeric \emph{priorities} in a search data structure called a treap~\cite{treappaper?aragon?}. The full details of this data structure
% are not needed to understand the example, but what is important to know
% is that each time a key is inserted into a treap, it must generate

Certain randomized data structures, such as treaps~\cite{SeidelA96},
need to generate random \emph{priorities} as operations are
performed.  One can view these priorities as an
abstract data type equipped with a total order supporting two operations: (1) a \emph{sample} function
that randomly generates a new priority according to some distribution, and (2) a \emph{comparison} operation
that takes a pair of priorities $(p_1, p_2)$ and returns $-1$ (if $p_1 < p_2$), $0$ (if $p_1 = p_2)$, or $1$ (if $p_2 < p_1$).
The full details of how priorities are used in such data structures are not relevant
here. Instead, what is important to know is that it is ideal to avoid
\emph{collisions}, that is, sampling the same priority multiple
times.

A simple way to implement priorities is to represent them as integers sampled from some fixed set $\{0, \dots, n\}$.
However, to minimize collisions, we may need to make $n$ very large.
But making $n$ large has a cost, because then priorities requires more random bits to generate and more space to store.
An alternative is to \emph{lazily} sample the integer that represents the priority.
Because we only need to compare priorities, we can delay sampling bits of the integer until they are needed to resolve ties during comparisons.
A lazily-sampled integer can be encoded as a pair of a tape label $\lbl$ and a linked list of length at most $N$, where each node in the list represents a \emph{digit} of the integer in base $B$, with the head of the list being the most significant digit.

In \appref{sec:app-additional-examples}, we describe such an implementation of lazily-sampled integers, with $N = 8$ and $B = 2^{32}$. Our Coq development contains a proof that this implementation is contextually
equivalent to code that eagerly samples a 256-bit integer by bit-shifting and adding
8 32-bit integers. Crucially, this contextual equivalence is at an \emph{abstract} existential type $\tau$. Specifically,
we define the type of abstract priorities
$ \tau \eqdef \Exists \alpha. (\tunit \rightarrow \alpha) \times ((\alpha \times \alpha) \rightarrow \tint) $.
Then we have the equivalence
$ \ctxeq{}{(\samplelazy, \cmplazy)}{(\sampleeager, \cmpZ)}{\tau} $
where $\cmpZ$ is just primitive integer comparison.
The proof uses tapes to presample the bits of the lazy integer and couples
these with the eager version. The $\cmplazy$ function traverses
and mutates the linked lists representing the integers being compared, which separation logic
is well-suited for reasoning about.
% The starting point for the proof is that when $\sampleeager$ samples
% the 8 32-bit integers needed to eagerly generate the 256-bit integer, we
% couple these samples with identical samples on the tape $\lbl$
% generated by $\samplelazy$.
% This establishes the following invariant:
% if we combine the digits of a lazy int that have already been
% sampled, plus the remaining digits on the tape $\lbl$, the result
% represents an integer that is equivalent to the corresponding one
% generated by $\sampleeager$. This invariant ensures that comparisons
% between two lazy integers have the same result as calls to $\cmpZ$ on the eager integers. Moreover, the invariant is preserved by
% calls to $\getnext$ during $\cmplist$, since $\getnext$ moves digits from the
% tape to the linked list representing the integer.

%%% Local Variables:
%%% mode: latex
%%% TeX-master: "../clutch-popl24"
%%% End:

\section{Counterexamples}\label{sec:counterexamples}
\newcommand{\lorprog}{\mathit{flip\_or}}
\newcommand{\flipprog}{\mathit{flip}}

This section justifies some design choices in \thereflog{} by presenting counterexamples showing the unsoundness of two variants of the logic.
In the first counterexample, we show that annotating sampling statements with tape labels is needed in our current formulation of the logic, since their omission leads to unsoundness.
In the second, we show that combining prophecy variables~\cite{DBLP:journals/pacmpl/JungLPRTDJ20} with the usual coupling rules of pRHL (without presampling) is unsound, implying that presampling cannot somehow be implemented in terms of prophecy variables.

% At first, the high-level relational rules for reasoning about presampling tapes may cause one to believe that once the idea of presampling arises, the rest is mostly straightforward to state and define.
% One may also be lead to believe that presampling tapes are just an application of prophecy variables \cite{DBLP:journals/pacmpl/JungLPRTDJ20}.
% But \emph{neither} is the case: a great deal of care goes into phrasing the rules of the operational semantics and the relational logic if you do not want to end up with an unsound system!
% We illustrate below the subtleties using two counterexamples.

\subsection{Syntactic restriction on presampling}
One may wonder whether it is necessary for tapes and labels to appear in the program and program state, but they do in fact play a subtle yet crucial role.
Consider the following program $\lorprog$ that applies a logical disjunction to two fresh samples:
\begin{align*}
  \lorprog \eqdef{}
  &\Let x = \Flip \TT in \\
  &\Let y = \Flip \TT in \\
  &x \mathbin{||} y
\end{align*}
and compare it to the program $ \flipprog \eqdef{} \Flip \TT$ that just samples a bit.
These two programs are obviously \emph{not} contextually equivalent: with probability $3/4$ the program $\lorprog$ will return $\True$ whereas the program $\flipprog$ only does so with probability $1/2$.
Yet, if we introduce a rule for $\Flip$ that could draw from \emph{any} presampling tape (\ie{}, without requiring sampling statements to be annotated with the tape they will draw from), the logic would allow one to ``prove'' that they are equivalent.

Assume the following (unsound!) rule
\begin{mathpar}
  \inferH{rel-tape-unsound}
  {
    \progtape{\lbl}{1}{b \cons \vec{b}} \\
    \progtape{\lbl}{1}{\vec{b}} \wand \refines{\Delta}[\mask]{\fillctx\lctx[b]}{\expr_2}{\type}
  }
  {\refines{\Delta}[\mask]{\fillctx\lctx[\Flip()]}{\expr_2}{\type}}
\end{mathpar}
that says that when sampling on the left-hand side, we may instead draw a bit $b$ from \emph{some} prover-chosen presampling tape $\lbl$.
To see why this rule cannot be sound, we will show $\refines{}{\flipprog}{\lorprog}{\tbool}$.

First, we introduce two tapes with resources $\progtape{\lbl_{1}}{1}{\nil}$ and $\progtape{\lbl_{2}}{1}{\nil}$ on the left-hand side, either explicitly allocated in code as in \thereflog{} or as pure ghost resources, if that is possible in our hypothetical logic.
Second, we couple the tape $\lbl_{1}$ with the $x$-sampling and $\lbl_{2}$ with the $y$-sampling using \ruleref{rel-couple-tape-l} such that we end up with $\progtape{\lbl_{1}}{1}{b_{1}}$ and $\progtape{\lbl_{2}}{1}{b_{2}}$ and the goal $\refines{}{\Flip\TT}{b_{1} \mathbin{||} b_{2}}{\tbool}$.
Finally, we do a case distinction on both $b_{1}$ and $b_{2}$: if both are $\True$, or both are $\False$, it does not matter which tape we use when applying \ruleref{rel-tape-unsound}.
If, on the other hand, only $b_{i}$ is $\True$, we choose $\lbl_{i}$ and apply \ruleref{rel-tape-unsound} which finishes the proof.

The crucial observation is that by labeling tapes in the program syntax, however, we prevent \emph{the prover} from doing case analysis on presampled values to decide which tape to read---the syntax will dictate which tape to use and hence which value to read.
Concretely, in \thelang{}, unlabeled $\Flip$s always reduce uniformly at random and only labeled sampling statements will read from presampling tapes which prevents us from proving the unsound \ruleref{rel-tape-unsound}.

Besides motivating why soundly allowing presampling is subtle, this counterexample also emphasizes why the fact that labels appear in the program and in the program syntax is important.
We do not claim that these annotations are absolutely necessary for some kind of presampling to be sound, as some very different formulation of the logic might be able to avoid them, but like for \emph{prophecy variables} \cite{DBLP:journals/pacmpl/JungLPRTDJ20} where similar ``ghost information'' is needed in the actual program code, it is not obvious how to do without it.
We remind the reader that presampling tapes nevertheless remain a proof-device as tapes can be erased through refinement as discussed in \cref{sec:key-ideas}.

\subsection{Incompatibility with prophecy variables}
\newcommand{\NewProph}{\operatorname{\langkw{NewProph}}}
\newcommand{\ResolveProph}[2]{\operatorname{\langkw{Resolve}} #1 \operatorname{\langkw{to}} #2}
\newcommand{\Proph}{\operatorname{\mathsf{Proph}}}
\newcommand{\andprog}{\mathit{flip\_proph}}

Presampling tapes bear \emph{some} resemblance to prophecy variables in that they give us the means to talk about the future.
However, prophecy variables, as previously developed in the context of Iris \cite{DBLP:journals/pacmpl/JungLPRTDJ20}, are unsound for the (synchronous) coupling logic as illustrated below.

Assume the existence of two operators $\NewProph$ and $\ResolveProph{p}{b}$ in our programming language and their (unsound for \thereflog{}!) Hoare-triple specifications found below.
\begin{mathpar}
  \inferH{wp-newproph-unsound}
  {}
  { \hoare{\TRUE}{\NewProph}{ p . \exists b \in \mathbb{B} . \Proph(p, b) }}

  \inferH{wp-resolve-unsound}
  {}
  { \hoare{ \Proph(p, b) \ast b' \in \mathbb {B} }{\ResolveProph{p}{b'}}{b = b'} }
\end{mathpar}
The specifications give us access to \emph{Boolean one-shot prophecies} \cite{DBLP:journals/pacmpl/JungLPRTDJ20}.
$\NewProph$ allocates a fresh prophecy variable $p$ and a resource $\Proph(p, b)$ that tracks its future resolution $b$.
Given ownership of $\Proph(p, b)$ then $\ResolveProph{p}{b'}$ resolves the prophecy variable $p$ to a value $b'$ and knowledge that $b = b'$ was the case all along.
To see why these operations and rules cannot be sound in the coupling logic, we will show $\refines{}{\andprog}{\flipprog}{\tbool}$ where
\begin{align*}
  \andprog \eqdef{}
  &\Let p = \NewProph in \\
  &\Let x = \Flip \TT in \\
  &\Let y = \Flip \TT in \\
  &\hspace{-0.15em}\ResolveProph{p}{y}; \\
  &x \mathbin{\&\&} y
\end{align*}
which cannot be the case as $\andprog$ returns $\True$ only with probability $1/4$.

We unfold the relational judgment and apply \ruleref{wp-newproph-unsound} which gives us a prophecy about $y$ and its future resolution $b$.
If $b$ is true, the evaluation on the left is predetermined to be $x \mathbin{\&\&} \True = x$.
By coupling the sampling of $x$ with the $\Flip\TT$ on the right using \ruleref{rel-couple-rands}, we finish using \ruleref{rel-rand-l} and \ruleref{wp-resolve-unsound}.
On the other hand, if $b$ is false, the evaluation on the left is predetermined to be $x \mathbin{\&\&} \False = \False$.
We apply \ruleref{rel-rand-l} first and couple the sampling of $y$ with the $\Flip\TT$ on the right using \ruleref{rel-couple-rands} and finish using \ruleref{wp-resolve-unsound}.

The counterexample shows that prophecy variables are unsound for the coupling logic, for the same reason that presampling is unsound without syntactic tape labels: If the prover can predict the outcomes of random samples ahead of time, it gives them too much power to choose which sampling they couple with.

%%% Local Variables:
%%% mode: latex
%%% TeX-master: "../clutch-popl24"
%%% End:

\section{Coq Formalization}\label{sec:coq}

All the results presented in the paper, including the background on probability theory, the formalization of the logic, and the case studies have been formalized in the Coq proof assistant~\cite{coq}.
The results about probability theory are built on top of the Coquelicot library~\cite{DBLP:journals/mics/BoldoLM15}, extending their results to real series indexed by countable types.

Although we build our logic on top of Iris~\cite{irisjournal}, significant work is involved in formalizing the operational semantics of probabilistic languages, our new notion of weakest precondition that internalizes the coupling-based reasoning, and the erasure theorem that allows us to conclude the existence of a coupling.
Our development integrates smoothly with the Iris Proof Mode \cite{DBLP:conf/popl/KrebbersTB17} and we have adapted much of the tactical support from ReLoC~\cite{DBLP:journals/lmcs/FruminKB21} to reason about the relational judgment.

%%% Local Variables:
%%% mode: latex
%%% TeX-master: "../clutch-popl24"
%%% End:

\section{Related Work}

\paragraph*{Separation logic}
Relational separation logics have been developed on top of Iris for a range of properties, such as contextual refinement \cite{DBLP:journals/lmcs/FruminKB21, DBLP:conf/popl/KrebbersTB17, DBLP:journals/pacmpl/TimanySKB18, DBLP:journals/pacmpl/TimanyB19}, simulation \cite{DBLP:conf/sosp/ChajedTKZ19, DBLP:journals/corr/abs-2109-07863, DBLP:journals/pacmpl/GaherSSJDKKD22}, and security \cite{DBLP:conf/sp/FruminKB21, DBLP:journals/pacmpl/GregersenBTB21,DBLP:journals/pacmpl/GeorgesTB22}.
The representation of the right-hand side program as a resource is a recurring idea, but our technical construction with run ahead is novel.
With the exception of \citet{DBLP:journals/pacmpl/TassarottiH19}, probabilistic languages have not been considered in Iris.
\citeauthor{DBLP:journals/pacmpl/TassarottiH19} develop a logic to show refinement between a probabilistic program and a semantic model, not a program.
The logic relies on couplings, but it requires synchronization of sampling.

In \citet{DBLP:journals/pacmpl/BatzKKMN19}, a framework in which logical assertions are functions ranging over the non-negative reals is presented.
% Reasoning is done through the use of predicate transformers, that semantically correspond to the expected value of these functions over the final distribution.
The connectives of separation logic are given an interpretation as maps from pairs of non-negative reals to the positive reals.
% While our work focuses on reasoning about contextual equivalence, they
This work focuses on proving quantitative properties of a single program, \eg{}, bounding the probability that certain events happen.
A variety of works have developed separation logics in which the separating conjunction models various forms of probabilistic independence~\cite{DBLP:journals/pacmpl/BartheHL20, DBLP:conf/lics/BaoDH021, DBLP:journals/pacmpl/BaoGHT22}.
For example, the statement $P \sep Q$ is taken to mean ``the distribution of $P$ is independent from the distribution of $Q$''.

Prophecy variables \cite{DBLP:conf/lics/AbadiL88, DBLP:journals/tcs/AbadiL91} have been integrated into separation logic in both unary \cite{DBLP:journals/pacmpl/JungLPRTDJ20} and relational settings \cite{DBLP:journals/lmcs/FruminKB21}.
The technical solution uses program annotations and physical state reminiscent of our construction with presampling tapes, but prophecy resolution is a physical program step, whereas presampling in our work is a logical operation.
Prophecies can also be erased through refinement \cite{DBLP:journals/lmcs/FruminKB21}.

\paragraph*{Probabilistic couplings}
Probabilistic couplings are a technique from probability theory that can be used to prove equivalences between distributions or mixing times of Markov chains~\cite{SPS_1983__17__243_0}.
In computer science, they have been used to reason about relational properties of programs such as equivalences~\cite{DBLP:conf/lpar/BartheEGHSS15} and differential privacy~\cite{DBLP:conf/ccs/BartheFGGHS16}.
%These techniques have also been extended to higher-order programs~\cite{DBLP:journals/pacmpl/AguirreBGGKS21}.
However, these logics requires the sampling points on both programs to be synchronized in order to construct couplings.
In a higher-order setting, the logic by~\citet{aguirre_relational_2018} establish so-called ``shift couplings'' between probabilistic streams that evolve at different rates, but these rules are ad-hoc and limited to the stream type.
Also in the higher-order setting, \citet{DBLP:journals/pacmpl/AguirreBGGKS21} use couplings to reason about adversarially-defined properties, however they only support synchronous couplings, first-order global state, and use a graded state monad to enforce separation of adversary memories.

\paragraph*{Logical relations}
Step-indexed logical relations have been applied to reason about contextual equivalence of probabilistic programs in a variety of settings.
\citet{DBLP:conf/fossacs/BizjakB15} develop logical relations for a language similar to ours, although only with first-order state.
This work has since been extended to a language with continuous probabilistic choice (but without state and impredicative polymorphism) \cite{DBLP:journals/pacmpl/WandCGC18}, for which equivalence is shown by establishing a measure preserving transformation between the sources of randomness for both programs.
Recently, this was further extended to support nested inference queries~\cite{DBLP:journals/pacmpl/ZhangA22}.

Another line of work~\cite{DBLP:journals/tcs/LagoG21,DBLP:journals/pacmpl/LagoG22} uses so called differential logical relations to reason about contextual distance rather than equivalence.
Programs are related using metrics rather than equivalence relations, which allows to quantify how similar programs are.

\paragraph*{Cryptographic frameworks}
% CertiCrypt
% - first order language, loops, ground state, sampling
% - tailored towards cryptographic game-playing proofs
% - pRHL for equivalence
% - swapping rules proven as part of pRHL based on syntactic independence
% - syntactic lazy/eager sampling support
% - failure event bound unary logic
CertiCrypt \cite{DBLP:conf/itp/BartheGB10, DBLP:conf/popl/BartheGB09} is a framework for cryptographic game-playing proofs written in a simple probabilistic first-order while-language (``pWhile'').
% 
% The pWhile language features loops, first-order procedures, ground state, and random sampling, and can be given a simple denotational semantics in terms of the subdistribution monad on program states.
% 
CertiCrypt formalizes a denotational semantics for pWhile in Coq and supports reasoning about the induced notion of program equivalence via a pRHL, and provides dedicated tactics for lazy/eager sampling transformations.
% due to their widespread use in cryptographic security proofs \cite{BellareR04,BellareR06}.
These kind of transformations are non-trivial for expressive languages like ours.
% 
% Statements can be swapped if they depend and act on disjoint sets of statically determined memory locations.
% 
% A sampling may be moved across program contexts which, according to syntactic analysis, do not depend on the sampled value.
% 
% An advantage of working in separation logic is that we get a native notion of ownership, while
% 
% Since pRHL is not a separation logic with a notion of ownership, care has to be taken to ensure that adversarial code does not read secrets from the program state.
% 
% Due to the lack of higher order functions, adversarial code cannot be simply represented as an abstract unknown computation, but forms a separate category of code with memory locations disjoint from the user program.
% 
% In \thereflog{}, we can use the usual frame rule to reason locally about resource use, including that of unknown code.
% 
% - also unary logic for `'failure events'' (later based on union bounds).
% Not sure we should mention this as we don't relate to it, but it is a distinguishing point.
% 
CertiCrypt also provides a quantitative unary logic.

% EasyCrypt
% - same core language, added module system
% - recent: support for query-counting for tracking adversarial complexity
% - not foundational but more automation
% Frankly I'm not sure we have much to say here. The wider EC crowd developed couplings as a proof method, but the EC tool seems to distinguish itself from CertiCrypt in the above points, none of which are particularly pertinent to what we do.
EasyCrypt \cite{DBLP:conf/fosad/BartheDGKSS13} is a standalone prover for higher-order logic building on CertiCrypt's ideas. It leverages the first-order nature of pWhile for proof automation via SMT solvers.
EasyCrypt extends pWhile with a module system \cite{DBLP:conf/ccs/BarbosaBGKS21} to support reasoning about abstract code as module parameters.
It integrates a quantitative unary logic with pRHL, and supports reasoning about complexity in terms of oracle calls \cite{DBLP:conf/ccs/BarbosaBGKS21}.
Both automation and these kind of properties are out of scope for our work but would be interesting future directions.

% FCF
% - shallowly embedded free monad
% - interpreted as subdistribution monad
% - operational semantics based on a list ℒ of presampled bits
% - adequacy wrt usual denotational semantics as |ℒ|→∞
% - swapping via comp_spec_eq_swap, evalDist_commute
In FCF \cite{PetcherM15}, programs are written as Coq expressions in the free subdistributions monad.
Proofs are conducted in a pRHL-like logic, where
%In this simpler denotational model,
successive sampling statements can be swapped thanks to the commutativity of the monad.
% The usual interpretation of programs in the subdistribution monad is shown equivalent to a deterministic semantics based on distributions over a list of presampled bits.
% Successive sampling statements can be swapped thanks to the commutativity of the monad.

% SSProve
% - shallow embedding: free monad st+prob as syntax to work in the model
% - gets interpreted into the standard model also used by CertiCrypt
% - swapping allowed so long as a semantic invariant is satisfied
SSProve \cite{DBLP:conf/csfw/AbateHRMWHMS21,ssprove:long} supports modular crypto proofs by composing ``packages'' of programs written in the free monad for state and probabilities.
The swap rule in SSProve allows exchanging commands which maintain a state invariant. Reasoning about dynamically allocated local state is not supported.

IPDL \cite{10.1145/3571223} is a process calculus for stating and proving cryptographic observational equivalences.
IPDL is mechanized in Coq and targeted at equational reasoning about interactive message-passing in high-level cryptographic protocol models, and hence considers a different set of language features.
% they do have channels which we don't, so let's not claim that we do more

%%% Local Variables:
%%% mode: latex
%%% TeX-master: "../clutch-popl24"
%%% End:

\section{Conclusion}\label{sec:conclusion}
We have presented \thereflog{}, a novel higher-order probabilistic relational separation logic with support for asynchronous probabilistic coupling-based proofs of contextual refinement and equivalence of probabilistic higher-order programs with local state and impredicative polymorphism.
We have proved the soundness of \thereflog{} formally in Coq using a range of new technical concepts and ideas such as \emph{left-partial couplings}, \emph{presampling tapes}, and a \emph{coupling modality}.
We have demonstrated the usefulness of our approach through several example program equivalences that, to the best of our knowledge, were not possible to establish with previous methods.

%Future work includes extending the ideas of \thereflog{} to concurrency and other probabilistic properties.

%%% Local Variables:
%%% mode: latex
%%% TeX-master: "../clutch-popl24"
%%% End:

\section*{Data Availability Statement}
The Coq formalization accompanying this work is available on Zenodo \cite{zenodo:clutch} and on GitHub at \href{https://github.com/logsem/clutch/tree/popl24}{https://github.com/logsem/clutch}.

%% Acknowledgments
\begin{acks}
  This work was supported in part by a Villum Investigator grant (no. 25804),
  Center for Basic Research in Program Verification (CPV), from the VILLUM
  Foundation.
  This work was co-funded by the European Union (ERC, CHORDS, 101096090).
  Views and opinions expressed are however those of the author(s) only and do not necessarily reflect
  those of the European Union or the European Research Council. Neither the European Union nor
  the granting authority can be held responsible for them.
  This material is based upon work supported in part by the \grantsponsor{NSF}{National Science Foundation}{} under Grant No.~\grantnum{NSF}{2318724}.
\end{acks}

%% Bibliography
\bibliography{refs}

% %% Appendix
%\pagebreak
\appendix

\section{Model of \thereflog{}}\label{sec:app-model}

The value interpretation of types is shown in \cref{fig:rel-interp}.

\begin{figure*}[t]
\centering
\begin{align*}
  \semInterp{\alpha}{\Delta}{\val_1,\val_2} &\triangleq \Delta(\alpha)(\val_{1}, \val_{2}) \\
  \semInterp{\tunit}{\Delta}{\val_1,\val_2} &\triangleq \val_1 = \val_2 = ()\\
  \semInterp{\tint}{\Delta}{\val_1,\val_2} &\triangleq \Exists z\in \integer . \val_1 = \val_2 = z \\
  \semInterp{\tnat}{\Delta}{\val_1,\val_2} &\triangleq \Exists n\in \nat . \val_1 = \val_2 = n \\
  \semInterp{\tbool}{\Delta}{\val_1,\val_2} &\triangleq \Exists b\in \bool . \val_1 = \val_2 = b\\
  \semInterp{\type\to\typeB}{\Delta}{\val_1,\val_2} &\triangleq
  \always{}(\All \valB_1, \valB_2 . \semInterp{\type}{\Delta}{\valB_1,\valB_2} \wand \refines{\Delta}{\val_1~\valB_1}{\val_2~\valB_2}{\typeB}{})\\
  \semInterp{\type\times\typeB}{\Delta}{\val_1,\val_2} &\triangleq \Exists \valB_1,\valB_1',\valB_2,\valB_2'. (\val_1=(\valB_1,\valB_1')) \sep  (\val_2=(\valB_2,\valB_2'))
    \sep \semInterp{\type}{\Delta}{\valB_1,\valB_2} \sep \semInterp{\typeB}{\Delta}{\valB_1',\valB_2'} \\
  \semInterp{\type+\typeB}{\Delta}{\val_1,\val_2} &\triangleq
 \Exists \valB_1,\valB_2 . (\val_1=\Inl(\valB_1) \sep \val_2=\Inl(\valB_2) \sep \semInterp{\type}{\Delta}{\valB_1,\valB_2})~\lor\\
                                           &{}\qquad (\val_1=\Inr(\valB_1) \sep \val_2=\Inr(\valB_2) \sep \semInterp{\typeB}{\Delta}{\valB_1,\valB_2}) \\
  \semInterp{\tmu \alpha. \type}{\Delta}{\val_1,\val_2} &\triangleq \left( \MU R .
                                                        \Lam (\val_1,\val_2) .
                                                        \Exists \valB_{1}, \valB_{2} . (\val_{1} = \fold \valB_{1}) \sep (\val' = \fold \valB_{2}) \sep \later \semInterp{\type}{\Delta, \alpha \mapsto R}{\valB_1,\valB_2}\right)(\val_1,\val_2)\\
  \semInterp{\All \alpha. \type}{\Delta}{\val_1,\val_2} &\triangleq \always{} \left(\All R.  (\refines{\Delta, \alpha \mapsto R}{\val_1~\_}{\val_2~\_}{\type}{} \right) \\
  \semInterp{\Exists \alpha. \type}{\Delta}{\val_1,\val_2} &\triangleq \Exists R, \valB_{1}, \valB_{2} . (\val_{1} = \Pack \valB_{2}) \sep (\val_{2} = \Pack \valB_{2}) \sep
                                                     \semInterp{\type}{\Delta, \alpha\mapsto R}{\valB_1,\valB_2} \\
  \semInterp{\tref{\type}}{\Delta}{\val_1,\val_2} &\triangleq \Exists \loc_1, \loc_2. (\val_1 = \loc_1) \sep (\val_2 = \loc_2) \sep
                                                    \knowInv{\namesp.\loc_1.\loc_2}{\Exists \valB_1,\valB_2. \progheap{\loc_1}{\valB_1} \sep \specheap{\loc_2}{\valB_2} \sep \semInterp{\type}{\Delta}{\valB_1,\valB_2} } \\
  \semInterp{\ttape}{\Delta}{\val_{1}, \val_{2}} &\triangleq \Exists \lbl_{1}, \lbl_{2}, \tapebound . (\val_{1} = \lbl_{1}) \sep (\val_{2} = \lbl_{2}) \sep
                                                  \knowInv{\namesp.\lbl_{1}.\lbl_{2}}{\progtape{\lbl_{1}}{\tapebound}{\nil} \sep \spectape{\lbl_{2}}{\tapebound}{\nil}}
\end{align*}
\caption{Relational interpretation of types}
\label{fig:rel-interp}
\end{figure*}

The full definition of the weakest precondition, including the fancy update modality and invariant masks, is the guarded fixpoint of the equation found below.
\begin{align*}
  \wpre{\expr_{1}}[\mask]{\Phi} \eqdef{}
  & (\expr_{1} \in \Val \land \pvs[\mask] \Phi(\expr_{1})) \lor{} \\
  & (\expr_{1} \not\in \Val \land \All \state_{1}, \cfg_{1} .
    \stateinterp(\state_{1}) \sep \specinterp(\cfg_{1}) \wand \pvs[\mask][\emptyset] \\
  & \quad \execCoupl(\expr_{1}, \sigma_{1}, \cfg_{1})(\Lam \expr_{2}, \state_{2}, \cfg_{2} .
    \later \pvs[\emptyset][\mask] \stateinterp(\state_{2}) \sep \specinterp(\cfg_{2}) \sep \wpre{\expr_{2}}[\mask]{\Phi}))
\end{align*}

We define an auxiliary partial execution distribution $\exec_{n}(\expr, \state) \in \Distr{\Conf}$.
\begin{align*}
  \exec_{n}(\expr, \state) \eqdef{}
  \begin{cases}
    \mret(\expr, \state) & \text{if}~\expr \in \Val~\text{or}~n = 0 \\
    \stepdistr(\expr, \state) \mbindi{} \exec_{(n - 1)} & \text{otherwise}
  \end{cases}
\end{align*}

The coupling modality used in the definition of the weakest precondition is defined below.
Simple rules or ``constructors'' that follow by unfolding are found in \cref{fig:execcoupl-rules}.
\begin{align*}
  &\execCoupl(\expr_{1}, \sigma_{1}, {\expr_{1}}', {\state_{1}}')(Z) \eqdef \MU \Psi : \Conf \times \Conf \to \iProp .  \\  
  & \quad (\Exists R .
    \begin{aligned}[t]
      &\red(\expr_{1}, \state_{1}) \sep \Rcoupl{\stepdistr(\expr_{1}, \state_{1})}{\stepdistr({\expr_{1}}', {\state_{1}}')}{R} \sep
        \All \cfg_{2}, {\cfg_{2}}' . R(\cfg_{2}, {\cfg_{2}}') \wand \pvs[\emptyset] Z (\cfg_{2}, {\cfg_{2}}')) \lor{} \\
    \end{aligned} \\
  & \quad (\Exists R .
    \begin{aligned}[t]
      &\red(\expr_{1}, \state_{1}) \sep \Rcoupl{\stepdistr(\expr_{1}, \state_{1})}{\mret({\expr_{1}}', {\state_{1}}')}{R} \sep
        \All \cfg_{2} . R(\cfg_{2}, ({\expr_{1}}', {\state_{1}}')) \wand \pvs[\emptyset] Z (\cfg_{2}, ({\expr_{1}}', {\state_{1}}'))) \lor{} \\
    \end{aligned} \\
  & \quad(\Exists R .
    \Rcoupl{\mret(\expr_{1}, \state_{1})}{\stepdistr({\expr_{1}}', {\state_{1}}')}{R} \sep{} 
    \All {\cfg_{2}}' . R((\expr_{1}, \state_{1}), {\cfg_{2}}') \wand \pvs[\emptyset] \Psi ((\expr_{1}, \state_{1}), {\cfg_{2}}')) \lor{} \\
  & \quad\hspace{-0.2em} % to fix the alignment...?
    \left(
    \bigvee\displaystyle_{\lbl \in \state_{1}}
    \Exists R  .
    \Rcoupl{\statestepdistr_{\lbl}(\state_{1})}{\stepdistr({\expr_{1}}', {\state_{1}}')}{R} \sep
    \All \state_{2}, {\cfg_{2}}' . R(\state_{2}, {\cfg_{2}}') \wand \pvs[\emptyset] \Psi ((\expr_{1}, \state_{2}), {\cfg_{2}}')
    \right) \lor{} \\
  & \quad\hspace{-0.2em} % to fix the alignment...?
    \left(
    \bigvee\displaystyle_{\lbl' \in \state_{2}}
    \Exists R  .
    \Rcoupl{\stepdistr(\expr_{1}, \state_{1})}{\statestepdistr_{\lbl'}({\state_{1}}')}{R} \sep
    \All \cfg_{2}, {\state_{2}}' . R(\cfg_{2}, {\state_{2}}') \wand \pvs[\emptyset] Z (\cfg_{2}, ({\expr_{1}}', {\state_{2}}'))
    \right) \lor{} \\  
    & \quad\hspace{-0.2em} % to fix the alignment...?
      \left(
      \bigvee\displaystyle_{(\lbl, \lbl') \in \state_{1} \times {\state_{1}}'}
      \Exists R  .
      \Rcoupl{\statestepdistr_{\lbl}(\state_{1})}{\statestepdistr_{\lbl'}({\state_{1}}')}{R} \sep
      \All \state_{2}, {\state_{2}}' . 
      (\state_{2}, {\state_{2}}') \wand \pvs[\emptyset] \Psi ((\expr_{1}, \state_{2}), ({\expr_{1}}', {\state_{2}}'))
      \right) 
\end{align*}
\begin{figure*}
  \centering
  \begin{mathpar}
    \infer
    {\red(\cfg_1) \\
      \Rcoupl{\stepdistr(\cfg_1)}{\stepdistr(\cfg_1')}{R} \\
      \All \cfg_{2}, {\cfg_{2}}' . R(\cfg_{2}, {\cfg_{2}}') \wand \pvs[\emptyset] Z(\cfg_{2}, {\cfg_{2}}')
    }
    {\execCoupl(\cfg_1, \cfg_1')(Z)}
    \and
    \infer
    { \red(\cfg_1) \\
      \Rcoupl{\stepdistr(\cfg_{1})}{\mret(\cfg'_1)}{R} \\
      \All \cfg_{2} . R(\cfg_{2}, \cfg'_{1}) \wand \pvs[\emptyset] Z(\cfg_{2}, {\cfg_{1}}')
    }
    {\execCoupl(\cfg_{1}, {\cfg_{1}}')(Z)}
    \and
    %     
    % \infer
    % { \Rcoupl{\mret(\cfg_1)}{\exec_n(\cfg'_{1})}{R} \\
    %   \All {\cfg_{2}}' . R(\cfg_{1}, {\cfg_{2}}') \wand \pvs[\emptyset] \execCoupl(\cfg_{1}, {\cfg_{2}}')(Z)
    % }
    % {\execCoupl(\cfg_{1}, {\cfg_{1}}')(Z)}
    % 
    % \and
    %
    \infer
    {\Rcoupl{\mret(\cfg_1)}{\stepdistr(\cfg'_{1})}{R} \\
      \All {\cfg_{2}}' . R(\cfg_{1}, {\cfg_{2}}') \wand \pvs[\emptyset] \execCoupl(\cfg_{1}, {\cfg_{2}}')(Z)
    }
    {\execCoupl(\cfg_{1}, {\cfg_{1}}')(Z)}
    \and
    \infer
    {\Rcoupl{\statestepdistr_{\lbl}(\state_{1})}{\stepdistr({\cfg_{1}}')}{R} \\
      \All \state_{2}, {\cfg_{2}}' . R(\state_{2}, {\cfg_{2}}') \wand \pvs[\emptyset] \execCoupl((\expr_{1}, \state_{2}), {\cfg_{2}}')(Z)
    }
    {\execCoupl((\expr_{1}, \state_{1}), {\cfg_{1}}')(Z)}
    \and
    \infer
    { \red(\cfg_1) \\
      \Rcoupl{\stepdistr(\cfg_{1})}{\statestepdistr_{\lbl}(\state'_{1})}{R} \\
      \All \cfg_{2}, {\state_{2}}' . R(\cfg_{2}, {\state_{2}}') \wand \pvs[\emptyset] Z(\cfg_2, ({\expr_{1}}', {\state_{2}}'))
    }
    {\execCoupl(\cfg_1, ({\expr_{1}}', {\state_{1}}'))(Z)}
    \and
    \infer
    {\Rcoupl{\statestepdistr_{\lbl}(\state_{1})}{\statestepdistr_{\lbl'}(\state'_{1})}{R} \\
      \All \state_{2}, {\state_{2}}' . R(\state_{2}, {\state_{2}}') \wand \pvs[\emptyset] \execCoupl((\expr_{1}, \state_2), ({\expr_{1}}', {\state_{2}}'))(Z)
    }
    {\execCoupl((\expr_1, \state_1, ({\expr_{1}}', {\state_{1}}'))(Z)}        
  \end{mathpar}  
  \caption{$\execCoupl$ unfolding rules.}
  \label{fig:execcoupl-rules}
\end{figure*}

%%% Local Variables:
%%% mode: latex
%%% TeX-master: "../appendix"
%%% End:

\section{On Case Studies and Additional Examples}
\label{sec:app-additional-examples}

\subsection{Sangiorgi and Vignudelli's ``copying'' example}
\label{sec:bisim}

Sangiorgi and Vignudelli prove a subtle contextual equivalence mixing probabilistic choice, local references, and recursion using environmental bisimulations
\cite{DBLP:conf/popl/SangiorgiV16}.
Under call-by-value evaluation, $\lambda$-abstraction fails to distribute over probabilistic choice. This is contrary to call-by-name, and can easily be seen by considering the terms $I$ and $L$ in \cref{fig:bisim}. When evaluated in context $(\Lam f. f () = f ()) [\,\cdot\,]$, $I$ returns $\True$ (and $\False$) with probability $\frac 1 2$, while $L$ returns $\True$ with probability $1$.
The non-linear use of $f$ in the context is characteristic of examples that behave differently under call-by-name and call-by-value. The equivalence of $K$ and $H$ is achieved by prohibiting such a ``copying'' use  by exploiting local state.

The environmental bisimulation technique developed in \cite{DBLP:conf/popl/SangiorgiV16} is sufficiently powerful to prove the equivalence as it works directly with the resulting distributions, but, to our knowledge, previous attempts at a proof working abstractly with programs via logical relations were not successful \cite[Sec.~1.5]{Bizjak:phd}.
%why? exactly because it was not possible to relate the different sampling points?

Intuitively, $K$ and $H$ should be equivalent despite the fact that abstraction does not distribute over probabilistic choice because the closures they return are protected by a counter that only allows them to be run once. On the first call, both have equal probability of returning $\True$ or $\False$. On subsequent calls, the counter $x$ ensures that they both diverge.

\begin{figure}
  \begin{minipage}[t]{0.5\linewidth}
    \begin{align*}
      L \eqdef{} & (\Lam \_. \True) \oplus (\Lam \_. \False) \\
      I \eqdef{} & \Lam \_. (\True \oplus \False) \\
      \\
      K \eqdef{} & \Let x = \Alloc 0 in (\Lam \_. M) \oplus (\Lam \_. N) \\
      H \eqdef{} & \Let x = \Alloc 0 in \Lam \_. (M \oplus N) \\
      H_\lbl \eqdef{}
                 & \Let x = \Alloc 0 in \\
                 & \Let \lbl = \AllocTape 1 in  \\
                 & \Lam \_. (M \oplus_\lbl N)
    \end{align*}
  \end{minipage}\hfill
  \begin{minipage}[t]{0.5\linewidth}
    \begin{align*}
      % \shortintertext{where}
      M \eqdef{} & \If \deref x = 0 then x \gets 1 ; \True \Else \Omega \\
      N \eqdef{} & \If \deref x = 0 then x \gets 1 ; \False \Else \Omega \\
      \Omega \eqdef{} & (\Rec f x = f x) () \\
      e_1 \oplus\phantom{_\lbl} e_2 \eqdef{} & \If \Flip () then e_1 \Else e_2 \\
      e_1 \oplus_\lbl e_2 \eqdef{} & \If \Flip(\lbl) then e_1 \Else e_2
    \end{align*}
  \end{minipage}
  \caption{\citeauthor{DBLP:conf/popl/SangiorgiV16}'s example.}
  \label{fig:bisim}
\end{figure}

The key insight that allows us to prove $K$ and $H$ contextually equivalent in \thereflog{} is to establish an asynchronous coupling between the two $\Flip$ operations.
Similarly to the proof of the lazy/eager coin example, we employ an intermediary version $H_\lbl$ of the program $H$ in which the sampling is delayed until the closure is run. The equivalence of $H_\lbl$ and $H$ follows from \ruleref{rel-rand-erase-r} and the symbolic execution rules.

The refinement $\refines{\emptyset}[\top]{H_\lbl}{K}{\tunit \to \tbool}$ is established by allocating a tape $\lbl$ and coupling the (eager) $\Flip ()$ in $K$ with the tape $\lbl$.
Because allocation of $\lbl$ is local to $H_\lbl$, we obtain exclusive ownership of the tape resource $\progtape \lbl 1 \nil$.
In particular, other parts of the program, \ie{} the context in which $H_\lbl$ is evaluated in, cannot sample to or consume bits from $\lbl$.
By \ruleref{rel-couple-tape-l}, we resolve the $\Flip()$ to $b$ in $K$ and obtain $\progtape{\lbl}{1}{b}$ for $H_\lbl$.
We then allocate the non-atomic invariant:
\begin{align*}
  (\progtape{\lbl}{1}{b} \sep \specheap{x}{0} \sep \progheap{x}{0}) \lor (\progtape{\lbl}{1}{\nil} \sep \specheap{x}{1} \sep \progheap{x}{1})
\end{align*}
The invariant describes the two possible states of the programs. Either the closures returned by $K$ and $H_\lbl$ have not been run yet, in which case the presampled bit $b$ is still on tape $\lbl$ and the counter $x$ is $0$ in both programs, or the bit has been consumed, and the counter is $1$ in both programs.
It is worth noting here that we will rely crucially on a form of local state encapsulation for tapes, which guarantees that once $b$ has been read from $\lbl$, the tape remains empty. We only consider the case where $b = \True$ ; the other case is analogous.

With the invariant in hand, we apply the proof rule for functions to work on the bodies of the two closures.
As a first step, we open our invariant, and are left to prove the equivalence in both cases of the disjunction.
By virtue of the non-atomic nature of the invariant, we can keep it open for several steps of evaluation, involving $\Flip$, pure reductions, and state-manipulating operations, until it is finally reestablished.

In the first case, we read $b$ from $\lbl$, yielding $\progtape{\lbl}{1}{\nil}$. We are left to prove the refinement of two structurally equal programs:
\begin{equation*}
  \If \deref x = 0 then x \gets 1 ; \True \Else \Omega
\end{equation*}
We take the first branch and set $x$ to $1$. We have now reproven the invariant, and both programs return $\True$ and conclude.

In the second case of the invariant, the closures have been invoked before, and we expect them to both diverge. However, before evaluation reaches $\Omega$ in $H_\lbl$, another $\Flip(\lbl)$ has to be resolved.
Here we exploit the fact that the $\lbl$ tape remains empty once we read $b$, as it is local to $H_\lbl$.
Logically, this observation manifests in the fact that after the allocation of $\lbl$, its ownership has been transferred into the invariant, and is now reclaimed.
We can thus use \ruleref{rel-rand-tape-empty-l} to resolve the $\Flip$ on an empty tape to a new random bit $b'$.
% NB: this is why it's important that \lbl is abstract: the tape is still empty.
Irrespectively of the value of $b'$, both programs diverge because we know that $\specheap{x}{1}$ and $\progheap{x}{1}$.
A diverging term refines any other term; in particular we appeal to \ruleref{rel-rec} to conclude the proof.

\subsection{ElGamal security proof}
% Copied from main paper, remove if paper and appendix in one document
\renewcommand{\keygen}{\mathit{keygen}}
\renewcommand{\encrypt}{\mathit{enc}}
\renewcommand{\decrypt}{\mathit{dec}}
\renewcommand{\query}{\mathit{query}}
\renewcommand{\dhreal}{\mathit{DH_{real}}}
\renewcommand{\dhrand}{\mathit{DH_{rand}}}
\renewcommand{\pkots}{\mathit{PK}}
\renewcommand{\tpkots}{\tau{}_{\mathit{\pkots}}}
\renewcommand{\tdh}{\tau{}_{\mathit{DH}}}
\renewcommand{\pkotsreal}{\pkots\mathit{_{real}}}
\renewcommand{\pkotsrealtape}{\pkots^{\mathit{tape}}_\mathit{real}}
\renewcommand{\pkotsrand}{\pkots\mathit{_{rand}}}
\renewcommand{\pkotsrandtape}{\pkots^{\mathit{tape}}_\mathit{rand}}

\begin{figure}
  \begin{minipage}[t]{0.33\linewidth}
    \makebox[.8\textwidth]{$\pkotsrand$} $\ctxeqrel$
    \begin{align*}
      \\
      \\
      &\smash{\Let sk = \Rand(n) in }\\
      &\smash{\Let pk = g^{sk} in }\\
      \\
      &\smash{\Let count = \Alloc 0 in }\\
      &\smash{\DumbLet \query = \Lam\ msg. }\\
      &\smash{\quad\If \deref{count} \neq 0 then }\\
      &\smash{\quad\quad\None }\\
      &\smash{\quad\Elsenospac }\\
      &\smash{\quad\quad count \gets 1 ; }\\
      &\smash{\quad\quad \Let b = \Rand(n) in }\\
      %&\smash{\quad\quad \Let B = g^b in }\\
      \\
      &\smash{\quad\quad \Let x = \Rand(n) in }\\
      \\
      %&\smash{\quad\quad \Let X = g^x in }\\
      &\smash{\quad\quad \Let (B, X) = (g^b, g^x) in }\\
      &\smash{\quad\quad \Some\ (B, X) }\\
      &\smash{\In (pk, \query)}
    \end{align*}
  \end{minipage}\hfill
  \begin{minipage}[t]{0.33\linewidth}
    \makebox[.8\textwidth]{$\pkotsrandtape{}$} $\ctxeqrel$
    \begin{align*}
      &\smash{\mhl{\Let \beta = \AllocTape(n) in} }\\
      &\smash{\mhl{\Let \gamma = \AllocTape(n) in} }\\
      &\smash{\Let sk = \Rand(n) in }\\
      &\smash{\Let pk = g^{sk} in }\\
      \\
      &\smash{\Let count = \Alloc 0 in }\\
      &\smash{\DumbLet \query = \Lam\ msg. }\\
      &\smash{\quad \If \deref{count} \neq 0 then }\\
      &\smash{\quad\quad \None }\\
      &\smash{\quad \Elsenospac }\\
      &\smash{\quad\quad count \gets 1 ; }\\
      &\smash{\quad\quad \Let b = \mhl{\Rand(n, \beta)} in }\\
      &\smash{\quad\quad \Let B = g^b in }\\
      &\smash{\quad\quad \Let c = \mhl{\Rand(n, \gamma)} in }\\
      &\smash{\quad\quad \mhl{\Let C = g^c in} }\\
      &\smash{\quad\quad \Let X = \mhl{msg \cdot{} C} in }\\
      &\smash{\quad\quad \Some\ (B, X) }\\
      &\smash{\In (pk, \query)}
    \end{align*}
  \end{minipage}\hfill
  \begin{minipage}[t]{0.33\linewidth}
    \makebox[.8\textwidth]{$\ctx[\dhrand]$}
    \begin{align*}
      &\smash{\DumbLet (pk,\mhl{B,C}) = }\\
      &\smash{\quad\Let a = \Rand(n) in }\\
      &\smash{\quad\mhl{\Let b = \Rand(n) in} }\\
      &\smash{\quad\mhl{\Let c = \Rand(n) in} }\\
      &\smash{\quad(g^a, \mhl{g^b, g^c}) \In }\\
      &\smash{\Let count = \Alloc 0 in }\\
      &\smash{\DumbLet \query = \Lam\ msg. }\\
      &\smash{\quad \If \deref{count} \neq 0 then }\\
      &\smash{\quad\quad \None }\\
      &\smash{\quad \Elsenospac }\\
      &\smash{\quad\quad count \gets 1 ; }\\
      \\
      \\
      \\
      \\
      &\smash{\quad\quad \Let X = msg \cdot{} C in }\\
      &\smash{\quad\quad \Some\ (B, X) }\\
      &\smash{\In (pk, \query)}
    \end{align*}
  \end{minipage}
  \caption{The ``rand'' direction of the security reduction.}
  \label{fig:rand-reduction}
\end{figure}

As stated in the main paper, the proof of $\ctxeq{}{\pkotsrand}{\ctx[\dhrand]}{\tpkots}$ is similar to that of $\ctxeq{}{\pkotsreal}{\ctx[\dhreal]}{\tpkots}$.
The sequence of games is displayed in \cref{fig:rand-reduction}.
Compared to $\pkotsrand$, the $\pkotsrandtape$ game samples from tapes $\beta$ and $\gamma$.
This difference is immaterial, but the tapes will be used to connect the lazily-sampling  $\pkotsrandtape$ to the eagerly-sampling $\ctx[\dhrand]$.
More interestingly, the $X$ component of the ciphertext is computed directly as a random group element $g^x$ in $\pkotsrand$, whereas $\pkotsrandtape$ multiplies the message $msg$ with a random group element $C$.
We have to justify that the two games are nonetheless equivalent.
Intuitively speaking, since $C$ is uniformly distributed, so is $msg \cdot C$, and hence $X$ is a group element sampled from a uniform distribution in both games.
The rigorous explanation for this argument hinges on the fact that multiplication with $msg$ induces a bijection on the set $\{0,1,\ldots,n\}$.

Recall that the group $G$ is of order $n+1$.
Since $G$ is generated by $g$, we can thus write $msg = g^k$ for some $k \in \{0,1,\ldots,n\}$.
We can therefore regroup the exponents in $msg\cdot{}C = g^k\cdot{}C = g^k \cdot g^c$ as $g^{k + c}$.
Since the function $f \eqdef \Lam x . (x - k) \bmod (n+1)$ is a bijection on $\{0,1,\ldots,n\}$, we can couple the sampling of $x$ in $\pkotsrand$ with the sampling of $c$ in $\pkotsrandtape$ to obtain some value $r$ for $x$ in $\pkotsrand$ and $f(r)$ for $c$ in $\pkotsrandtape$.
By definition of $f$, the value computed for $X$ in $\pkotsrandtape$ is then equal to $X = msg\cdot{}C = g^{k + c} = g^{k + f(r)} = g^{k + r - k} = g^r$, just as in $\pkotsrand$, and the two programs are equivalent.
This argument is a standard ingredient in the security proof of ElGamal, and part of our formalization.

The main difference between $\pkotsrandtape$ and $\ctx[\dhrand]$ is that the $\query$ function in $\pkotsrandtape$ samples $b$ and $c$ lazily, whereas in $\ctx[\dhrand]$, the sampling of $b$ and $c$ occurs eagerly in the beginning.
Once again, the proof proceeds in a manner similar to the lazy-eager coin example.

\subsection{Eager/Lazy Hash Function}
% Copied from main paper, remove if combined
\renewcommand{\eagerhash}{\mathit{eager\_hash}}
\renewcommand{\lazyhash}{\mathit{lazy\_hash}}
\renewcommand{\initmap}{\mathit{init\_map}}
\renewcommand{\getmap}{\mathit{get}}
\renewcommand{\setmap}{\mathit{set}}
\renewcommand{\sampleall}{\mathit{sample\_all}}
\renewcommand{\alloctapes}{\mathit{alloc\_tapes}}
\renewcommand{\inithashrng}{\mathit{init\_hash\_rng}}
\renewcommand{\maxint}{\texttt{MAX}}
\renewcommand{\initboundedrng}{\mathit{init\_bounded\_rng}}

\newcommand{\eagerhashcode}{
  \begin{minipage}[t]{.49\linewidth}
    \begin{align*}
      \sampleall \eqdef{}
      &\Rec f m\spac n = \\
      &\Let n' = n - 1 in\\
      &\If n' < 0 then \TT \Else \\
      &\quad \Let b = \Flip \TT in \\
      &\quad \setmap\ m\ n'\ b; \\
      &\quad f\ m\ n' \\
    \end{align*}
  \end{minipage}
  \hfill
  \begin{minipage}[t]{.49\linewidth}
    \begin{align*}
      \eagerhash \eqdef{}
      &\Lam n. \\
      &\Let m = \initmap\ () in \\
      &\sampleall\ m\ (n + 1); \\
      &(\Lam k.
        \MatchML \getmap\ m\ k with
        \Some(b) => b
        | \None => \False
        end) {}
    \end{align*}
  \end{minipage}
}

\newcommand{\lazyhashcode}{
\begin{align*}
  \alloctapes \eqdef{}
  &\Rec f m\spac n = \\
  &\Let n' = n - 1 in\\
  &\If n' < 0 then \TT \Else \\
  &\quad \Let \lbl = \AllocTape in \\
  &\quad \setmap\ m\ n'\ \lbl; \\
  &\quad f\ m\ n' \\
  \lazyhash \eqdef{}
  &\Lam n. \\
  &\Let vm = \initmap\ \TT in \\
  &\Let tm = \initmap\ \TT in \\
  &\alloctapes\ tm\ (n + 1); \\
  &(\Lam k.
  \MatchMLnarrow \getmap\ vm\ k with
  \Some(b) => b
  |\None =>
   {\MatchMLnarrow \getmap\ tm\ k with
    \Some(\lbl) =>
      {\begin{array}[t]{l}
        \Let b = \Flip(\lbl) in \\
        \setmap\ vm\ b; \\
        b
      \end{array}}
    | \None => \False
    end {}}
  end) {} \\
\end{align*}
}

\newcommand{\eagerhashfigure}{
 \begin{figure}
   \eagerhashcode
   \caption{Eager hash function.}
   \label{fig:eager-hash-app}
 \end{figure}
}
\newcommand{\lazyhashfigure}{
 \begin{figure}
   \lazyhashcode
   \caption{Lazy hash function.}
   \label{fig:lazy-hash-app}
 \end{figure}
}

As explained in the body of the paper, it is
common to model the hash function as if it satisfies the so-called
\emph{uniform hash assumption} or the \emph{random oracle model}.  That is, a hash function $h$ from a
set of keys $K$ to values $V$ behaves as if, for each key $k$, the
hash $h(k)$ is randomly sampled from a uniform distribution over $V$,
independently of all the other keys.

\eagerhashfigure{}

\cref{fig:eager-hash-app} gives the complete code for the eager
hash function that was excerpted earlier.
Given a non-negative integer $n$, executing $\eagerhash\ n$ returns a hash
function with $K = \{0, \dots, n\}$ and $V = \bool$. To do so, it
first initializes a mutable map $m$, and then calls $\sampleall$ on
$m$. For each key $k \in K$, the function $\sampleall$ samples a
boolean $b$ with $\Flip$ and stores the value $b$ for the key $k$ in
the map $m$. This sampled boolean serves as the hash for $b$. The
function returned by $\eagerhash$ uses this map to look up the hash
values of keys. On input $k$, it looks up $k$ in the map and returns
the resulting value if one is found, and otherwise returns
$\False$. Since $\sampleall$ adds every key in $K$ to the map, this
latter scenario only happens if $k$ is not in $K$.

\lazyhashfigure{}

\cref{fig:lazy-hash-app} gives the full code for the lazy sampling version of the random hash generator.
Given a non-negative integer $n$, executing $\lazyhash\ n$ returns a
hash function for the key space $K = \{0, \dots, n\}$. For its
internal state, it uses two physical maps, the tape map $tm$, stores
tapes to be used for random sampling, and the value map $vm$, stores
the previously sampled values for keys that have been hashed. After
initializing these maps, it calls $\alloctapes$, which allocates a
tape for each key $k \in K$ and stores the associated tape in $tm$.
The hash function returned by $\lazyhash$ determines the hash for a
key $k$ in two stages. It first looks up $k$ in $vm$ to see if $k$
already has a previously sampled hash value, and if so, returns the
found value. Otherwise, it looks up $k$ in the tape map $tm$. If no
tape is found, then $k$ must not be in $K$, so the function returns
$\False$. If a tape $\lbl$ is found, then the code samples a boolean
$b$ from this tape with $\Flip$, stores $b$ for the key $k$ in $vm$,
and then returns $b$.

We
also prove that for any non-negative number $n$, the eager and lazy
versions are contextually equivalent, that is,
$\ctxeq{}{\eagerhash\ n}{\lazyhash\ n}{\tint \rightarrow \tbool}$.
The core idea behind this contextual equivalence proof is to maintain
a particular invariant between the internal state of the two hash functions.
Let $m$ be the internal map used by the eager hash and let $tm$ and $vm$ be the tape
and value maps, respectively, for the lazy hash. Then, at a high level, the invariant
maintains the following properties:
\begin{enumerate}
  \item $\dom(m) = \dom(tm) = \{0, \dots, n\}$.
  \item For all $k \in \{0, \dots, n\}$, if $m[k] = b$ then either
    \begin{enumerate}
    \item $vm[k] = b$, or
    \item $vm[k] = \bot$ and  $tm[k] = \lbl$ for some tape label $\lbl$ such that $\progtape{\lbl}{[b]}$.
    \end{enumerate}
\end{enumerate}
Case (a) and (b) of the second part of this invariant capture the two possible states each key $k$ can be in.
Either hash of $k$ has been looked up before (case a), and so the sampled value stored in $vm$ must match that of $m$, or it has not been looked up (case b) and the tape for the key must contain the same value as $m[k]$ for its next bit.

To establish this invariant when the hashes are initialized, we
asynchronously couple the eager hash function's $\Flip$ for key $k$
with a tape step for the tape $\lbl$ associated with $k$ in the lazy
table.  The invariant ensures that the values returned by the two hash
functions will be the same when a key $k$ is queried.  The cases of
the invariant corresponding to the branches of the lazy function's
match statements: if the key $k$ is in $K$ and has been queried
before, the maps will return the same values found in $m$ and $vm$. If
it has not been queried before, then the $\Flip$ statement in the lazy
version will be draw the value on the tape for the key, which matches
$m[k]$. Moreover, the update that writes this sampled value to $vm$
preserves the invariant, switching from case (b) to case (a) for
queried key.

\subsection{Random Generators from Hashes}

Here we provide further details on generating a random boolean sampler from a lazy hash function.
The following function, $\inithashrng$, returns
a function that can be used to generate random booleans:
\begin{align*}
  \inithashrng \eqdef{}
  &\Lam \_. \\
  &\Let f = \lazyhash\ \maxint in \\
  &\Let c = \Alloc{0} in \\
  &(\Lam \_. \Let n = \deref{c} in \\
  &\quad \Let b = f\ n in \\
  &\quad c \gets n + 1; \\
  &\quad b)
\end{align*}
When run, $\inithashrng$ generates a lazy hash function $f$ for the key space $K = \{0, \dots, \maxint\}$
for some fixed constant $\maxint$. It then allocates a counter $c$ as a mutable reference initialized to $0$.
The returned function, let us call it $h$, uses $f$ and $c$ to generate random booleans. Each time
$h$ is called, it loads the current value $n$ form $c$, hashes $n$ with $f$ to get
a boolean $b$. It then increments $c$ and returns the boolean $b$, which serves as a random boolean.
Repeated calls to $h$ return independent, uniformly sampled booleans, so long as we make no more than $\maxint$ calls.
The reason this works is that we have assumed the hash function $f$ is uniformly random, so the hashes of different keys are independently sampled. So long as we make fewer than $\maxint$ calls to $h$, each call will hash a distinct number $n$ (the current counter value), so it will be independent of all previous and future calls. After $\maxint$ calls, $c$ will exceed $\maxint$ and so we will hash a key outside of $f$'s key space. Recall from the previous example that $\lazyhash$ returns $\False$ on inputs outside its key space.

This example might at first seem artificial, but using cryptographic
primitives such as hashes or block ciphers to generate pseudorandom
numbers is in fact commonly done~\cite{Barker2015}. Although the example
here is simplified compared to real implementations, it captures one
of the core verification challenges common to real implementations.
Namely, for correctness, one must show that the ``key'' or ``counter'' being
hashed or encrypted (here the values of $n$ obtained from $c$) are not
re-used.

To capture the guarantees of $\inithashrng$ more formally, we prove
that $\inithashrng$ is contextually equivalent to the following
``bounded'' random number generator that directly calls $\Flip$:
\begin{align*}
  &\initboundedrng \eqdef{} \\
  &\qquad \Lam \_. \\
  &\qquad \Let c = \Alloc{0} in \\
  &\qquad (\Lam \_. \Let n = \deref{c} in \\
  &\qquad \quad \Let b = \If n \leq \maxint then \Flip \TT 
    \Else \False in \\
  &\qquad \quad c \gets n + 1; \\
  &\qquad \quad b)
\end{align*}
With $\initboundedrng$, the returned generator function again uses a
counter $c$, however the value of this counter is just used to track
the number of samples generated. If the number of calls is less than or equal to $\maxint$,
it returns a boolean generated by a call to $\Flip$. Otherwise,
it just returns $\False$. (In \thelang, integers are unbounded, so there is no
issue with overflow.)

At a high level, the proof proceeds by maintaining the following
invariant relating the generator functions returned by both
$\inithashrng$ and $\initboundedrng$. Let $f$ be the hash function in the hash-based generator, $n_h$ be its counter value, and $n_b$ be the value of the bounded generator's counter. Then:
\begin{enumerate}
  \item $n_h = n_b$.
  \item The key space of $f$ is $\{0, \dots, \maxint\}$
  \item For all $k \in \{0, \dots, \maxint\}$, if $k \geq n_h$ then $f$ has not yet hashed $k$.
\end{enumerate}
The first and second parts of the invariant guarantees that once $n_h$
exceeds $\maxint$, and thus falls outside $f$'s key space, both
generators will return the same value of $\False$. In addition, based on the third part of
the invariant, so long as $n_h \in \{0, \dots, \maxint\}$
when the hash-based generator evaluates $f\ n_h$,  we will be able
to couple the hash value it samples with the $\Flip$ command in the bounded generator, ensuring
that both generators return the same value.

The $\inithashrng$ function above generates a single random generator
from a hash function. However, in some scenarios, it is necessary to
be able to generate multiple independent streams of random
numbers. For example, in a language with parallelism or concurrency,
using a single generator returned by $\inithashrng$ in multiple
threads would mean sharing mutable access to the counter $c$, so that
synchronization primitives would be needed to prevent racy accesses.
Related issues have motivated the need to ``split'' a random number generator
into two streams in the context of a lazy language like Haskell~\cite{ClaessenP13}.

While \thelang\ is sequential, we can still explore the question of
how to create multiple independent random generators. One approach
would be to call $\inithashrng$ multiple times. But that assumes that
we have the ability to initialize multiple random oracle hash
functions. In practice, if we instantiate the random oracle model with
a particular concrete hash function, like SHA-256, we cannot feasibly
use different hash functions each time $\inithashrng$ is
called. Instead, we would like a way to generate multiple independent
random number generators from a single hash function.

The solution is to get the illusion of multiple independent hash functions out of a single hash function by partitioning the key space of the hash. Specifically, we use a wrapper around the hash function so that it now takes \emph{two} integers as input, instead of one, to obtain a so-called \emph{keyed} hash\footnote{In cryptographic settings, a similar construction is called a hash message authentication code (HMAC). However, HMACs typically use a different way of combining the two arguments to avoid certain vulnerabilities.}:
\newcommand{\lazykeyedhash}{\mathit{lazy\_keyed\_hash}}
\begin{align*}
  \lazykeyedhash \eqdef{}
  &\Lam \_. \\
  &\Let f = \lazyhash\ (2^{(p_k + p_v)} - 1) in \\
  &(\Lam k\ v. f\ (k \cdot 2^{p_v} + v))
\end{align*}
The returned keyed hash function takes two inputs, a key $k$ and a
value $v$ to be hashed, combines them into a single integer, and calls
the lazy hash function $f$ on that single integer.  The values $p_k$
and $p_v$ are fixed constants that determine the range of the keys and
values that can be hashed.  If $h$ is the returned hash function, we
can treat the partially-applied functions $h\ k_1$ and $h\ k_2$ for
$k_1 \neq k_2$ as if they were two independent hash functions, so long
as we do not apply the functions to values $v$ that are larger than
$2^{p_v} - 1$.

Using this, we have verified a version of the hash-based random number
generator that supports splitting multiple independent generators out
of a single hash. Each generator has a distinct key $k$ its own
internal counter $c$. When a sample is requested, it reads the value
$n$ from the counter and computes the keyed hash of $k$ and $n$ to get
a boolean. To ensure that the keys used by the generators are
distinct, we assign keys using a shared counter that is incremented
every time a new generator is initialized. The complete details can be
found in the accompanying Coq development.

\subsection{Lazily Sampled Big Integers}
% Copied from main paper, remove if combined
\renewcommand{\samplelazy}{\mathit{sample\_lazy\_int}}
\renewcommand{\sampleint}{\mathit{sample32}}
\renewcommand{\sampleeager}{\mathit{sample256}}
\renewcommand{\getnext}{\mathit{get\_next}}
\renewcommand{\cmplist}{\mathit{cmp\_list}}
\renewcommand{\cmplazy}{\mathit{cmp\_lazy}}
\renewcommand{\cmpZ}{\mathit{cmp}}

Our last example is motivated by a data structure called a
treap~\cite{SeidelA96}.  A treap is a binary search tree
structure that relies on randomization to ensure with high probability
that the tree will be balanced.  One of the key aspects of the treap
is that every key that is inserted into the treap is first assigned a
random numerical \emph{priority}. During the insertion process, this
priority value is compared with the priorities of keys already in the
treap.  The exact details of this comparison process are not relevant
here; what is important to know is that, ideally, all of the assigned
priority values are different (that is, there are no \emph{collisions}
of priorities).  Thus, in analyzing the treap, it is common to treat
these priorities as if they are sampled from a continuous
distribution, such as the uniform distribution on the interval
$[0,1]$, to ensure that the probability of collisions is 0.
Eberl et al.~\cite{EberlHN20} have previously mechanized such an analysis
of treaps in Isabelle/HOL.

In actual implementations, the priorities are instead typically
represented with some fixed precision, say as an approximate floating
point number sampled from $[0, 1]$, or as an integer sampled uniformly
from some set $\{0, \dots, n\}$, so that there is some probability of
collision. However, in the latter case, as long as $n$ is big enough
relative to the number of keys added to the tree, the probability of a
collision can be kept low, and the performance properties of the treap
are preserved. The probability of a collision is an instance of the
well-known ``birthday problem''.

But in some scenarios, we may need to decide on $n$ without knowing in
advance how many keys will end up being added to the treap. If we err
on the conservative side by making $n$ very large, say $2^{256} - 1$,
the probability of a collision will be very low, but we will need to
use $256$ bits to store the priorities, which is wasteful if we end up
only storing a moderate number of nodes.

An alternative is to \emph{lazily} sample the integer that represents
the priority. The insight is that the actual numerical value of the priorities
is not relevant: the only operation that they must support is comparing two priorities
to determine if they are equal, and if not, which one is larger.
\cref{fig:lazy-int} gives an implementation of a lazily-sampled integer
A lazily-sampled integer is encoded as a pair of a tape label $\lbl$ and a linked list of length at most $N$,
where each node in the list represents a \emph{digit} of the integer in
base $B$, with the head of the list being the most significant digit.
For concreteness, here we consider $N = 8$ and $B = 2^{32}$, so that the encoded numbers
can be at most $2^{256}-1$. Rather than sampling all digits up front, we instead
only sample digits when needed as part of comparing a lazy integer to another.
\begin{figure}
  \begin{align*}
    \samplelazy \eqdef{} & \Lam \_. (\AllocTape, \Alloc \None) \\
    \\
    \getnext \eqdef{}
    &\Lam\ \lbl\ r. \\
    &\MatchML \deref{r} with
      \Some\ v => v
      | \None =>
      {\begin{array}[t]{l}
      \Let z = \sampleint\ \lbl in \\
      \Let next = \Alloc \None in \\
      r \gets \Some\ (z, next); \\
      (z, next) 
      \end{array}}
      end {} \\
    \\
    \cmplist \eqdef{}
    &\Rec f n\ \lbl_1\ l_1\ \lbl_2\ l_2 = \\
    &\If n = 0 then 0 \\
    &\Elsenospac \\
    &\quad\Let (z_1, l_1') = \getnext\spac \lbl_1\spac l_1 in \\
    &\quad\Let (z_2, l_2') = \getnext\spac \lbl_2\spac l_2 in \\ 
    &\quad\Let res = \cmpZ z_1\spac z_2 in \\
    &\quad\If res = 0 then \\
    &\quad\quad f\ (n-1)\ \lbl_1\ l_1'\ \lbl_2\ l_2' \\
    &\quad\Elsenospac\ res\\
    \\
    \cmplazy \eqdef{}
    &\Lam\ (x_1, x_2). \\
    &\Let (\lbl_1, l_1) = x_1 in \\
    &\Let (\lbl_2, l_2) = x_2 in \\
    &\If l_1 = l_2 then 0 \\
    &\Elsenospac \cmplist\ 8\ \lbl_1\ l_1\ \lbl_2\ l_2
  \end{align*}
  \caption{Implementation of lazily-sampled integers.}
  \label{fig:lazy-int}
\end{figure}

The function $\samplelazy$ samples a lazy integer by generating a tape
and a reference to an empty linked list.  At this point, the sampled
integer is entirely indeterminate. Given a tape label $\lbl$ and a
reference $r$ to a digit in a lazy integer's list, evaluating
$\getnext\ \lbl\ r$ returns the integer $z$ for that digit and a
reference $next$ to the digit after $r$ in the list. There are two
alternatives when getting the digit: either (1) the digit for $r$ has
already been sampled, so that $\deref{r}$ will be $\Some\ v$, where $v$
is a pair of the form $(z, next)$; or (2) the digit for $r$ has not
yet been sampled, so that $\deref{r}$ will be $\None$. In case 1,
$\getnext\ \lbl\ r$ just returns $v$.  In case 2, $\getnext$ will
first sample the value $z$ for the digit by calling
$\sampleint\ \lbl$, which generates a 32 bit integer by sampling it
bit-by-bit with repeated calls to $\Flip \lbl$.  It then allocates a
new reference $next$ for the next digit in the list, initialized to a
value $\None$.  Before returning $(z, next)$ it stores this pair in
the reference $r$.

Evaluating $\cmplist\ n\ \lbl_1\ l_1\ \lbl_2\ l_2$ compares the two
lazy integers $(\lbl_1, l_1)$ and $(\lbl_2, l_2)$ by doing a
digit-by-digit comparison. It returns $-1$ if the first integer is
smaller, $0$ if the integers are equal, and $1$ if the first integer
is larger. The first argument $n$ tracks the number of remaining
digits in the integers. Let us consider the case that $n > 0$ first
(the else branch). In that case, $\cmplist$ will call $\getnext$ on
each integer to get the next digit.  Let $z_1$ and $z_2$ be these
digit values, respectively.  These digits are compared using $\cmpZ$,
which returns $-1$ (if $z_1 < z_2$), $0$ (if $z_1 = z_2$), or $1$
($z_1 > z_2$). Because the lazy integers are stored with
most-significant digits earlier in the list, if $\cmpZ\ z_1\ z_2$ is
non-zero we already know which lazy integer is larger, and the result
of $\cmpZ\ z_1\ z_2$ gives the correct ordering of the whole lazy
integer. On the other hand if $\cmpZ$ returns $0$, then $z_1 = z_2$,
in which case we cannot yet tell which lazy integer is larger. Thus,
$\cmplist$ recursively calls itself to compare the next digits in the
lists, decrementing the $n$ argument to track that there is one fewer
digit remaining.  In the base case of the recursion, when $n =
0$, that means all digits of the integers have been equal, hence
the value of the integers are equal, so we return $0$.
Because $\getnext$ conveniently encapsulates the sampling of unsampled digits, $\cmplist$
looks like a normal traversal of the two linked lists, as if they were
eagerly sampled.

Note that if $x$ is a lazy integer, then comparing $x$ with itself
using $\cmplist$ unfortunately forces us to sample all of the
unsampled digits of $x$. The routine $\cmplazy$ is a wrapper to
$\cmplist$ that implements a small optimization to avoid this. The
function $\cmplazy$ takes as input a pair of lazy integers $(x_1,
x_2)$. Before calling $\cmplist$, it first checks whether the pointers
to the heads of $x_1$ and $x_2$'s lists are equal; if they are the two
integers must be equal, so it returns $0$ immediately without calling
$\cmplist$.

We prove that this implementation of lazily-sampled integers is contextually
equivalent to code that eagerly samples an entire 256-bit integer by bit-shifting and adding
8 32-bit integers. This contextual equivalence is at an \emph{abstract} existential type $\tau$. Specifically,
we define
\[
\tau \eqdef \Exists \alpha. (\tunit \rightarrow \alpha) \times ((\alpha \times \alpha) \rightarrow \tint)
\]
Then we have the following equivalence:
\[ \ctxeq{}{(\samplelazy, \cmplazy)}{(\sampleeager, \cmpZ)}{\tau} \]
The starting point for the proof is that when $\sampleeager$ samples
the 8 32-bit integers needed to assemble the 256-bit integer, we
couple these samples with identical samples on the tape $\lbl$
generated by $\samplelazy$.  Then, the key invariant used in the proof
says that if we combine the digits of a lazy int that have already been
sampled, plus the remaining digits on the tape $\lbl$, the result
represents an integer that is equivalent to the corresponding one
generated by $\sampleeager$. This holds initially and is preserved by
calls to $\getnext$ during $\cmplist$, since it moves digits from the
tape to the linked list representing the integer.

%%% Local Variables:
%%% mode: latex
%%% TeX-master: "../appendix"
%%% End:

\section{Rules}
\label{sec:app-rules}

We repeat the rules presented in the main paper for ease of reference.

\figstrucrules{}
\figreflogrules{}
\begin{figure*}
%  \small
  \centering
  \begin{mathpar}
    \relranderasel{}
    \and
    \relranderaser{}
  \end{mathpar}
  \caption{Tape erasure rules.\label{fig:erasure-rules}}
\end{figure*}
%NB: The erasure rules are derivable from the coupling + symbolic execution rules.
\figreflogtaperules{}
\figreflogcouplingrules{}
\figrefloginvariantrules{}

%%% Local Variables:
%%% mode: latex
%%% TeX-master: "../appendix"
%%% End:

\end{document}